\documentclass[reprint,superscriptaddress,amsmath,amssymb, aps,prd]{revtex4-2}

\usepackage{graphicx}
\usepackage{dcolumn}
\usepackage{bm}
\usepackage{amssymb}
\usepackage{pifont}
\usepackage{xcolor}
\usepackage{orcidlink}

\renewcommand{\arraystretch}{1.5} 

\begin{document}

\title{Correlation and Data-Analysis Distinctiveness of Time-Delay Interferometry Configurations}

\author{Gang Wang\,\orcidlink{0000-0002-9668-8772}}
\email[Gang Wang: ]{gwanggw@gmail.com, gwang@nbu.edu.cn}
\affiliation{Institute of Fundamental Physics and Quantum Technology, Ningbo University, Ningbo, 315211, China}
\affiliation{Department of Physics, School of Physical Science and Technology, Ningbo University, Ningbo, 315211, China}
\affiliation{Shanghai Astronomical Observatory, Chinese Academy of Sciences, Shanghai 200030, China}

\date{\today}

\begin{abstract}

Time-Delay Interferometry (TDI) is essential for space-based gravitational wave (GW) missions, as it suppresses laser frequency noise and achieve the required sensitivity. Beyond the standard Michelson configuration, a variety of second-generation TDI schemes have been proposed, each utilizing different combinations of inter-spacecraft laser links. In this work, we conduct a comparative study of several representative TDI configurations with different time spans, and show that while their (quasi-)orthogonal channels are highly correlated, their performance in data analysis can differ among these schemes. In the low-frequency regime, the performance of different TDI configurations are nearly identical. Their distinctions emerge primarily at high frequencies, where the GW wavelength becomes comparable to the arm length. In this regime, shorter TDI time spans with minimal null frequencies facilitate more accurate waveform modeling and parameter recovery in frequency domain. In contrast, configurations with longer time spans and more null frequencies, such as the Michelson, are more susceptible to frequency aliasing and waveform modulation effects, which degrade inference accuracy. However, if signal modeling and analysis are performed in the time domain, the optimal science channels of these TDI configurations exhibit consistent performance in parameter inference. Considering the usability in both frequency and time domain, the short-span PD4L scheme, which exhibits minimal nulls and superior performance in high frequencies, emerges as a promising candidate for future space-based GW mission designs.

\end{abstract}


\maketitle

\section{Introduction}

Time-delay interferometry (TDI) was developed to suppress the dominant laser frequency noise and thereby achieve the sensitivity requirements for space-borne gravitational wave (GW) interferometers \cite{1997SPIE.3116..105N,1999ApJ...527..814A,2000PhRvD..62d2002E}. It is indispensable for both milli-Hz missions such as LISA \cite{LISA:2017pwj,Colpi:2024xhw}, TAIJI \cite{Hu:2017mde}, and TianQin \cite{TianQin:2015yph} and sub-milli-Hz missions like ASTROD-GW \cite{Ni:2012eh}. The basic principle of TDI is to construct synthesized interferometric observables by appropriately delaying and combining laser phase measurements from different spacecraft (S/C), thereby emulating equal-arm interferometry.

First-generation TDI was designed for static, unequal-arm configurations and can effectively cancel laser noise in the absence of relative motion \cite{1999ApJ...527..814A}. However, in real mission scenarios, the orbital dynamics of the S/C cause time-varying arm lengths, requiring the development of second-generation TDI schemes to accommodate S/C motion \cite{Tinto:2003vj,Shaddock:2003dj}.
A wide variety of second-generation TDI configurations have been developed, involving different inter-S/C link combinations and delay operations \cite{Shaddock:2003dj,Tinto:2003vj,Vallisneri:2005ji,Wang:2011,Wang:2012fqs,Wang:2020pkk,Hartwig:2021mzw,Tinto:2022zmf}. Each configuration yields three ordinary observables through cyclic permutations of the S/C indices.
It is well understood that these observables are not independent. In the static equal-arm limit, the first-generation TDI variables can be expressed in terms of four Sagnac generators $(\alpha, \beta, \gamma, \zeta)$ \cite{1999ApJ...527..814A}, and this generator-based decomposition has been extended to second-generation TDI under the assumption of constant unequal arm lengths \cite{Hartwig:2021mzw}. Within this framework, correlations between different TDI constructions are therefore expected on general grounds.

Despite this underlying correlation, it is common to construct three optimal observables (A, E, and T) from three ordinary channels of a TDI configuration via an orthogonal transformation \cite{Prince:2002hp,Vallisneri:2007xa}. In the idealized limit, these observables are expected to be statistically independent, and the corresponding sets from different TDI configurations are often regarded as having equivalent performance. In particular, the two science channels (A and E) from different configurations can be related through a rotation in the orthogonal basis, while the null channels are fully correlated.
However, in realistic LISA-like scenarios, the arm lengths are dynamically unequal due to the relative motion between the spacecraft ($\sim 8\,\mathrm{m/s}$) and the rotation of the constellation (leading to differences between counter-propagating links of order $|L_{ij} - L_{ji}| \simeq 1.7\,\mathrm{ms}$). These effects break the exact symmetries assumed in the analytic construction and introduce residual correlations among the nominally orthogonal observables.
Moreover, different TDI configurations involve different effective time spans. While generator-based arguments neglect the delay range and suggest that analytically correlated configurations have equivalent capabilities, their practical performance in data analysis can differ once realistic effects, such as finite data duration, and waveform evolution, are taken into account.

The Michelson TDI configuration is widely adopted as a baseline for noise suppression and data analysis. However, under realistic orbital motion, its performance degrades due to the appearance of overmuch null frequencies and unstable noise properties \cite{Wang:2024alm,Wang:2024hgv}. To mitigate the effect of null frequencies, we previously implemented a hybrid Relay configuration with a time span of $8L$ (where $L$ is the light-travel time along one arm), which improves the signal extraction. However, this long time span becomes disadvantageous in the high-frequency regime, where it induces strong frequency aliasing and long signal tails that affect the accuracy of GW signal modeling in frequency domain.
To address these issues, we investigate the PD4L configuration, which has a reduced time span of $4L$. This shorter delay structure alleviates aliasing effects, shortens tails at the end of signal, and reduce data margins at the boundaries, thereby improving high-frequency signal modeling and parameter inference.

In this work, we examine a representative set of second-generation TDI configurations with varying time spans. Our aim is to demonstrate that optimal observables derived from different TDI schemes are highly correlated, especially in the low-frequency regime. In contrast, their high-frequency performance is sensitive to the time-delay structure and the presence of null frequencies. Longer time spans lead to more severe frequency aliasing and extended signal tails, while the null frequencies induces significant amplitude modulation. Therefore, TDI schemes with short time span and minimal null frequencies are preferable for accurate signal modeling and efficient parameter estimation. Among the candidates considered, we find that the PD4L configuration offers robust performance for data analysis. In contrast, when operating in the time domain, all optimal channel sets become effectively equivalent.

This paper is organized as follows: 
Section~\ref{sec:tdi} introduces the second-generation TDI configurations examined in this work.
In Section~\ref{sec:correlations}, we analyze the correlations in both GW responses and noise spectra of different TDI schemes, evaluate the noise spectral stabilities, and compare average sensitivity of orthogonal channels and their cross-correlation.
Section~\ref{sec:inference} investigates the response to a specific GW signal and highlights differences in high-frequency performance, focusing on signal fidelity and parameter recovery. 
Conclusions and discussion are presented in Section \ref{sec:conclusions}.
(We set $G=c=1$ in this work except where specified otherwise in the equations.)

\section{Time delay interferometry} \label{sec:tdi}

\subsection{Second-generation configurations}

\begin{figure*}[htb]
\includegraphics[width=0.18\textwidth]{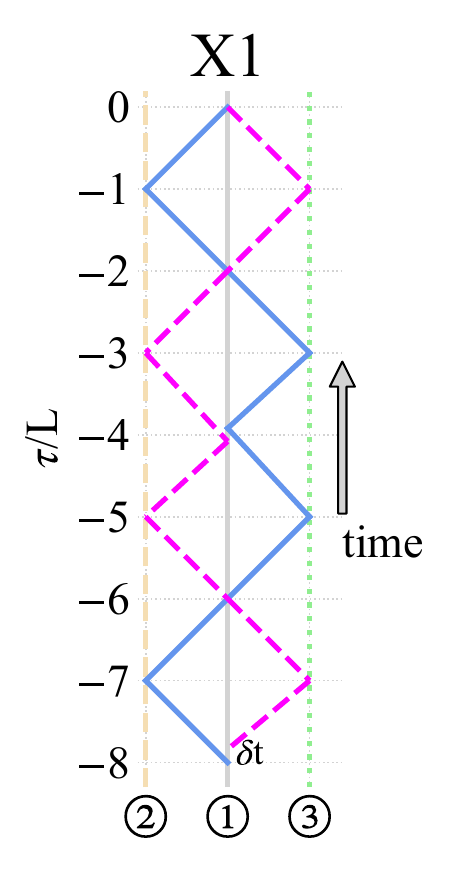}
\includegraphics[width=0.23\textwidth]{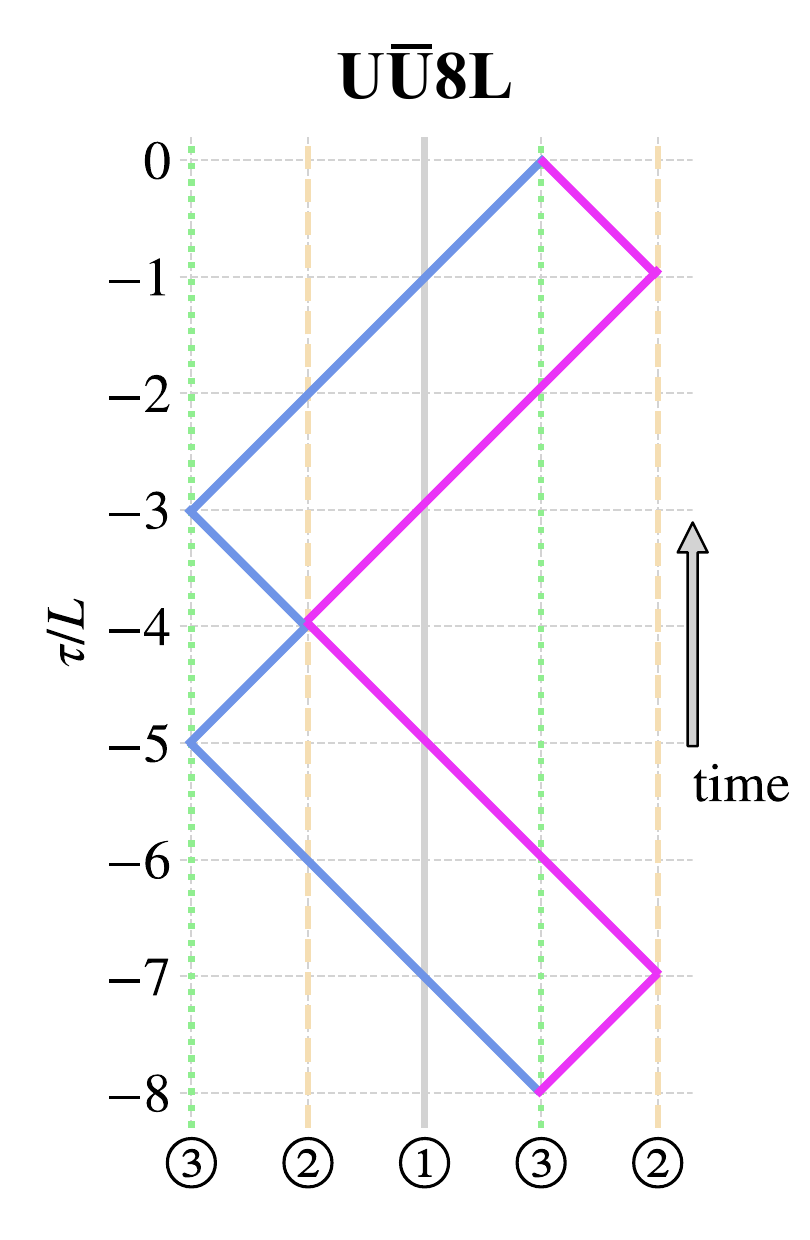}
\includegraphics[width=0.23\textwidth]{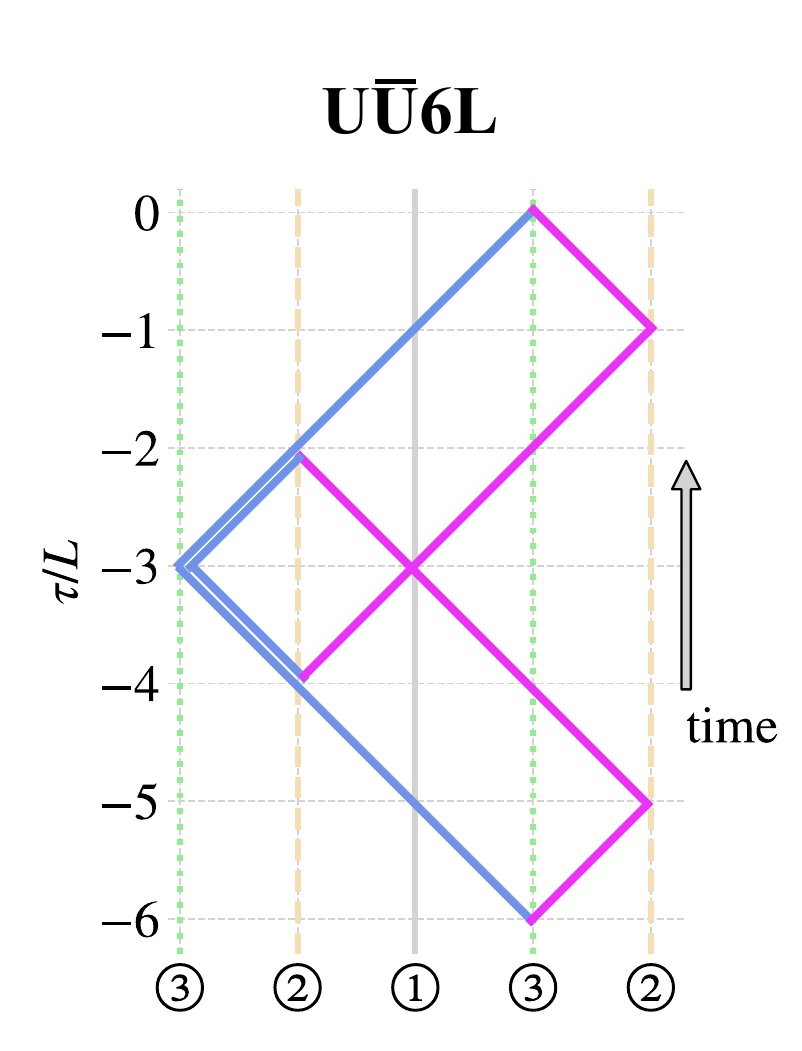} \\
\includegraphics[width=0.28\textwidth]{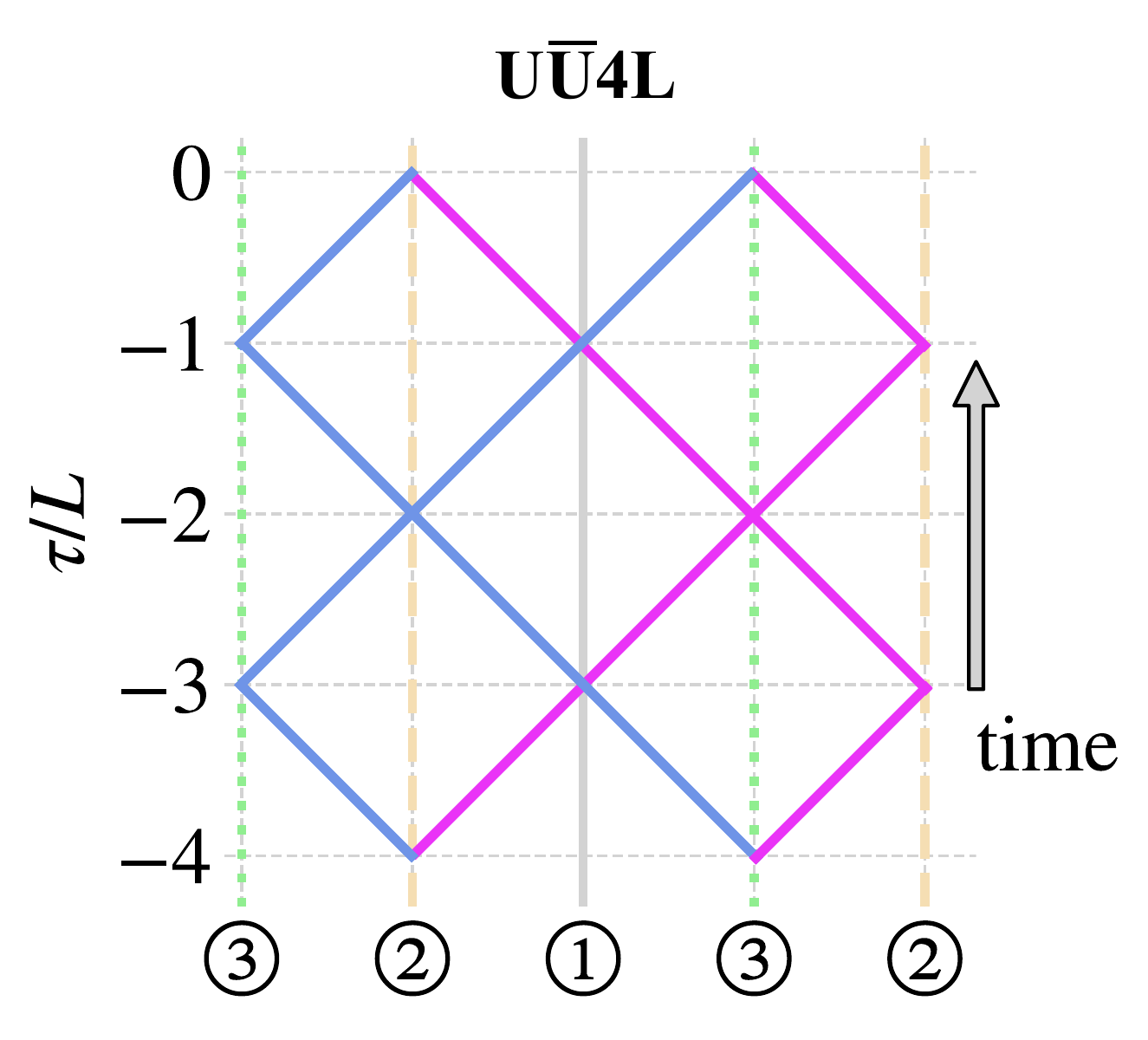}
\includegraphics[width=0.28\textwidth]{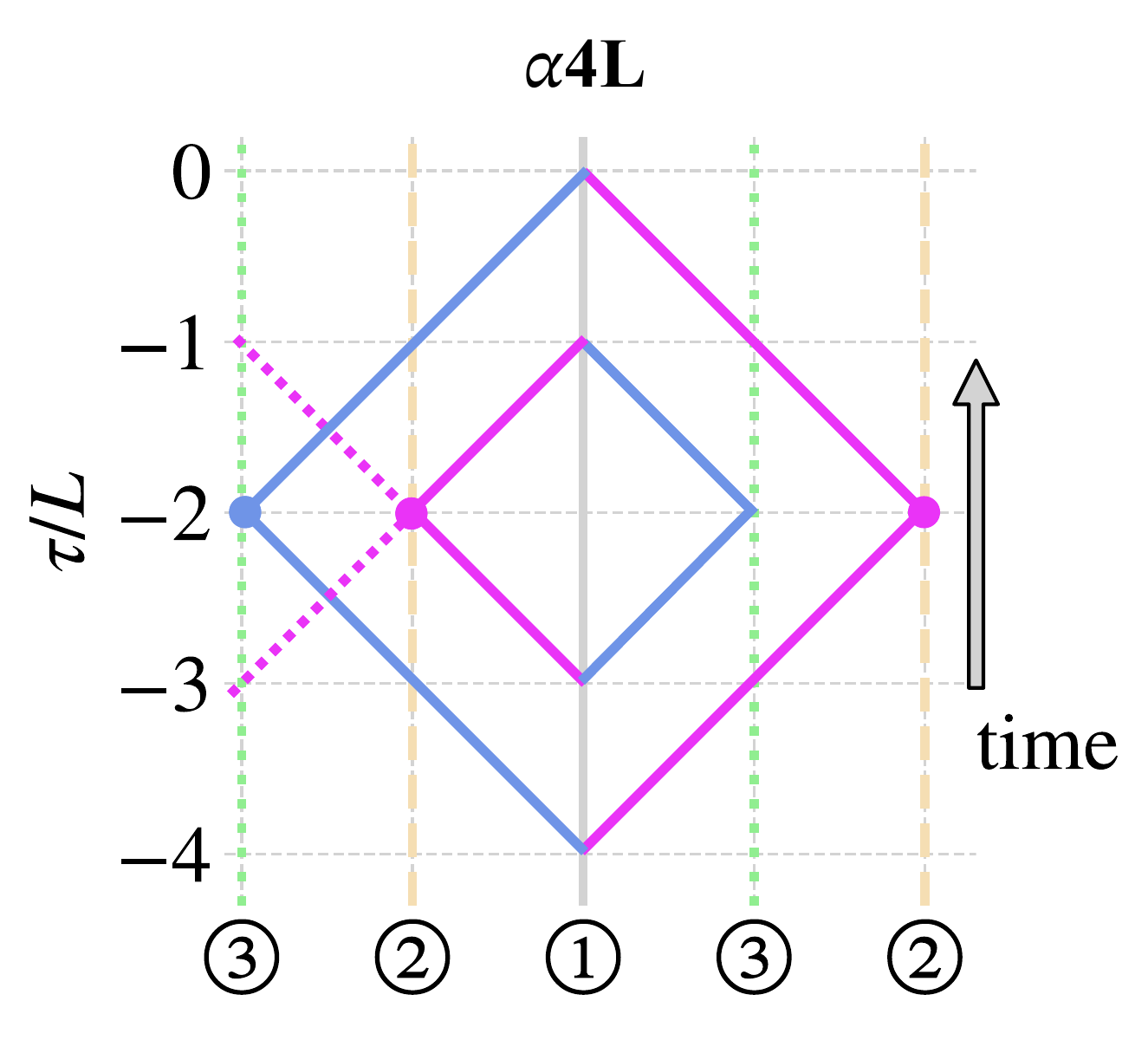}
\includegraphics[width=0.28\textwidth]{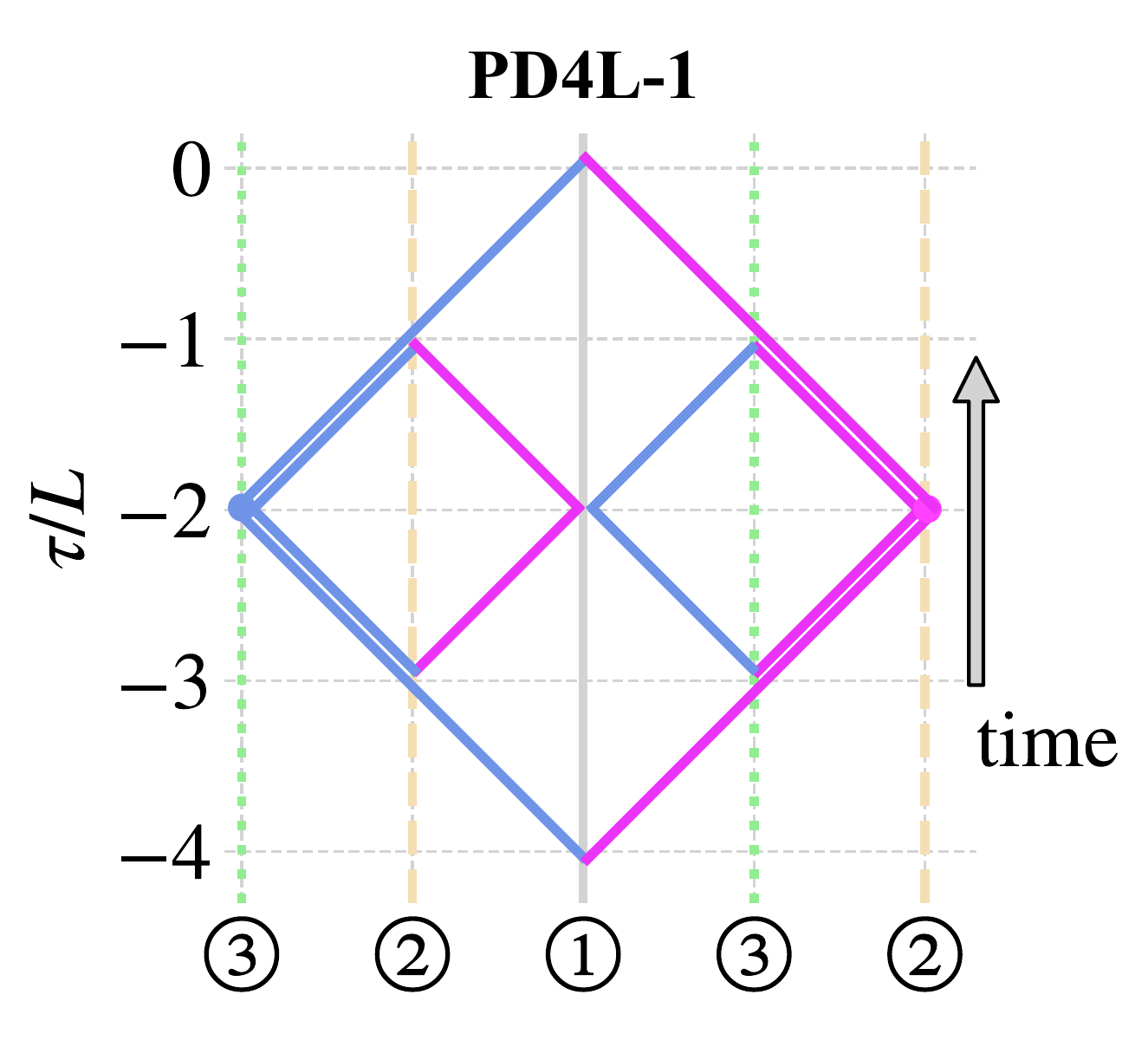}
\caption{\label{fig:tdi_diagram_8L6L4L}  Geometric diagrams of representative TDI channels spanning $8L$ (X1 upper left, U$\overline{\mathrm{U}}$8L upper center, \cite{Wang:2024alm}), $6L$ span (U$\overline{\mathrm{U}}$6L upper right), and $4L$ span (U$\overline{\mathrm{U}}$4L lower left, $\alpha$4L lower center, PD4L-1 lower right) \cite{Wang:2011}. S/C trajectories are shown as vertical lines, with an extra line for S/C2 (\textcircled{2}) for clarity. Except for X1, other five channels have nulls only at $f_\mathrm{null}=m/L, \ (m=1,2,3, \dots)$.  }
\end{figure*}

Most classical second-generation TDI are constructed either by synthesizing first-generation TDI or by employing other methods \cite[and references therein]{Shaddock:2003dj,Tinto:2003vj,Vallisneri:2005ji,Wang:2011,Wang:2012fqs,Wang:2020pkk,Hartwig:2021mzw,Tinto:2022zmf,Wu:2022qov}. A fiducial example is the second-generation Michelson TDI. Each observable in this configuration utilizes four laser links spanning two arms. By selecting different initial S/C and propagation sequence, three observables (X1, Y1, Z1) can be constructed. The first of these, X1, is expressed as:
\begin{eqnarray} \label{eq:X1_measurement}
{\rm X1:} \ & \overrightarrow{121313121} \ \overleftarrow{131212131}.
\end{eqnarray}
This expressions follows the notation introduced in \cite{Vallisneri:2005ji}, where arrows indicate the direction of time: ``$\rightarrow$'' denotes forward time (from earlier to later times, read left to right), while ``$\leftarrow$'' denotes backward time propagation (from later to earlier times, also read left to right). The digits correspond to the indices of the S/C. For instance, the sequence $\overrightarrow{121313121}$ represents the forward-in-time virtual laser propagation path as illustrated by the solid blue lines in the upper left panel of Fig.~\ref{fig:tdi_diagram_8L6L4L}. Conversely, the term $\overleftarrow{131212131}$ describes the reverse time order as indicated by dashed magenta lines. The paths for the other two channels (Y1 and Z1) can be obtained by cyclically permuting the S/C indices $1 \rightarrow 2 \rightarrow 3$ in the expression for X1.

Another widely used second-generation TDI is the Sagnac-type configuration \cite{Shaddock:2003dj}, and three observables are denoted as ($\alpha$6L, $\beta$6L, $\gamma$6L). The first observable is written as:
\begin{eqnarray}
\mathrm{\alpha 6L:} \ & \overrightarrow{1231321} \ \overleftarrow{1321231}.
\end{eqnarray}
The suffix "6L" is added to the channel name to emphasis the configuration span and distinguish it from other configurations. For brevity, we refer to all TDI observables in this section as second-generation and without explicitly repeating the term "second-generation".

Original Michelson observables exhibit nulls at frequencies $f_\mathrm{null}=m/(4L), \ (m=1,2,3, \dots)$, and Sagnac configuration exhibits nulls at $f_\mathrm{null}=m/(3L)$. \citet{Vallisneri:2005ji} developed a bunch of alternative configurations, which reduce the null frequencies to $f_\mathrm{null}=m/(2L)$ for modified Michelson-type observables, and as low as $f_\mathrm{null}=m/L$ for U-type, E-type and P-type configurations. We analyze and compare these schemes with the PD4L configuration (defined later in Eq \eqref{eq:PD4L_path}) in Appendix \ref{sec:tdi_in_Vallisneri_2005}. We refer to $f_\mathrm{null}=m/L$ as the \textit{minimal null frequency} for a TDI configuration, as such nulls naturally arise from the differential operation $d(t) - d(t- nL)$, where $n=1,2,3,\dots$. This operation leads to nulls at frequencies $f=m/(nL)$ in the frequency domain, with $n=1$ corresponding to the lowest or minimal null frequencies. A notable exception is the ordinary channel of the first-generation Sagnac configuration, whose has no null frequencies. We explain the reason for this feature and examine the null structures of other first-generation TDI configurations in Appendix \ref{sec:null_in_1st_TDI}.

Building on the geometric method developed in \cite{Vallisneri:2005ji}, we have design new TDI schemes, and some achieve minimum nulls at $f_\mathrm{null}=m/L$ \cite{Wang:2011,Wang:2012fqs}. For instance, self-splicing the first-generation Saganc configuration yields three configurations, two of them span $4L$ and are labeled as $\alpha$4L. Since these two are equivalent, we examine one of them here, and the first observable is expressed as:
\begin{eqnarray}
\mathrm{\alpha 4L:}    \ &  \overrightarrow{1 2 3 2 1} \ \overleftarrow{1 3 2 1} \ \overrightarrow{1 3 1 } \ \overleftarrow{1 2 3 1}.
\end{eqnarray}
Its geometric path is illustrated in the lower center panel of Fig. \ref{fig:tdi_diagram_8L6L4L}. Further TDI configurations can be constructed by synthesizing two asymmetric Relay-type observables with time shifts. We refer to these as hybrid Relay configurations. One such example spans $8L$ and is labeled ($\mathrm{U\overline{U}}$8L, $\mathrm{V\overline{V}}$8L, $\mathrm{W\overline{W}}$8L). The first channel is expressed as:
\begin{eqnarray}
\mathrm{U\overline{U}8L:} \ & \overrightarrow{312323213} \ \overleftarrow{323121323}.
\end{eqnarray}
Similarly, hybrid Relay configurations with $6L$ and $4L$ spans are denoted as ($\mathrm{U\overline{U}}$6L, $\mathrm{V\overline{V}}$6L, $\mathrm{W\overline{W}}$6L) and ($\mathrm{U\overline{U}}$4L, $\mathrm{V\overline{V}}$4L, $\mathrm{W\overline{W}}$4L), respectively. Examples:
\begin{eqnarray}
\mathrm{U\overline{U}6L:} \ & \overrightarrow{3123213} \ \overleftarrow{ 32312 } \ \overrightarrow{232}  \ \overleftarrow{21323}, \\
\mathrm{U\overline{U}4L:} \ &  \overrightarrow{31} \ \overleftarrow{12} \ \overrightarrow{23213} \ \overleftarrow{3231} \ \overrightarrow{1232} \  \overleftarrow{21323}, 
\end{eqnarray}
The geometries paths of first channel of three hybrid Relay configurations are shown in Fig. \ref{fig:tdi_diagram_8L6L4L} \cite{Wang:2011,Wang:2020pkk}. An additional TDI configuration, PD4L, combines the first-generation Monitor and Beacon types. The three corresponding observables are labeled PD4L-1, PD4L-2, and PD4L-3. The first is written as:
\begin{eqnarray}
\textrm{PD4L-1:}\ & \overrightarrow{1 2 3 2} \ \overleftarrow{2 1 2} \ \overrightarrow{2 3 2 1} \ \overleftarrow{1 3 2 3} \ \overrightarrow{3 1 3} \ \overleftarrow{3 2 3 1} \label{eq:PD4L_path}.
\end{eqnarray}
Its geometry is shown in the lower right panel of Fig.~\ref{fig:tdi_diagram_8L6L4L}. As analyzed in our previous work \cite{Wang:2025mee}, the PD4L configuration demonstrates excellent performance with minimal null frequencies. In this study, we compare the correlations and variations among these representative TDI configurations to highlight the redundancy present across various TDI designs.

\subsection{(Quasi-)Orthogonal TDI observables}

\begin{figure}[htbp]
\includegraphics[width=0.48\textwidth]{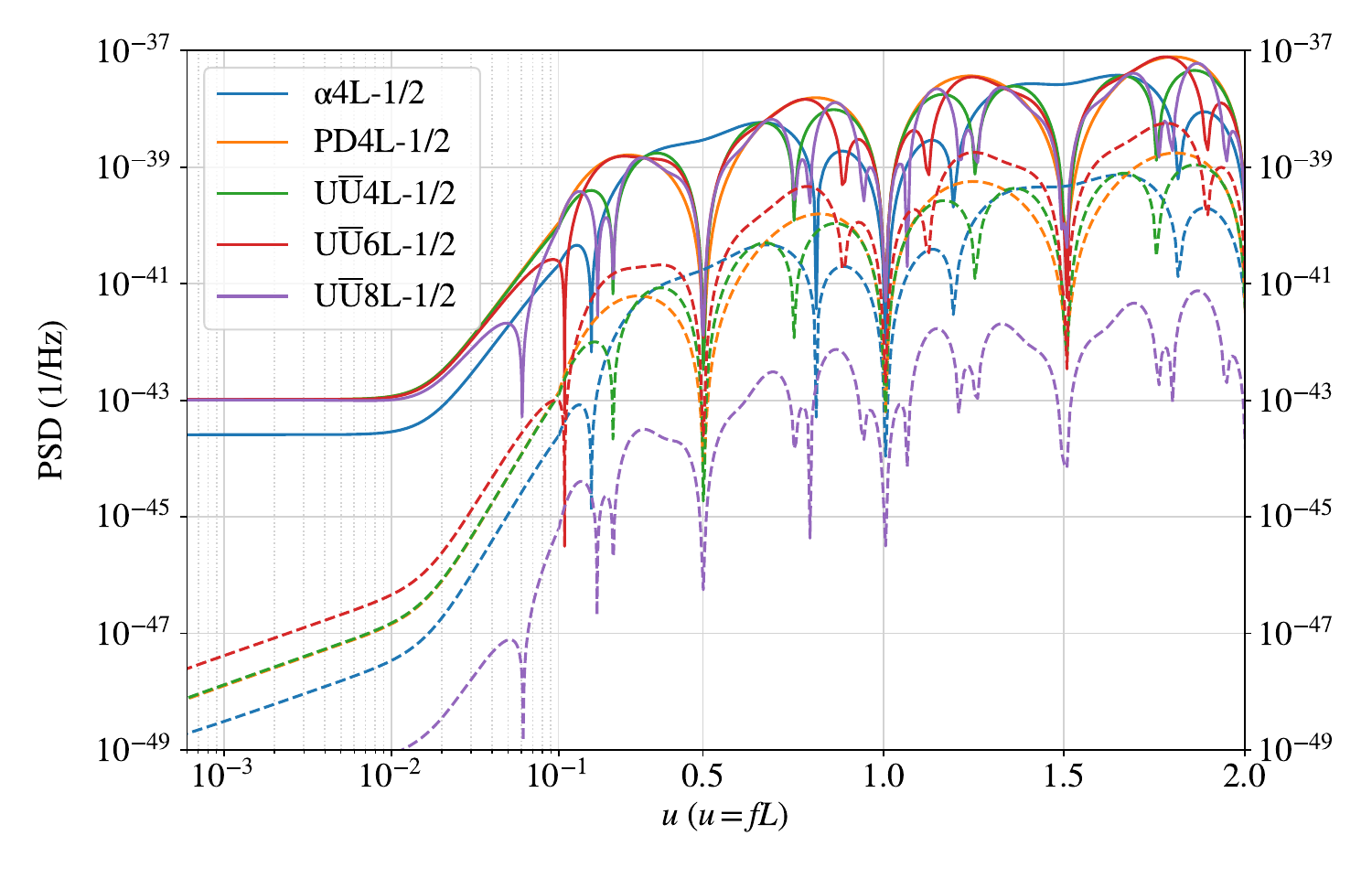}
\caption{\label{fig:TDI_csd_real_imag} Real (solid curves) and imaginary (dashed curves with same colors) parts of CSDs between two ordinary TDI channels from selected configurations. The dominance of the real parts justifies the approximation in Eq.~\eqref{eq:covmat_approximation} and supports the use of (quasi-)orthogonal transformations.
}
\end{figure}

For each TDI configuration, three ordinary channels $(a, b, c)$ can be derived by cyclically permuting the S/C indices. The noise covariance matrix of these channels is given by
\begin{equation} \label{eq:covmat_approximation}
\begin{bmatrix}
S_\mathrm{a} & S_\mathrm{ab} & S_\mathrm{ac} \\
S_\mathrm{ba} & S_\mathrm{b} & S_\mathrm{bc} \\
S_\mathrm{ca} & S_\mathrm{cb} & S_\mathrm{c}
\end{bmatrix}
\simeq 
\begin{bmatrix}
S_\mathrm{a} & S_\mathrm{ab} & S_\mathrm{ab} \\
S_\mathrm{ab} & S_\mathrm{a} & S_\mathrm{ab} \\
S_\mathrm{ab} & S_\mathrm{ab} & S_\mathrm{a}
\end{bmatrix}, 
\end{equation}
where $S_\mathrm{a}$ is the power spectral density (PSD) of channel $a$, and $S_\mathrm{ab}$ is the cross spectral density (CSD) between channels $a$ and $b$. Under the assumption of equal arm lengths and identical noise budgets, the imaginary parts of the CSDs often vanish in certain TDI schemes, resulting in identical off-diagonal elements in Eq. \eqref{eq:covmat_approximation}. In this case, the covariance matrix becomes symmetric, and a set of orthogonal or optimal observables (A, E, T) can be obtained by diagonalizing the covariance matrix. The transformation is given by \cite{Prince:2002hp,Vallisneri:2007xa}:
\begin{equation} \label{eq:abc2AET}
\begin{bmatrix}
\mathrm{A}_a  \\ \mathrm{E}_a  \\ \mathrm{T}_a
\end{bmatrix}
 =
\begin{bmatrix}
-\frac{1}{\sqrt{2}} & 0 & \frac{1}{\sqrt{2}} \\
\frac{1}{\sqrt{6}} & -\frac{2}{\sqrt{6}} & \frac{1}{\sqrt{6}} \\
\frac{1}{\sqrt{3}} & \frac{1}{\sqrt{3}} & \frac{1}{\sqrt{3}}
\end{bmatrix}
\begin{bmatrix}
a \\ b  \\ c
\end{bmatrix}.
\end{equation}
In practice, small asymmetries in arm lengths or instrumental noise can introduce deviations from the ideal case, including nonzero imaginary components in the CSDs. Nevertheless, the orthogonal transformation remains approximately valid when the real parts of the CSDs significantly dominate over the imaginary parts. Figure~\ref{fig:TDI_csd_real_imag} demonstrates that, for five selected TDI configurations, the real components (solid lines) of the CSDs are orders of magnitude larger than their imaginary components (dashed lines with same colors), thus validating the approximation in Eq.~\eqref{eq:covmat_approximation}.
The noise spectra used in this analysis include both acceleration noise and optical metrology system (OMS) noise, modeled according to LISA mission designs \cite{LISA:2017pwj,Colpi:2024xhw},
\begin{equation} \label{eq:noise_budgets}
\begin{aligned}
& \sqrt{ \mathrm{ S_{acc} } } = 3 \frac{\rm fm/s^2}{\sqrt{\rm Hz}} \sqrt{1 + \left(\frac{0.4 {\rm mHz}}{f} \right)^2 }  \sqrt{1 + \left(\frac{f}{8 {\rm mHz}} \right)^4 }, \\
& \sqrt{ \mathrm{ S_{oms} } } = 15 \frac{\rm pm}{\sqrt{\rm Hz}} \sqrt{1 + \left(\frac{2 {\rm mHz}}{f} \right)^4 }.
 \end{aligned}
\end{equation}
The calculation is performed using a random selected time point from a numerical orbit \footnote{https://github.com/gw4gw/LISA-Like-Orbit} \cite{Wang:2017aqq}, yielding the instantaneous arm length $[ L_{12}, L_{21}, L_{13}, L_{31}, L_{23}, L_{32}] \simeq [8.296, 8.297, 8.275, 8.277, 8.313, 8.313]$ s.
Throughout this work, we label optimal observables (A, E, T) with subscripts to indicate their associated TDI configurations. For instance, the Michelson configuration yields observables A$_\mathrm{X1}$, E$_\mathrm{X1}$, and T$_\mathrm{X1}$. The complete list of optimal channels for all configurations examined is summarized in Table~\ref{tab:optimal_tdi_list}.

\begin{table}[tbhp]
\caption{\label{tab:optimal_tdi_list} List of optimal observables for selected second-generation TDI configurations.}
\begin{ruledtabular}
\begin{tabular}{c|c c}
TDI configuration & optimal channels  \\
\hline
Michelson &  (A$_\mathrm{X1}$, E$_\mathrm{X1}$, T$_\mathrm{X1}$) \\
Sagnac &  (A$_\mathrm{\alpha 6L}$, E$_\mathrm{\alpha 6L}$, T$_\mathrm{\alpha 6L}$) \\
U$\overline{\mathrm{U}}$8L  & (A$_\mathrm{U\overline{U}8L}$, E$_\mathrm{U\overline{U}8L}$, T$_\mathrm{U\overline{U}8L}$) \\
U$\overline{\mathrm{U}}$6L  & (A$_\mathrm{U\overline{U}6L}$, E$_\mathrm{U\overline{U}6L}$, T$_\mathrm{U\overline{U}6L}$) \\
U$\overline{\mathrm{U}}$4L  &  (A$_\mathrm{U\overline{U}4L}$, E$_\mathrm{U\overline{U}4L}$, T$_\mathrm{U\overline{U}4L}$) \\
$\alpha$4L &  (A$_\mathrm{\alpha 4L}$, E$_\mathrm{\alpha 4L}$, T$_\mathrm{\alpha 4L}$) \\
PD4L  &  (A$_\mathrm{PD4L}$, E$_\mathrm{PD4L}$, T$_\mathrm{PD4L}$) 
\end{tabular}
\end{ruledtabular}
\end{table}

From a geometric perspective, the TDI channels can be interpreted as equivalent virtual interferometers. The general response of an interferometer to GWs can be expressed as \cite{Freise:2008dk}:
\begin{equation}
\begin{aligned}
h(\kappa) =& \sin \zeta \left[ \left( K_1 \sin 2 \kappa + K_2 \cos 2 \kappa \right) h_{+} \right. \\ 
 &  \left. + \left( K_3 \sin 2 \kappa + K_4 \cos 2 \kappa \right) h_{\times} \right]
\end{aligned}
\end{equation}
where $\zeta$ is the interferometer’s opening angle, $\kappa$ is the rotation angle with respect to the baseline, and $K_n$ are coefficients determined by the other parameters. If three ordinary channels with orientations $a = h(0^\circ)$, $b = h(240^\circ)$, and $c = h(120^\circ)$, the corresponding optimal channels can be interpreted as \cite{Wang:2020fwa}:
\begin{align}
\mathrm{A} = &  \frac{ h(120^\circ) - h(0^\circ) }{\sqrt{2}} = \sqrt{ \frac{3}{2} } h(105^\circ),  \label{eq:equivalent_A} \\
\mathrm{E} = & \frac{ h(0^\circ) - 2 h(240^\circ) + h(120^\circ) }{\sqrt{6}} = \sqrt{ \frac{3}{2} } h(150^\circ),  \label{eq:equivalent_E} \\
\mathrm{T} = &  \frac{ h(0^\circ) + h(240^\circ) + h(120^\circ) }{\sqrt{3}} = 0.
\end{align}
The A and E channels are sensitive to GWs and are equivalent to two orthogonal interferometers, rotated by $45^\circ$ with respect to each other. In the long-wavelength limit ($L/\lambda_\mathrm{GW} < 0.1$), they span the measurement plane of the detector constellation. By contrast, the T channel is insensitive to GWs below $\sim$50 mHz and is therefore dominated by instrumental noise in low frequency regime (as it will shown in Fig. \ref{fig:sensivity_8L6L4L}).

\section{Correlation between TDI configurations} \label{sec:correlations}

In the static limit, both first- and second-generation TDI variables can be expressed in terms of a small set of generators $(\alpha, \beta, \gamma, \zeta)$ \cite{1999ApJ...527..814A,Hartwig:2021mzw}, implying that correlations between different TDI constructions are expected on general grounds. In this section, we quantify these correlations using numerical orbits that incorporate dynamically unequal arm lengths.

\subsection{Correlation of sky-averaged GW response and instrumental noise}

\begin{figure*}[thbp]
\includegraphics[width=0.48\textwidth]{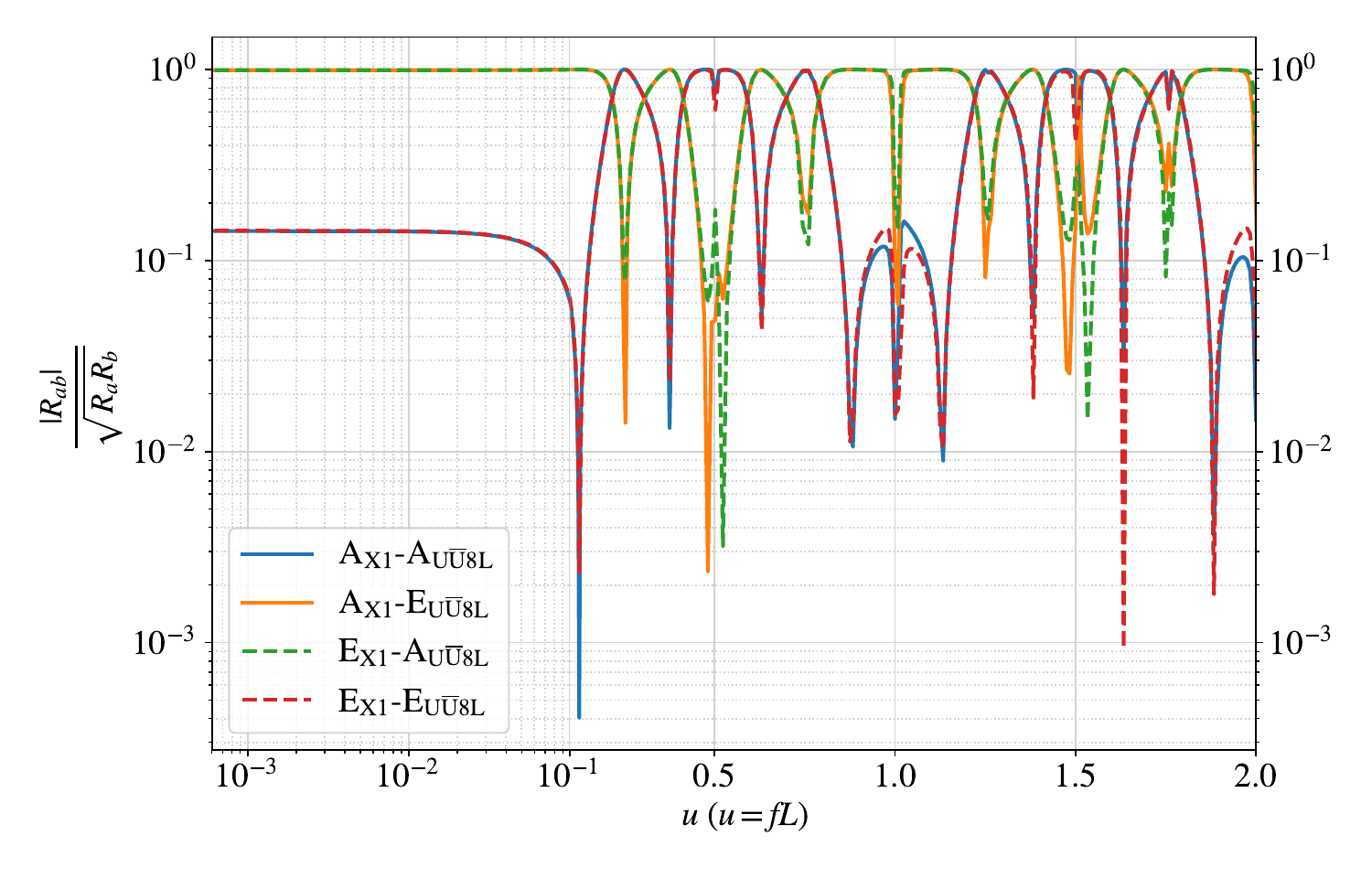}
\includegraphics[width=0.48\textwidth]{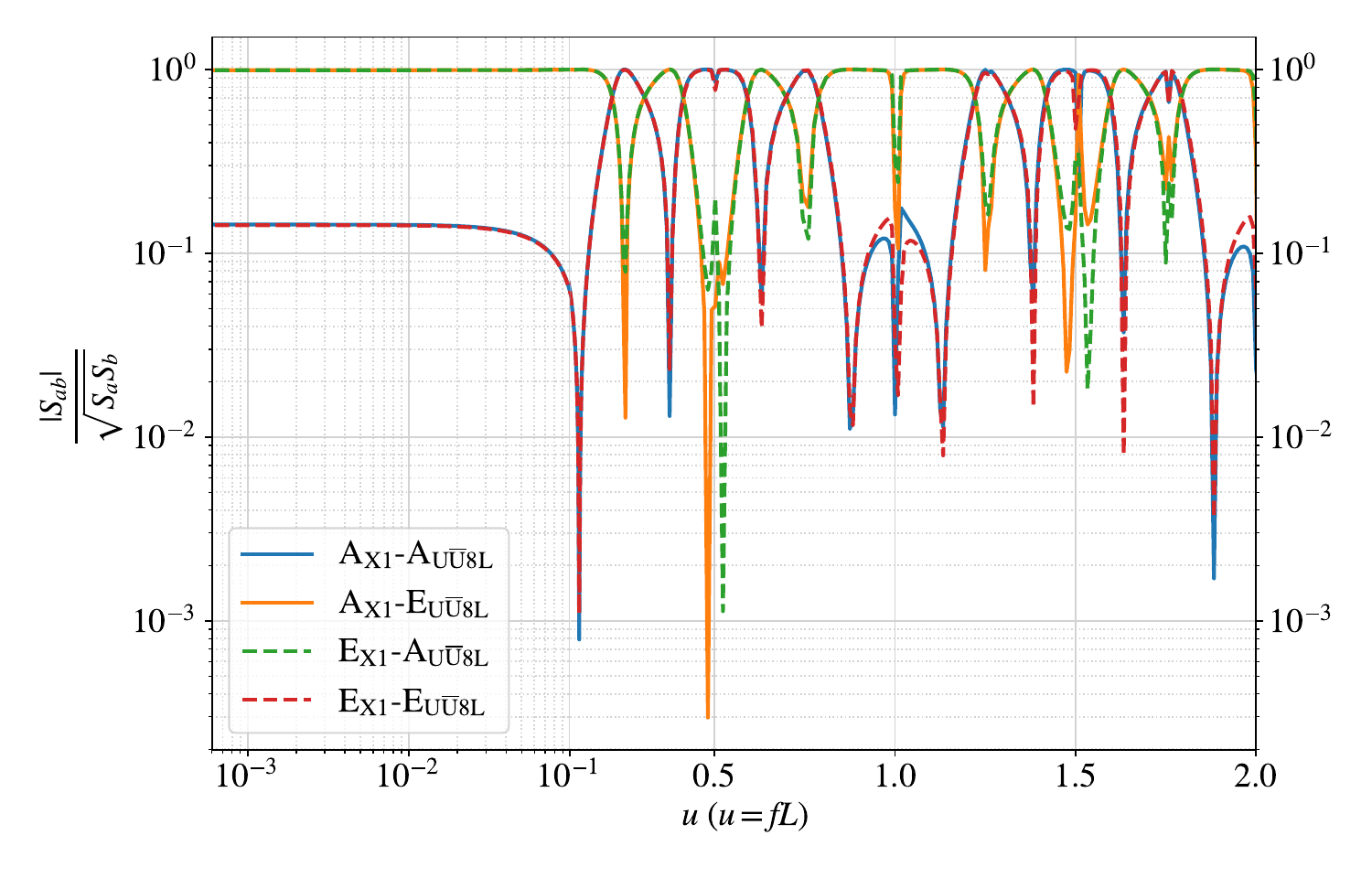}
\includegraphics[width=0.48\textwidth]{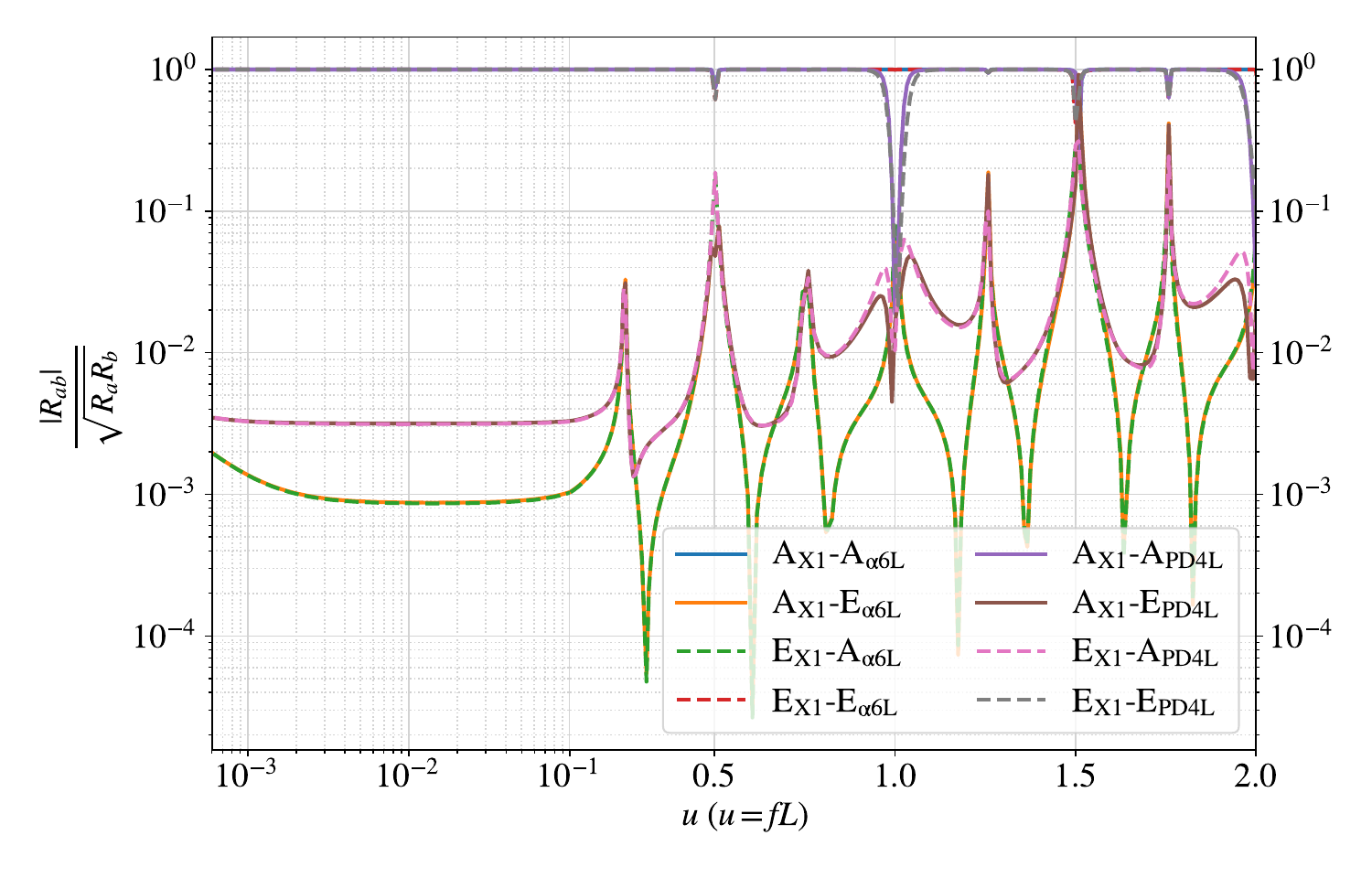}
\includegraphics[width=0.48\textwidth]{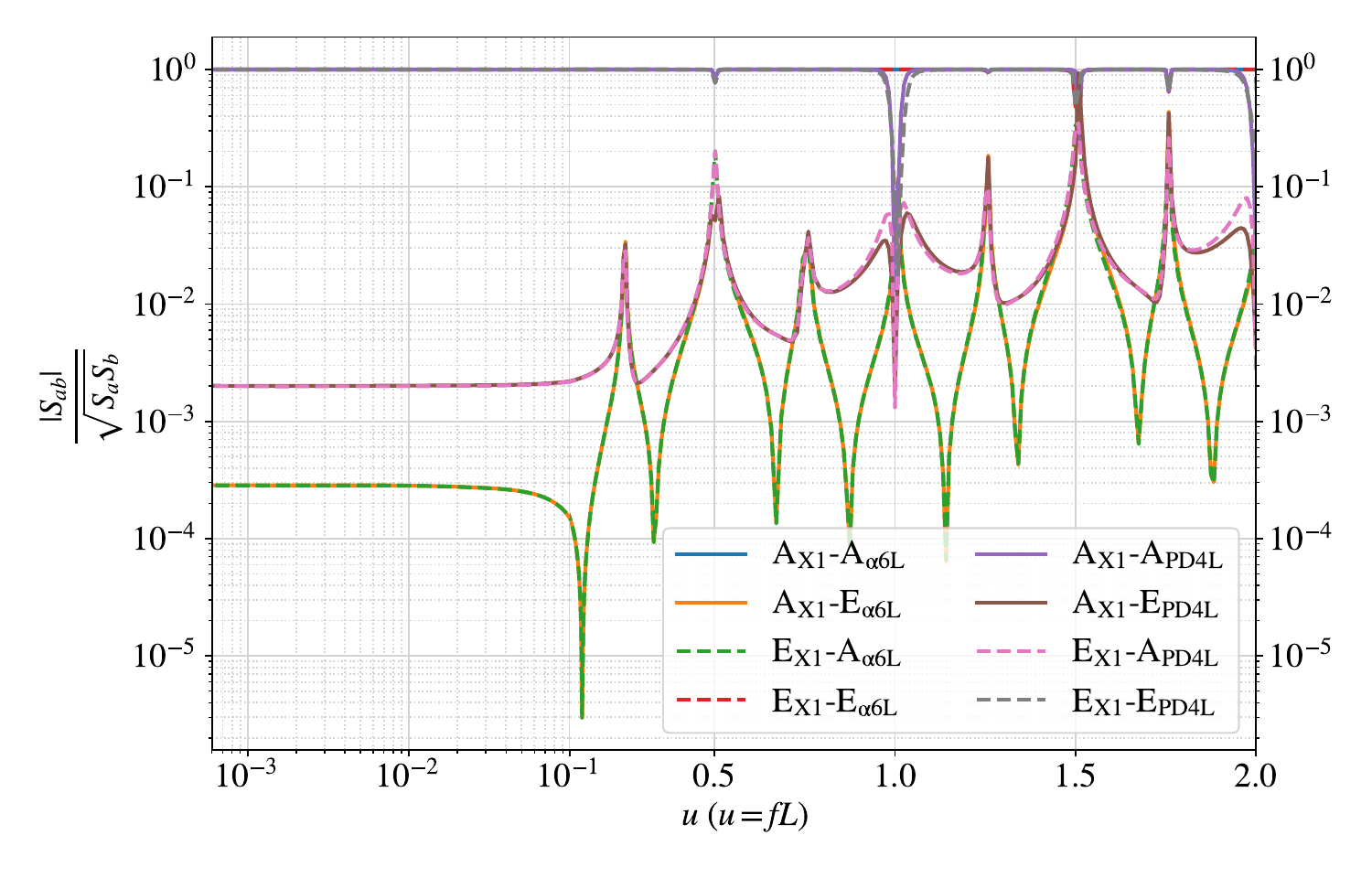}
\includegraphics[width=0.48\textwidth]{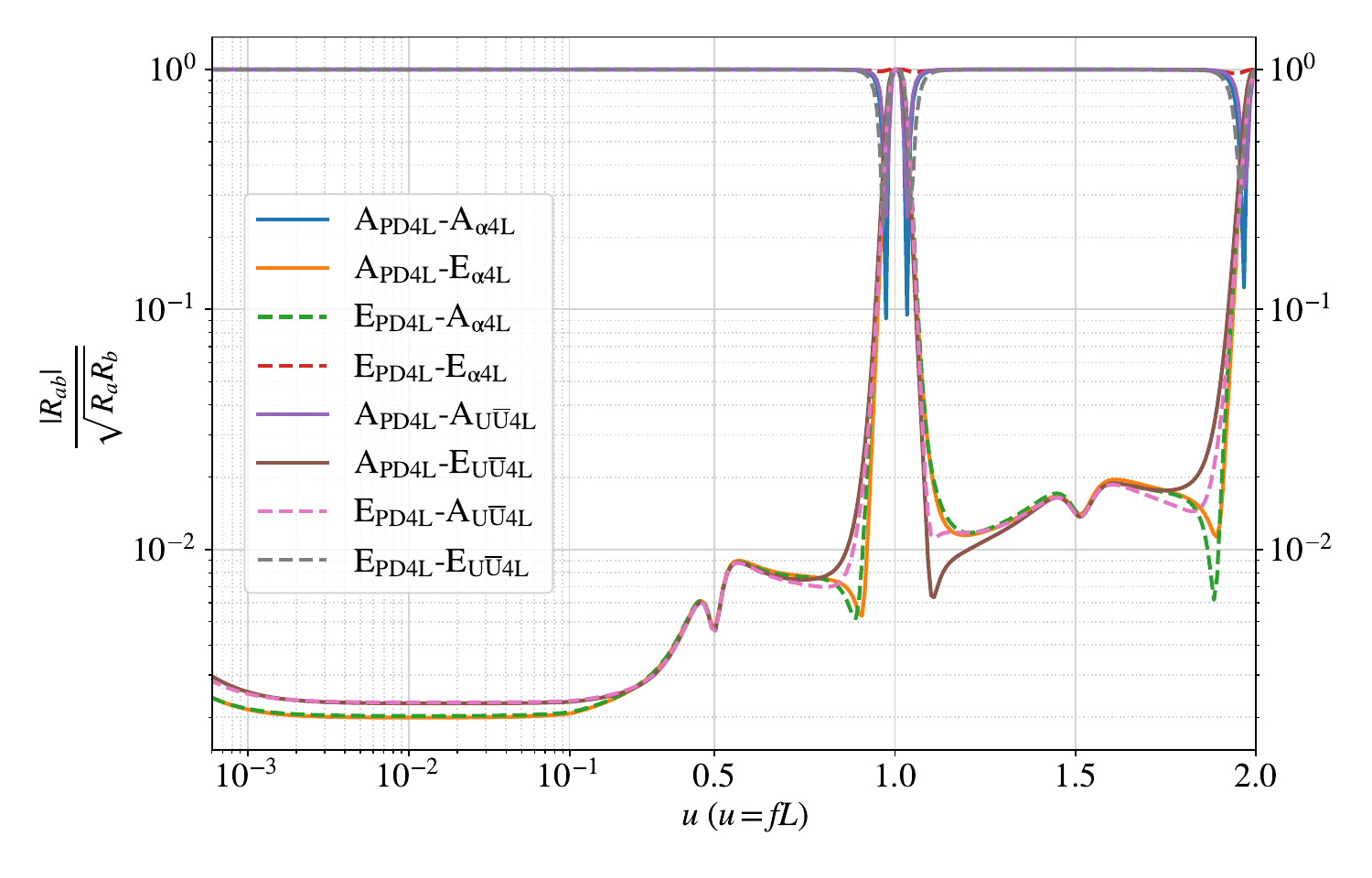}
\includegraphics[width=0.48\textwidth]{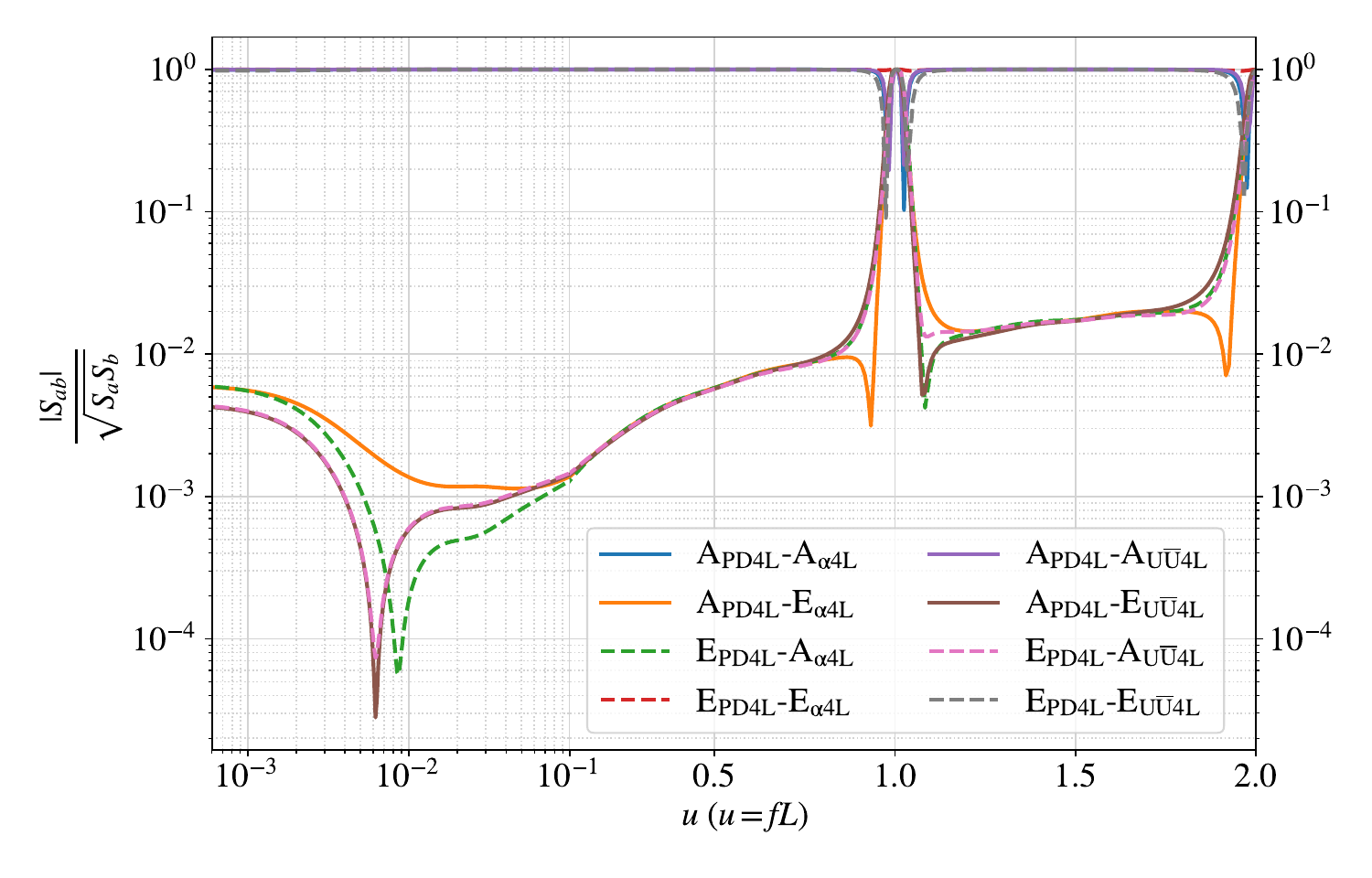}
\includegraphics[width=0.48\textwidth]{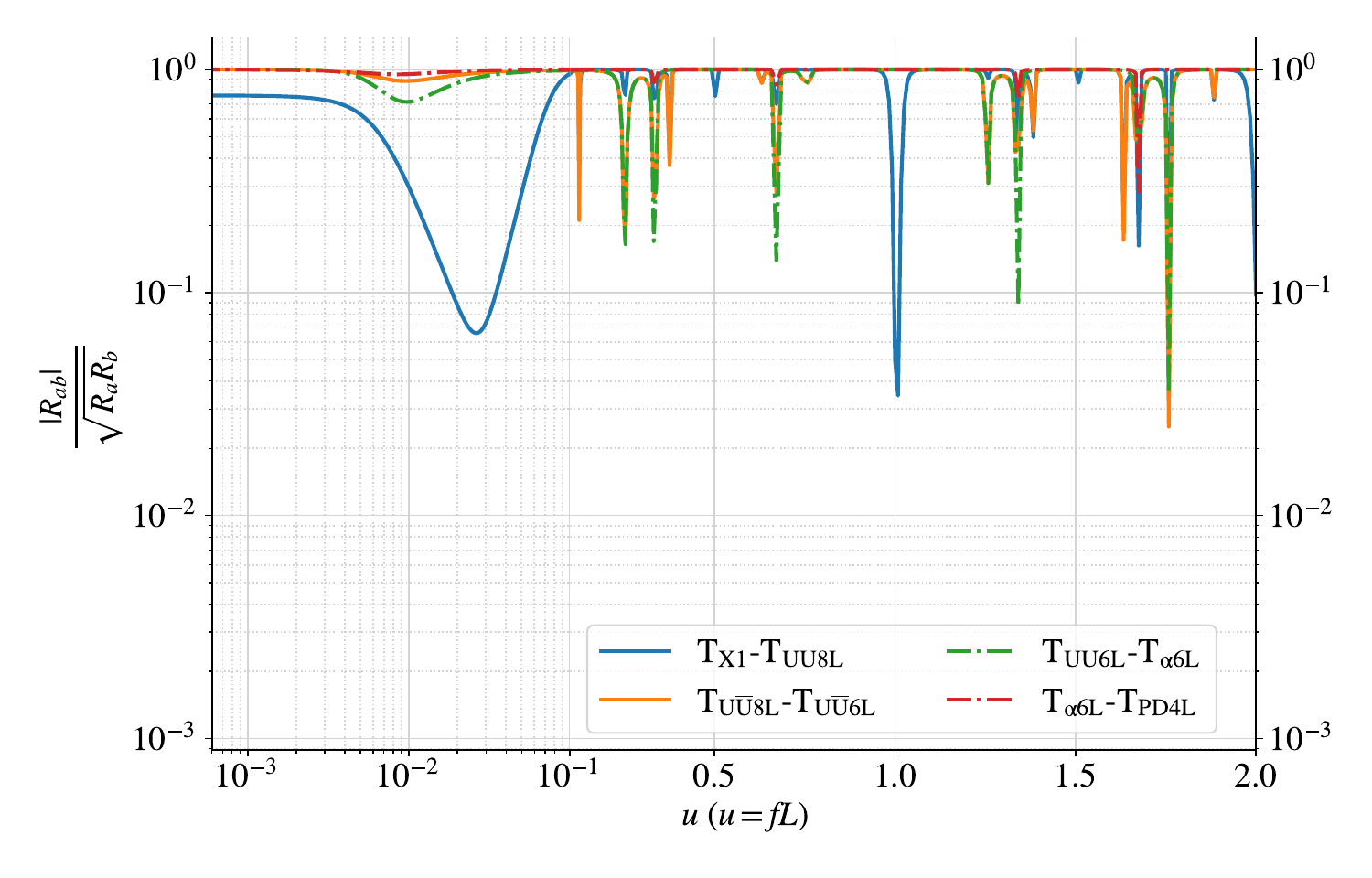}
\includegraphics[width=0.48\textwidth]{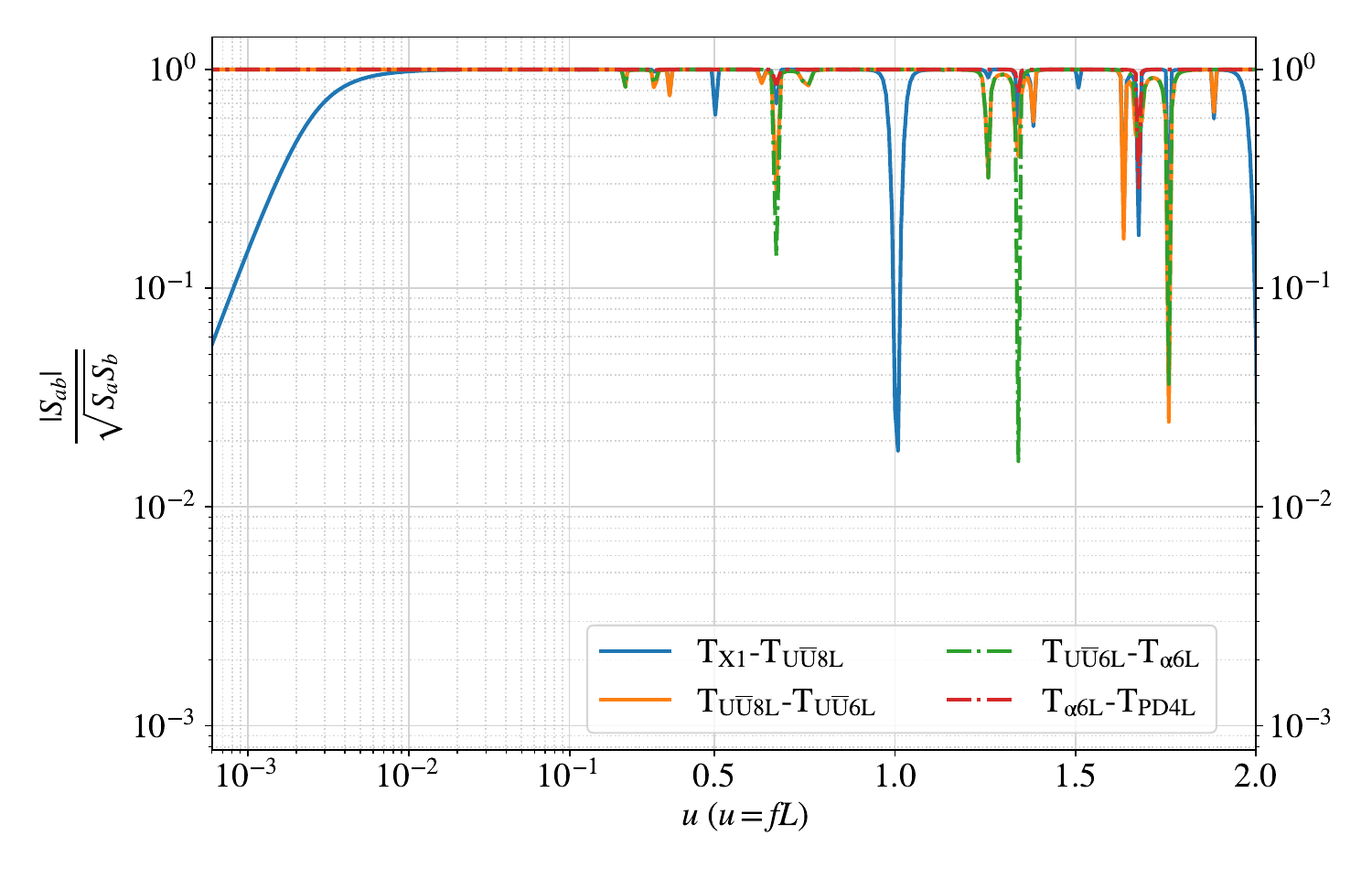}
\caption{\label{fig:TDI_resp_csd_correlation} 
Cross-correlation between the A and E science channels (first to third rows) and the T channels (bottom row) in selected TDI configurations. Left column: sky-averaged cross-correlation of GW responses. Right column: cross-correlation of instrumental noise. The horizontal axis shows the dimensionless frequency $u = fL$, where $f$ is the frequency and $L$ is the nominal arm length in seconds.
}
\end{figure*}

The sky-averaged GW response of a TDI channel at a given frequency is computed as:
\begin{equation} \label{eq:resp_averaged}
\begin{aligned}
 R_{a} (f) =& \frac{1}{4 \pi}  \int^{2 \pi}_{0} \int^{\frac{\pi}{2}}_{-\frac{\pi}{2}} |F_{a} (f, \beta, \lambda)|^2 \cos \beta {\rm d} \beta {\rm d} \lambda.
\end{aligned}
\end{equation}
where $F_{a} (f, \beta, \lambda)$ is GW response function of TDI channel $a$, and ($\lambda$, $\beta$) denote the ecliptic longitude and latitude in the solar-system barycentric coordinate system, respectively.
The average cross-response between two TDI channel $a$ and $b$ is given by:
\begin{equation} \label{eq:resp_cross_averaged}
\begin{aligned}
 R_{ab} (f) =& \frac{1}{4 \pi}  \int^{2 \pi}_{0} \int^{\frac{\pi}{2}}_{-\frac{\pi}{2}} F^\ast_{a} (f, \beta, \lambda)  F_{b} (f, \beta, \lambda)  \cos \beta {\rm d} \beta {\rm d} \lambda.
\end{aligned}
\end{equation}
The normalized cross-correlation GW response between two channels is defined as $\frac{|R_{ab} |}{ \sqrt{R_a R_b} }$. These correlations for orthogonal channels in selected TDI configurations are shown in the left column of Fig. \ref{fig:TDI_resp_csd_correlation}.
Similarly, the noise correlation between two TDI channels is defined as $\frac{|S_{ab} |}{ \sqrt{S_a S_b} }$. These correlations are displayed in the right column of Fig. \ref{fig:TDI_resp_csd_correlation}.
The horizontal axis in all plots uses the dimensionless quantity $u = fL$, where $f$ is the frequency and $L$ is the nominal arm length in seconds, to clearly display the values at characteristic frequencies. A logarithmic scale is adopted for $u < 0.1$, transitioning to a linear scale for higher frequencies to better highlight features near null frequencies.

Since the A and E observables from any TDI configuration are (quasi-)orthogonal and span the constellation plane, the A and E channels from two different configurations can be related through a rotation:\begin{equation} \label{eq:AET2AET}
\begin{bmatrix}
\mathrm{A}_{x}  \\ \mathrm{E}_{x}
\end{bmatrix}
 = \mathcal{C}_{y \rightarrow x}
\begin{bmatrix}
\cos \xi & - \sin \xi \\
\sin \xi & \cos \xi
\end{bmatrix}
\begin{bmatrix}
\mathrm{A}_{ y}  \\ \mathrm{E}_{y}
\end{bmatrix},
\end{equation}
where $\mathcal{C}_{y \rightarrow x}$ are scaling coefficients between configuration $y$ and $x$, and $\xi$ is an effective rotation angle.
As shown in the top row of Fig. \ref{fig:TDI_resp_csd_correlation}, for the Michelson-type configuration (X1) and the hybrid Relay configuration (U$\overline{\text{U}}$8L), we find $\xi \simeq 3 \pi / 2$. This implies that A$_\mathrm{X1}$ is highly correlated with E$_\mathrm{U\overline{U}8L}$ (solid orange), while E$_\mathrm{X1}$ correlates with A$_\mathrm{U\overline{U}8L}$ (dashed green), particularly at low frequencies $u < 0.1$. At higher frequencies, the rotation angle becomes frequency-dependent. The downward spikes at $u=m/4, \ (m=1,2,3...)$ are caused by Michelson's nulls, while other spikes arise from sign changes when taking the absolute values. Overall, the trends in GW response correlation match closely with those in noise correlation.

Correlations between A and E channels for additional configuration pairs are shown in the second and third rows of Fig. \ref{fig:TDI_resp_csd_correlation}, and their associated rotation angles are listed in Table \ref{tab:optimal_tdi_roration angle}. With the exception of regions near their null frequencies (where $u$ is an integer), the A and E observables from different TDI configurations remain highly correlated. We thus conclude that the science channels (A, E) are generally correlated across TDI schemes. 
The cross-correlations between the T channels of four TDI configurations are presented in the bottom row of Fig.~\ref{fig:TDI_resp_csd_correlation}. Among them, T$_\mathrm{X1}$ exhibits moderate decorrelation with T$_\mathrm{U\overline{U}8L}$ in the low-frequency regime, as indicated by the solid blue curve. In contrast, the remaining null streams show strong correlations across the spectrum, except for the sharp numerical dips near their respective null frequencies.

\begin{table}[htbp]
\caption{\label{tab:optimal_tdi_roration angle} Rotation angles $\xi$ between optimal A and E observables in Eq. \eqref{eq:AET2AET} at low frequencies.}
\begin{ruledtabular}
\begin{tabular}{c c c }
optimal channels $x$ & optimal channels $y$ & $\xi$ \\
\hline
 (A$_\mathrm{X1}$, E$_\mathrm{X1}$) &  (A$_\mathrm{U\overline{U}8L}$, E$_\mathrm{U\overline{U}8L}$) & $3 \pi / 2$  \\
 (A$_\mathrm{X1}$, E$_\mathrm{X1}$) &  (A$_\mathrm{\alpha 6L}$, E$_\mathrm{\alpha 6L}$) & $ \pi $ \\
  (A$_\mathrm{X1}$, E$_\mathrm{X1}$) &  (A$_\mathrm{U\overline{U}6L}$, E$_\mathrm{U\overline{U}6L}$) & $ 3 \pi /2$ \\
  (A$_\mathrm{X1}$, E$_\mathrm{X1}$) &  (A$_\mathrm{PD4L}$, E$_\mathrm{PD4L}$) & $0$ \\
 (A$_\mathrm{PD4L}$, E$_\mathrm{PD4L}$)  &  (A$_\mathrm{\alpha 4L}$, E$_\mathrm{\alpha 4L}$) & $\pi $ \\
(A$_\mathrm{PD4L}$, E$_\mathrm{PD4L}$)  &  (A$_\mathrm{U\overline{U}4L}$, E$_\mathrm{U\overline{U}4L}$) & $\pi $
\end{tabular}
\end{ruledtabular}
\end{table}

\subsection{Stability of noise spectra}

\begin{figure*}[htbp]
\includegraphics[width=0.45\textwidth]{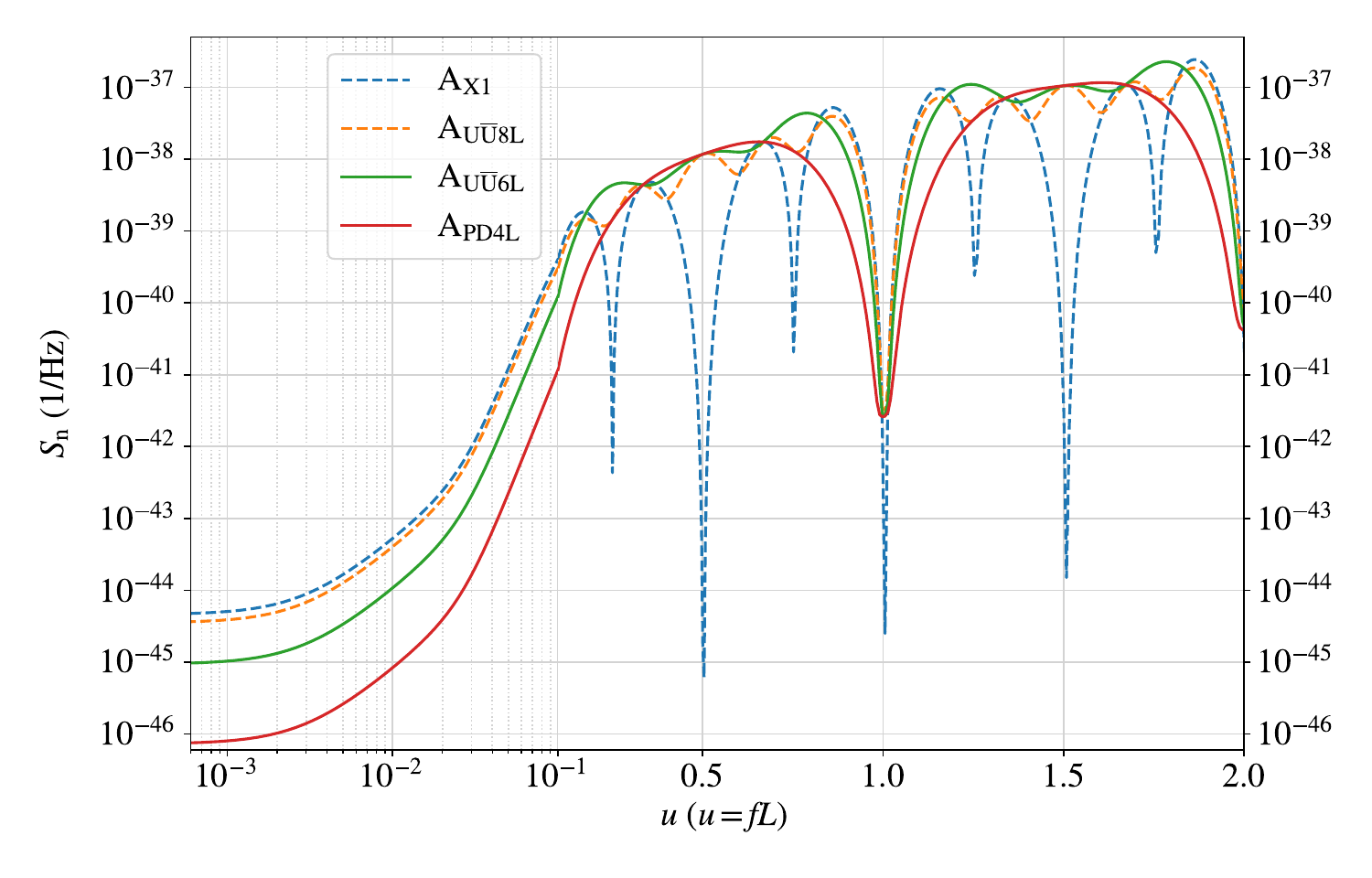}
\includegraphics[width=0.45\textwidth]{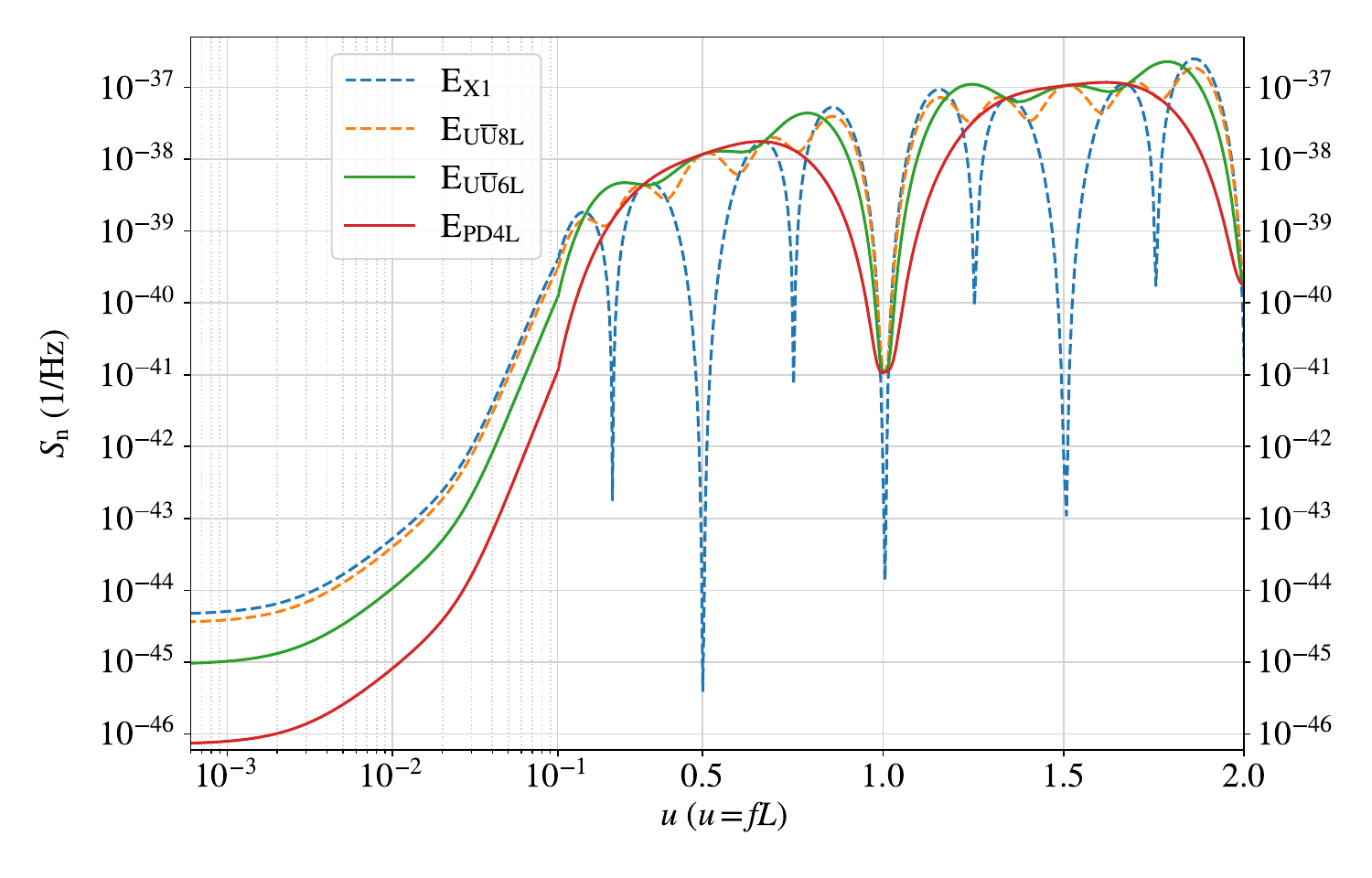}
\includegraphics[width=0.45\textwidth]{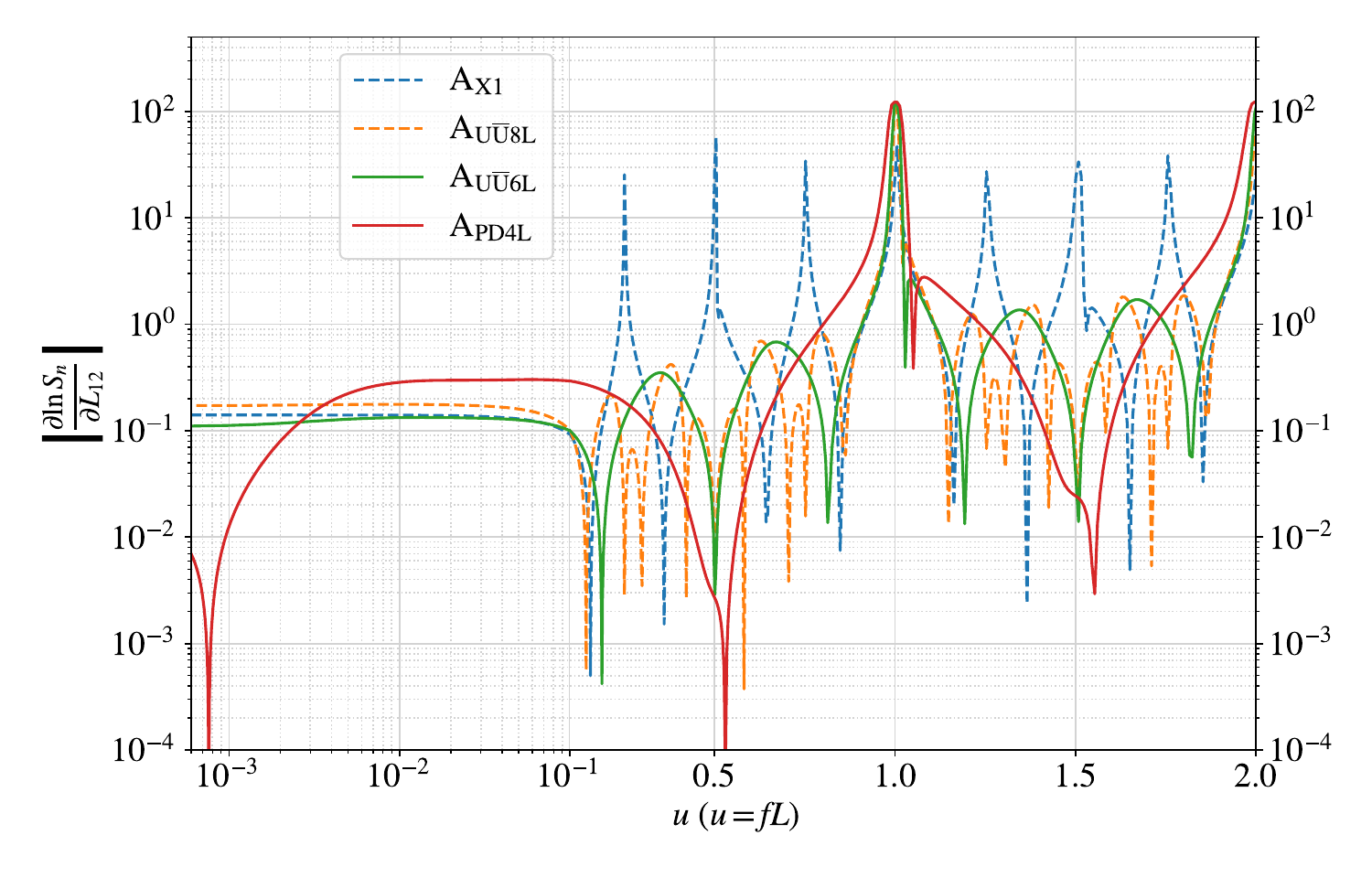}
\includegraphics[width=0.45\textwidth]{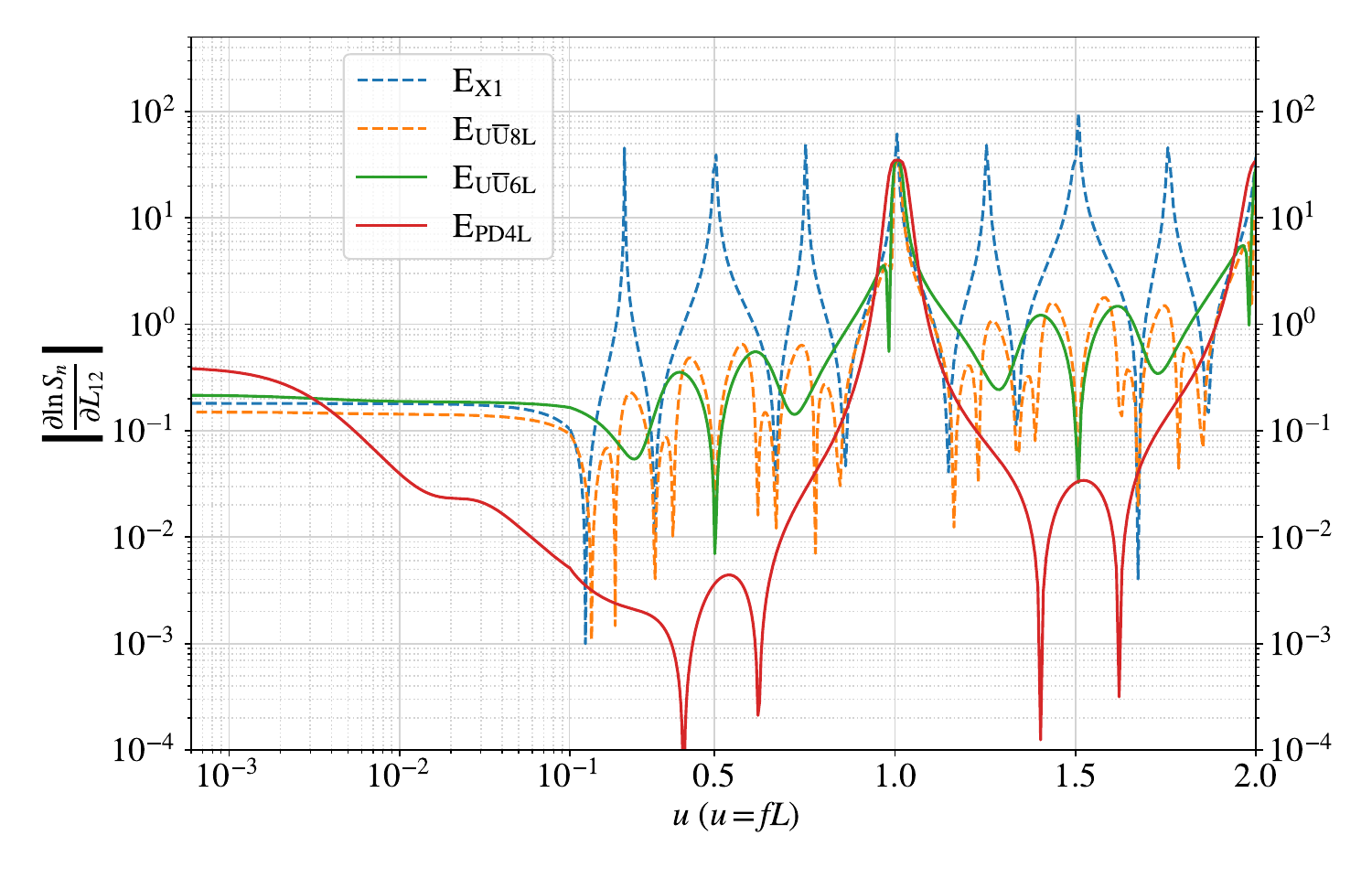}
\includegraphics[width=0.45\textwidth]{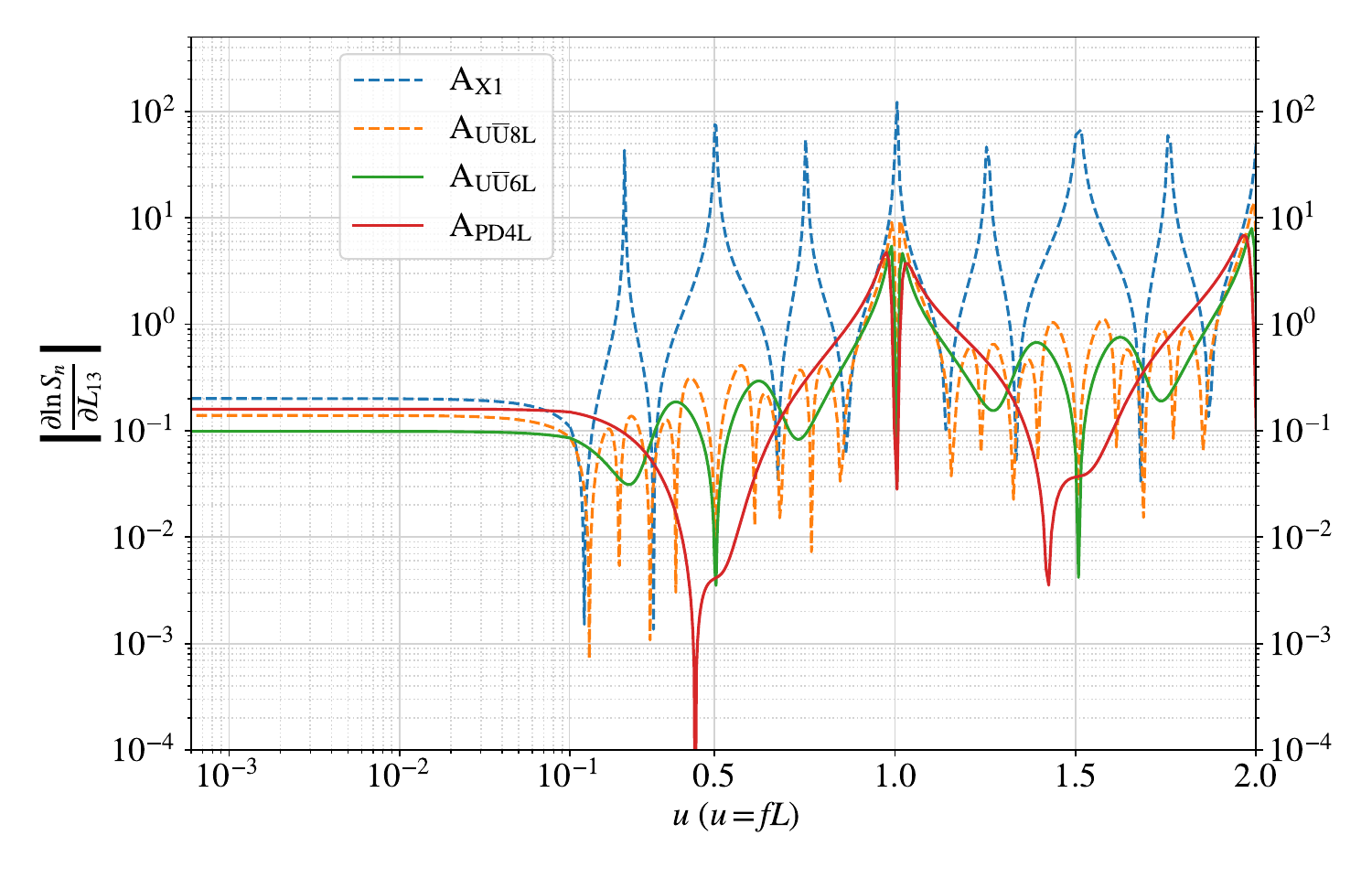}
\includegraphics[width=0.45\textwidth]{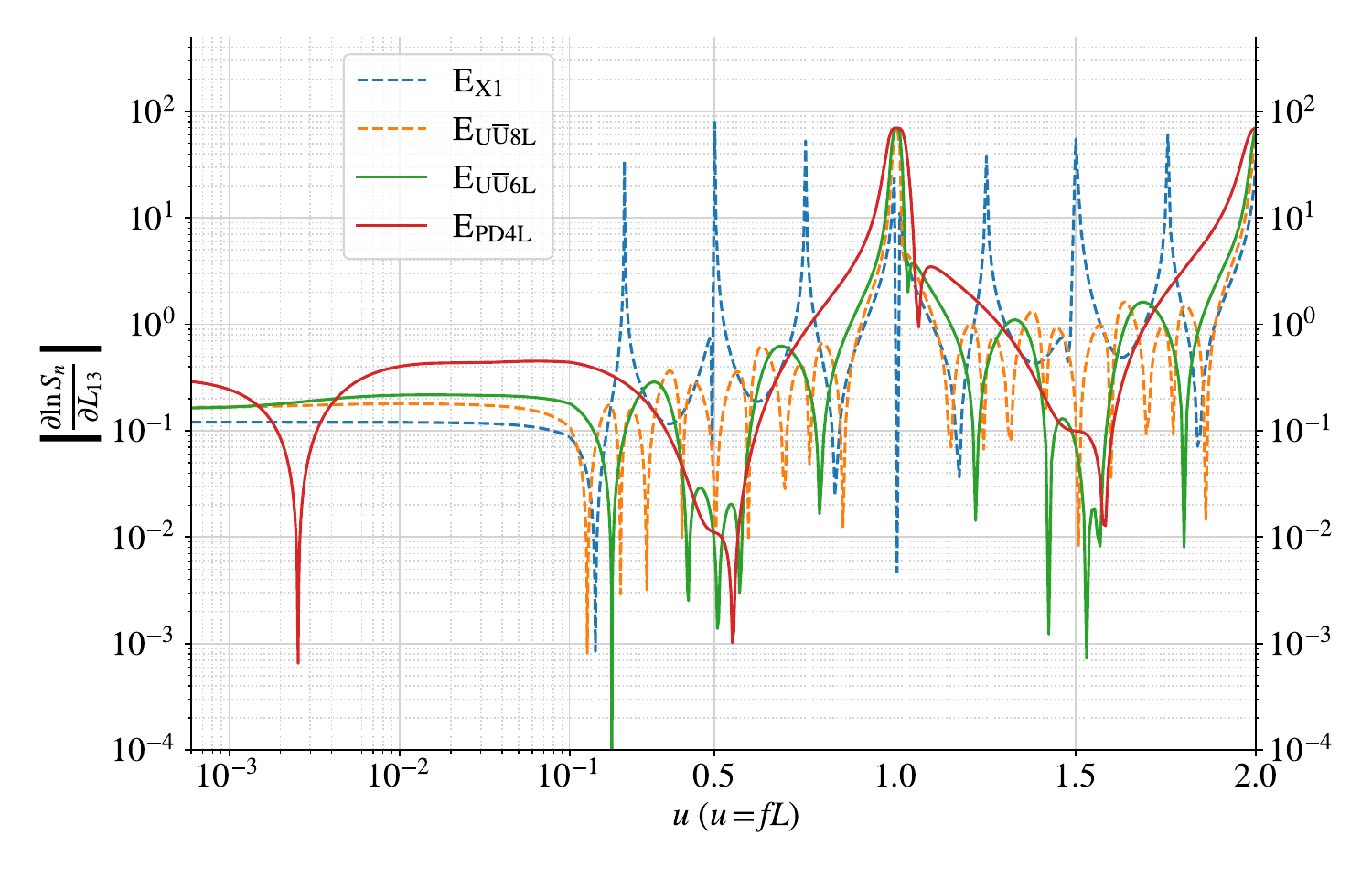}
\includegraphics[width=0.45\textwidth]{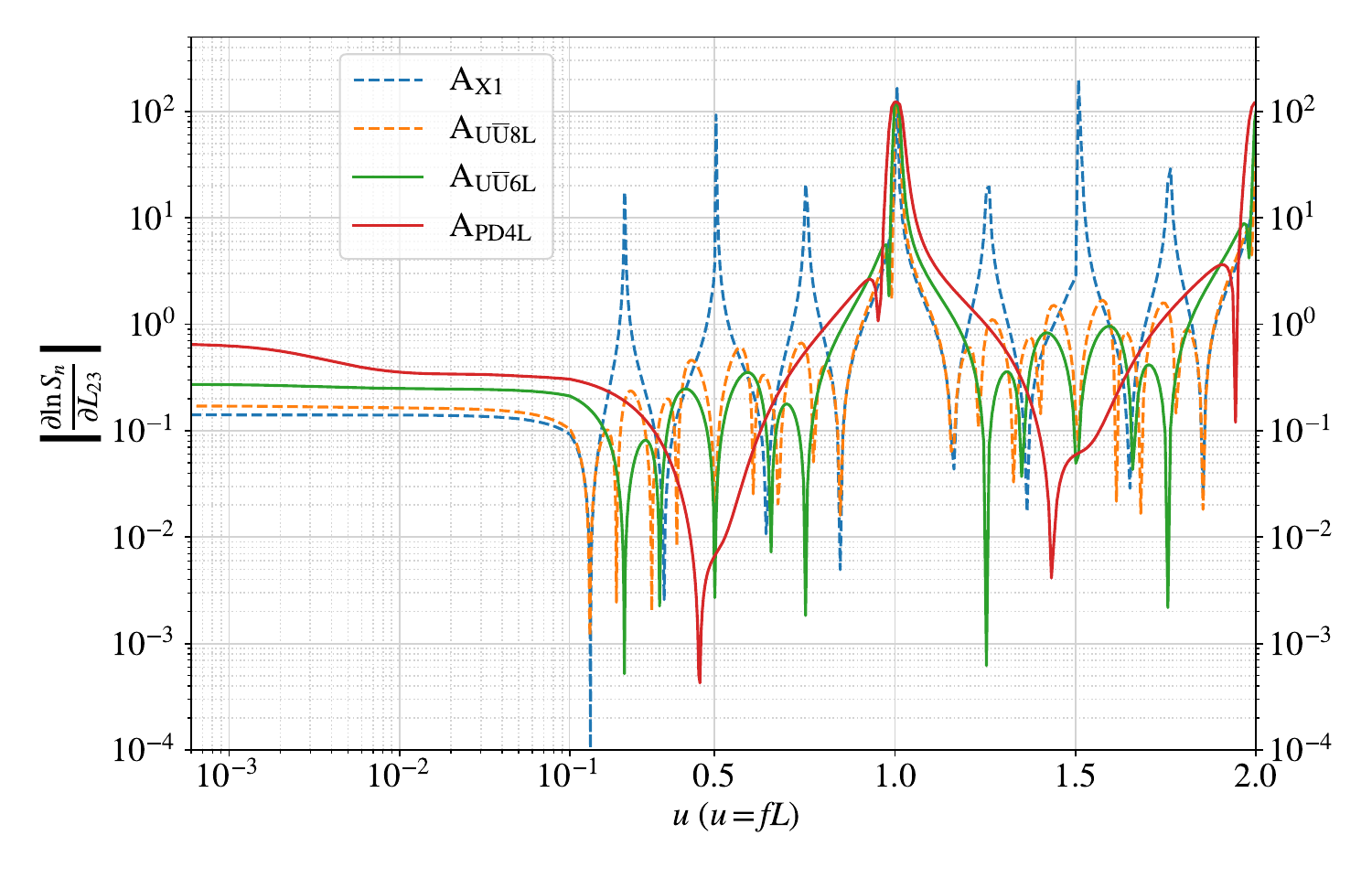}
\includegraphics[width=0.45\textwidth]{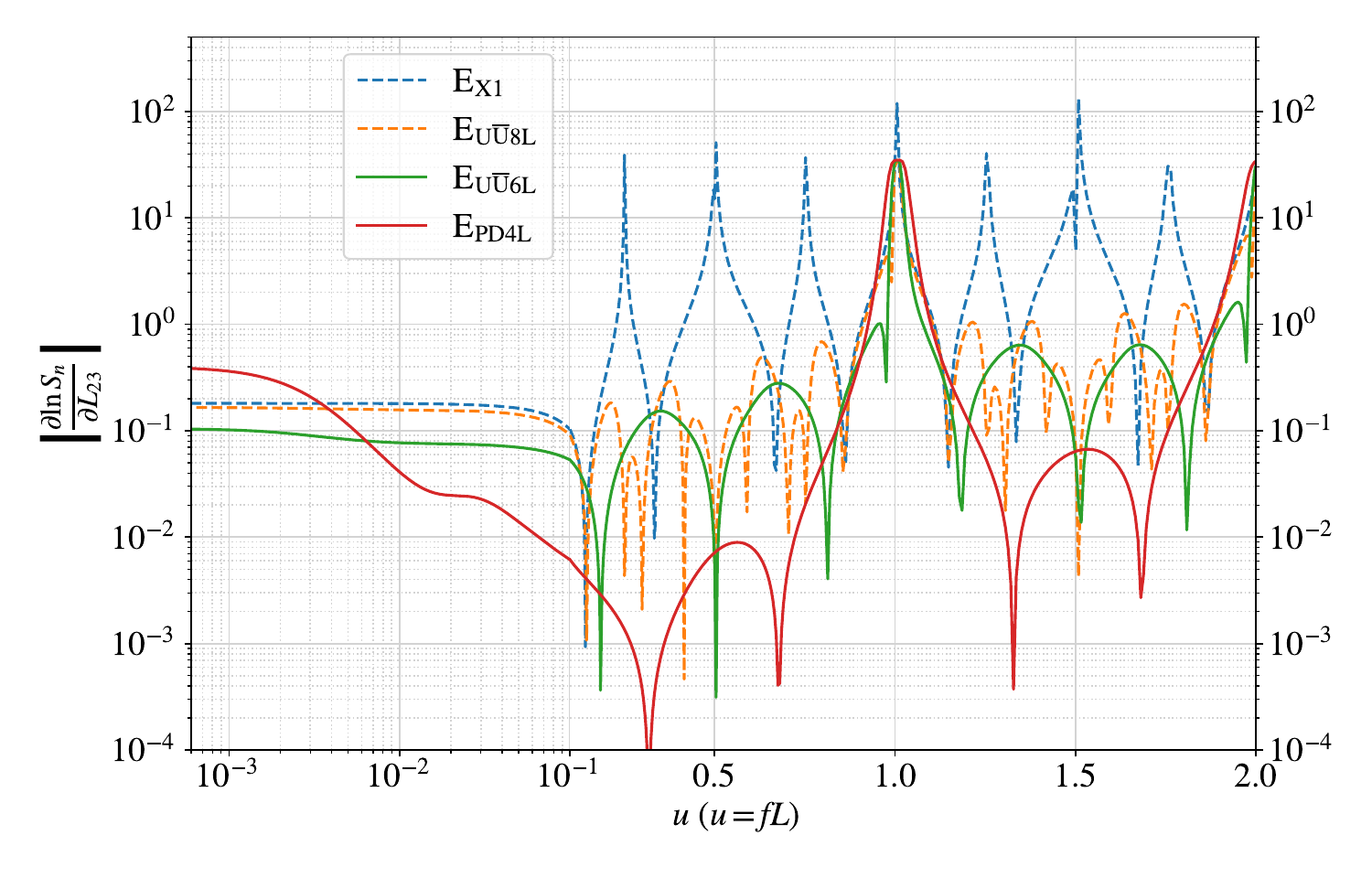}
\caption{\label{fig:dSn_dLij_AE_8L6L4L}
Noise PSDs (top row) and their logarithmic derivatives with respect to the arm lengths $L_{12}$, $L_{13}$, and $L_{23}$ (second to fourth rows), for optimal TDI channels A (left) and E (right) from selected configurations. Although A and E are expected to have identical PSDs, they respond differently to arm-length variations and S/C-dependent noise. Here we assume identical acceleration and OMS noises for all three S/C. The most prominent derivatives are near null frequencies, where PSDs are suppressed and more sensitive to arm variations. In particular, A$_\mathrm{X1}$ and E$_\mathrm{X1}$ channels exhibit the greatest instability at high frequencies among all cases due to their most frequent nulls. (Results partially reproduced from \cite{Wang:2025mee}.)
}
\end{figure*}

Although the GW response and noise spectra are highly correlated between paired science channels across different TDI configurations, their noise PSDs could differ due to dependence on specific arms lengths and instrumental noises from each S/C. Even assuming constant instrumental noise, the PSDs vary over time as a result of arm-length evolution induced by orbital dynamics.
To assess this effect, we compare three alternative TDI configurations with time spans of $8L$, $6L$, and $4L$ to the Michelson configuration. Figure \ref{fig:dSn_dLij_AE_8L6L4L} shows the noise PSDs for the A (left column) and E (right column) channels, along with their derivatives with respect to the three arm lengths. The top row presents the noise spectra of the science channels. Except Michelson’s A$_\mathrm{X1}$ and E$_\mathrm{X1}$ showing nulls at $u = m/4$ ($m = 1,2,3,\ldots$), all other observables exhibit nulls at integer values of $u = m$. 

The second to fourth rows of Fig. \ref{fig:dSn_dLij_AE_8L6L4L} display the logarithmic derivatives of the noise PSDs with respect to $L_{12}$, $L_{13}$, and $L_{23}$, respectively. In these computations, variations in $L_{ij}$ and $L_{ji}$ are treated symmetrically, and $\partial \ln S_n / \partial L_{ij}$ captures contributions from both. The curves show sharp peaks near null frequencies due to the steep PSD drops at these locations, making them especially sensitive to small arm-length changes. Negative spikes often indicate sign changes when taking absolute values.
In the low-frequency regime ($u < 0.1$), the noise spectra from the 8L configurations (Michelson and hybrid Relay $8L$) are more stable, with derivatives generally less than 0.2 across all arms (dashed lines). The U$\overline{\mathrm{U}}$6L configuration shows moderate variations, while PD4L exhibits noticeable differences among the arms. A likely reason is that TDI configuration with shorter time span yield lower PSDs in the low-frequency band. Thus, even a same change appears more significant when normalized by a smaller baseline.
In the high-frequency range ($u > 0.1$), the stabilities of different TDI configurations diverges more substantially. The Michelson configuration becomes the most unstable due to its most frequent null frequencies. In contrast, U$\overline{\mathrm{U}}$8L and U$\overline{\mathrm{U}}$6L maintain more consistent fluctuations across all arms. The PD4L channels exhibit stronger arm-dependent instability, especially near its null frequencies.

\begin{figure*}[htbp]
\includegraphics[width=0.48\textwidth]{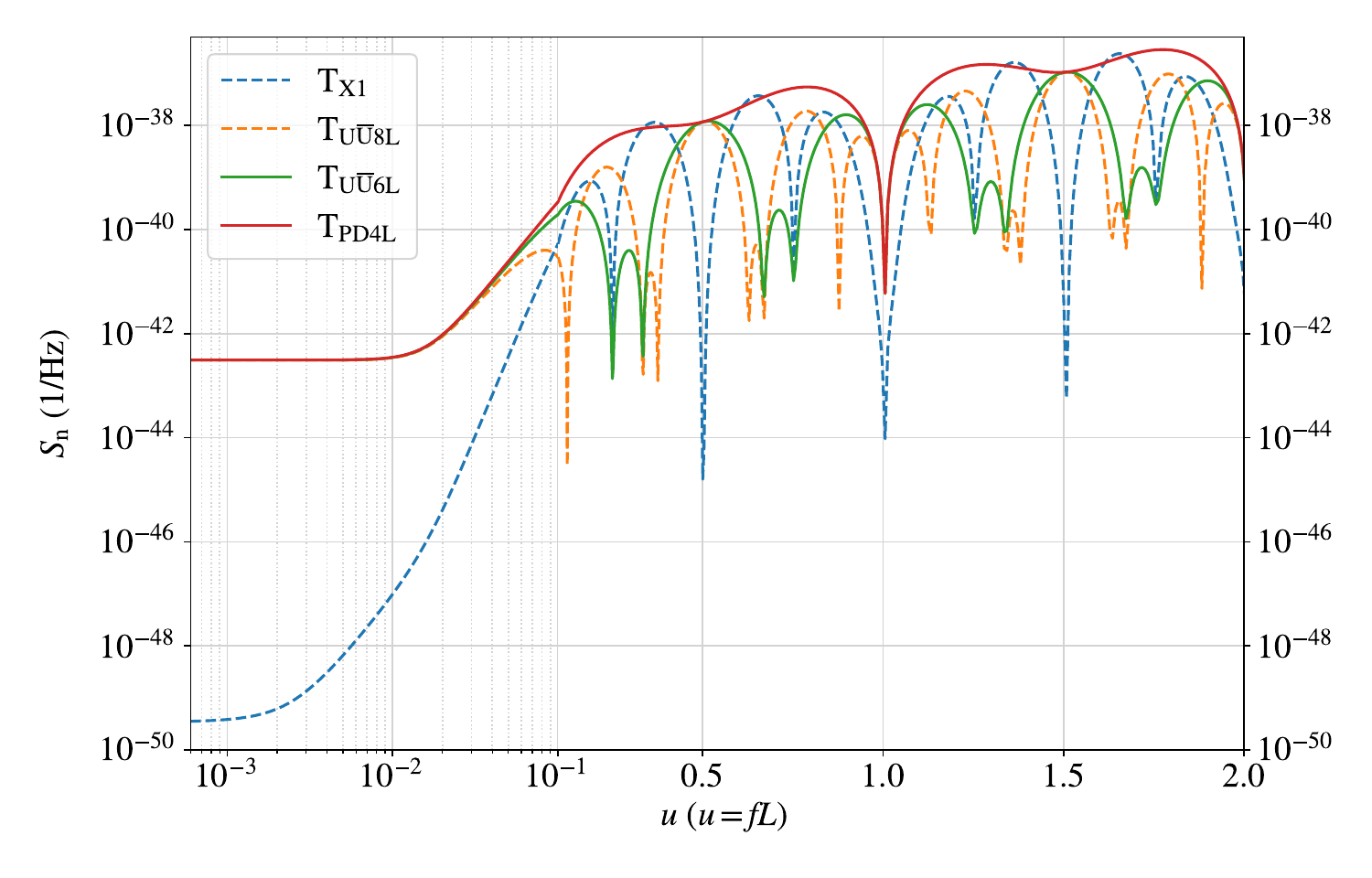}
\includegraphics[width=0.48\textwidth]{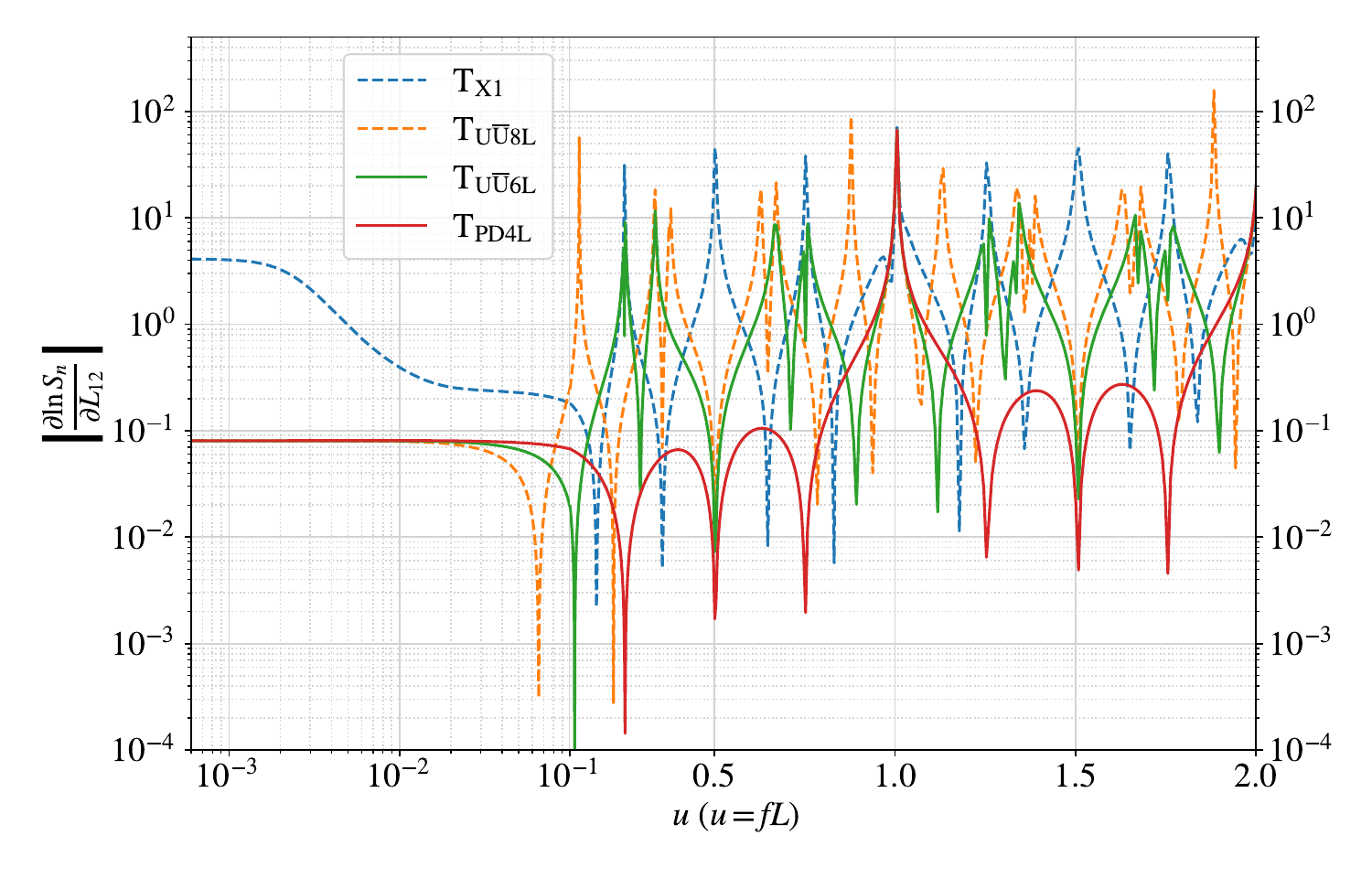}
\includegraphics[width=0.48\textwidth]{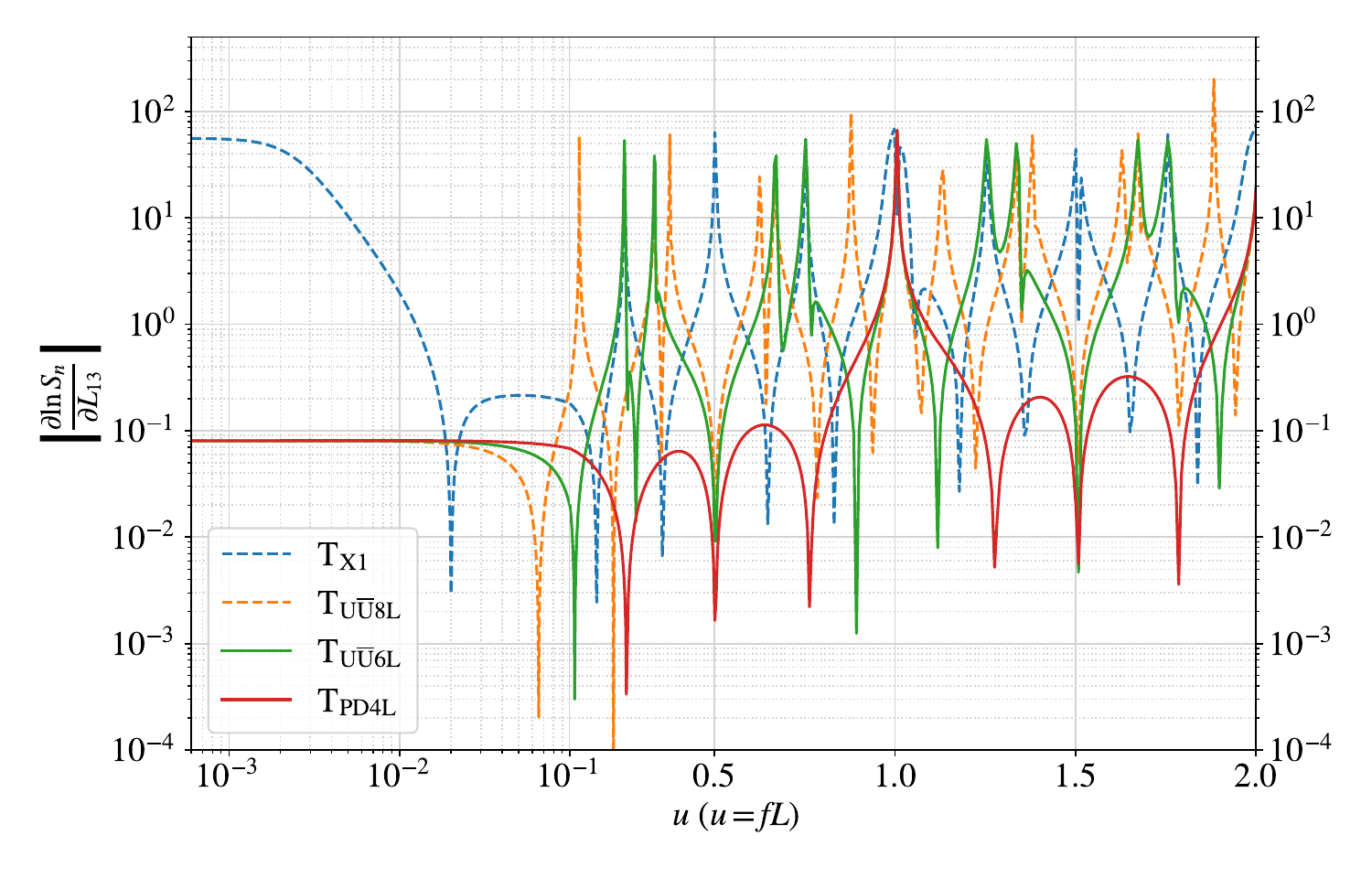}
\includegraphics[width=0.48\textwidth]{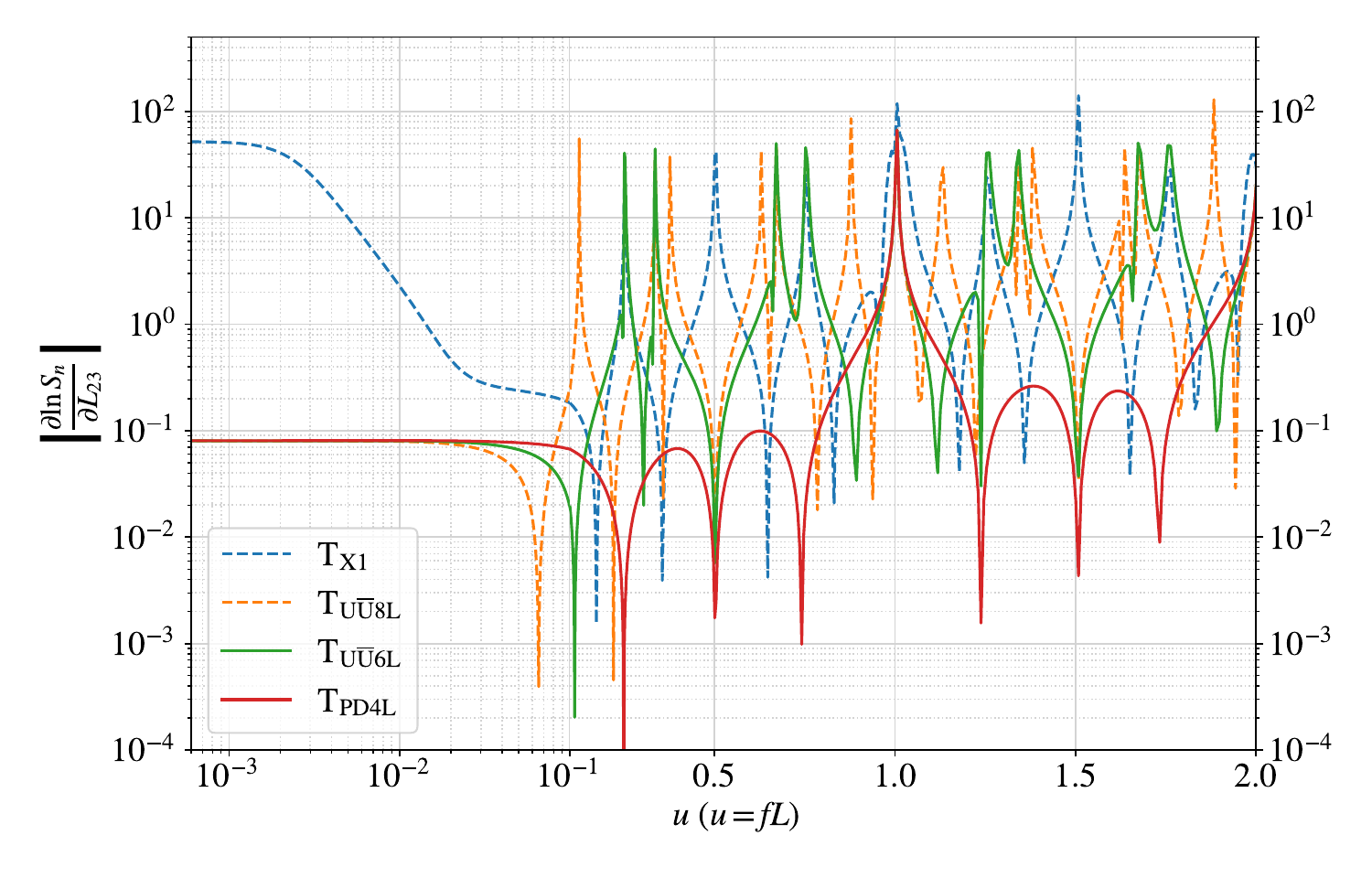}
\caption{\label{fig:dSn_dLij_T}
Noise PSDs (top-left) of the null channels and their derivatives with respect to $L_{12}$, $L_{13}$, and $L_{23}$ (remaining panels), for four TDI configurations. Among them, T$_\mathrm{PD4L}$ has the fewest null frequencies (only at $u = m$) and the highest PSD level, making it the most stable null stream (see red solid curves). In contrast, T$_\mathrm{X1}$ has nulls at $u = m/4$ and shows the lowest PSDs at $u < 0.1$, making it the most unstable in this band. Due to symmetric usage of all arms, the derivatives are nearly identical across arm lengths, except for T$_\mathrm{X1}$, which sensitive to arm-length asymmetry at low frequencies. T$_\mathrm{U\overline{U}6L}$ and T$_\mathrm{U\overline{U}8L}$ remain relatively stable at low frequencies but become less stable as frequency increases. (Results partially reproduced from \cite{Wang:2025mee}.)
}
\end{figure*}

Figure \ref{fig:dSn_dLij_T} focuses on the null channels. The top-left panel shows the PSDs of null streams from the four TDI configurations, while other panels display their logarithmic derivatives with respect to each arm length. Among these, T$_\mathrm{PD4L}$ exhibits the fewest null frequencies (at $u = m$) and the highest overall PSD among the null channels. These characteristics contribute to its superior spectral stability, as reflected by the small derivatives across all arms. Conversely, T$_\mathrm{X1}$ has notably low PSDs for $u < 0.1$ and nulls at $u = m/4$, which make it particularly unstable, as indicated by its large derivatives.
T$_\mathrm{U\overline{U}6L}$ and T$_\mathrm{U\overline{U}8L}$ exhibit more complex null patterns, leading to unstable spectra in the frequency band $u>0.1$.
From the three derivative plots, it's evident that the derivatives of a null stream with respect to each arm length are similar. This similarity stems from the symmetric structure of null streams, where all three arms are combined using identical delay operations. Assuming equal arm lengths and identical noise characteristics for each S/C, the PSD of a null channel can be written as:
\begin{equation}
S_\mathrm{T, TDI} = \sum_{ij} \left[ \mathcal{C}_\mathrm{acc, TDI} S_{\mathrm{acc}, ij} + \mathcal{C}_\mathrm{OMS, TDI} S_{\mathrm{OMS}, ij} \right].
\end{equation}
where $\mathcal{C}_\mathrm{acc, TDI}$ and $\mathcal{C}_\mathrm{OMS, TDI}$ are respective coefficients for the acceleration noise and OMS noise, and they are (closely) identical across six components. This symmetry also results in a strong correlation among null channels, as discussed in Fig. \ref{fig:TDI_resp_csd_correlation}. The T$_\mathrm{X1}$ observable is an exception, where the asymmetry between three arms takes effect at low frequencies.

\subsection{Average sensitivity of optimal channels}

\begin{figure*}[htbp]
\includegraphics[width=0.45\textwidth]{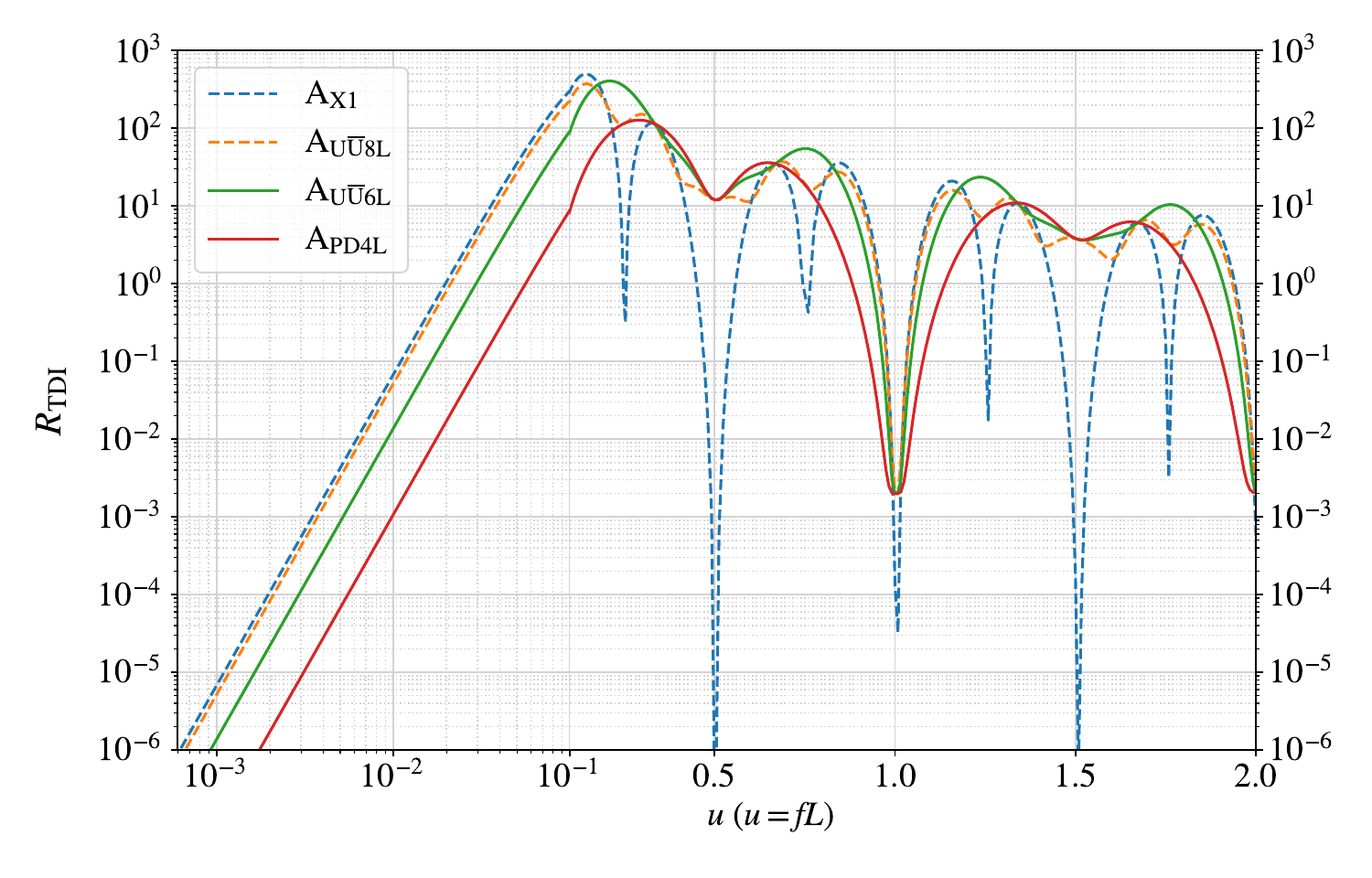}
\includegraphics[width=0.45\textwidth]{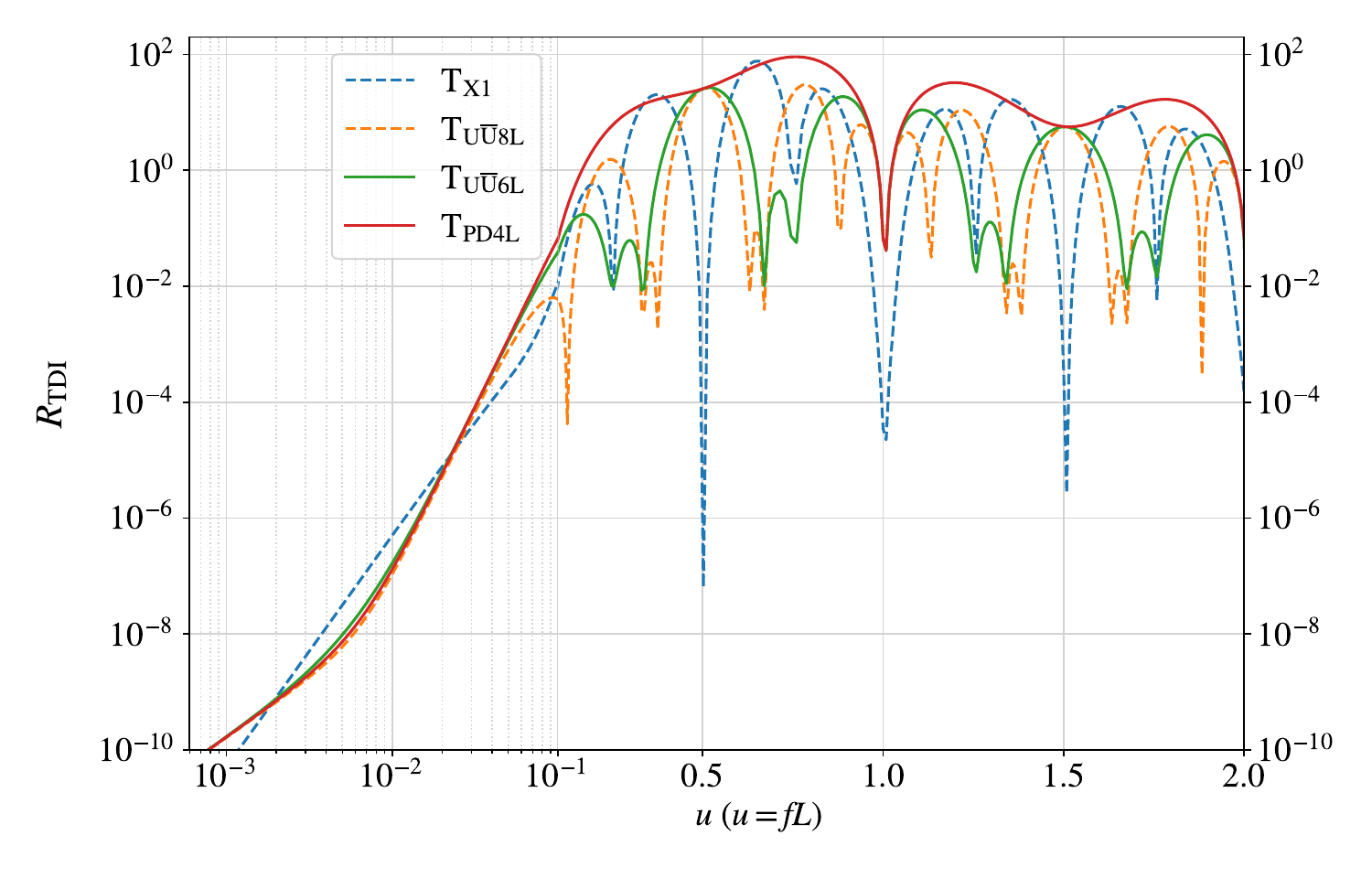}
\caption{\label{fig:resp_avg}
Sky-averaged GW response of selected TDI channels. Left panels show the A (science) channels, while right panels show the T (null) channels. At low frequencies ($u < 0.1$), all science channels exhibit nearly constant responses, consistent with the long-wavelength approximation. At higher frequencies, strong frequency dependence emerges, particularly near response nulls. The spikes observed around the nulls are primarily due to numerical artifacts.
}
\end{figure*}

The sky-averaged GW responses of selected science and null TDI channels are illustrated in Fig.~\ref{fig:resp_avg}. For the science channels (upper left panel), A$_\mathrm{X1}$ exhibits response nulls at frequencies $u = m/4$ consistent with its PSD. And other three science channels exhibit nulls only at $u=m$, resulting in smoother response curves at high frequencies for $u>0.1$.
At low frequencies ($u < 0.1$), the average response follows a power-law trend. A$_\mathrm{X1}$ and A$_\mathrm{U\overline{U}8L}$ have comparable response levels, while A$_\mathrm{U\overline{U}6L}$ is slightly lower due to its shorter effective arm length to cumulate signal. A$_\mathrm{PD4L}$ yields the lowest response among the four, reflecting its shortest TDI time span. However, this shorter span also leads to lower noise level in the same band, as evidenced in the PSDs shown earlier in Fig.~\ref{fig:dSn_dLij_AE_8L6L4L}.
For the null channels (right panel of Fig.~\ref{fig:resp_avg}), T$_\mathrm{PD4L}$ exhibits the smoothest response across the full frequency band. Other null channels are affected by their respective null frequencies in the high-frequency regime. Nonetheless, all four null channels display comparable average responses in the low-frequency region ($u < 0.1$), which are significantly lower than those of the science channels.
Mollweide sky maps of the GW response at three representative dimensionless frequencies, $u = [0.1, 0.72, 1.25]$, are presented in Appendix~\ref{sec:Mollweide_maps} to illustrate the frequency-dependent evolution of the antenna pattern. The corresponding statistical measures (variance, skewness, and kurtosis) for these TDI channels are shown in Fig.~\ref{fig:resp_variance_skewness_kurtosis}.

\begin{figure*}[thbp]
\includegraphics[width=0.48\textwidth]{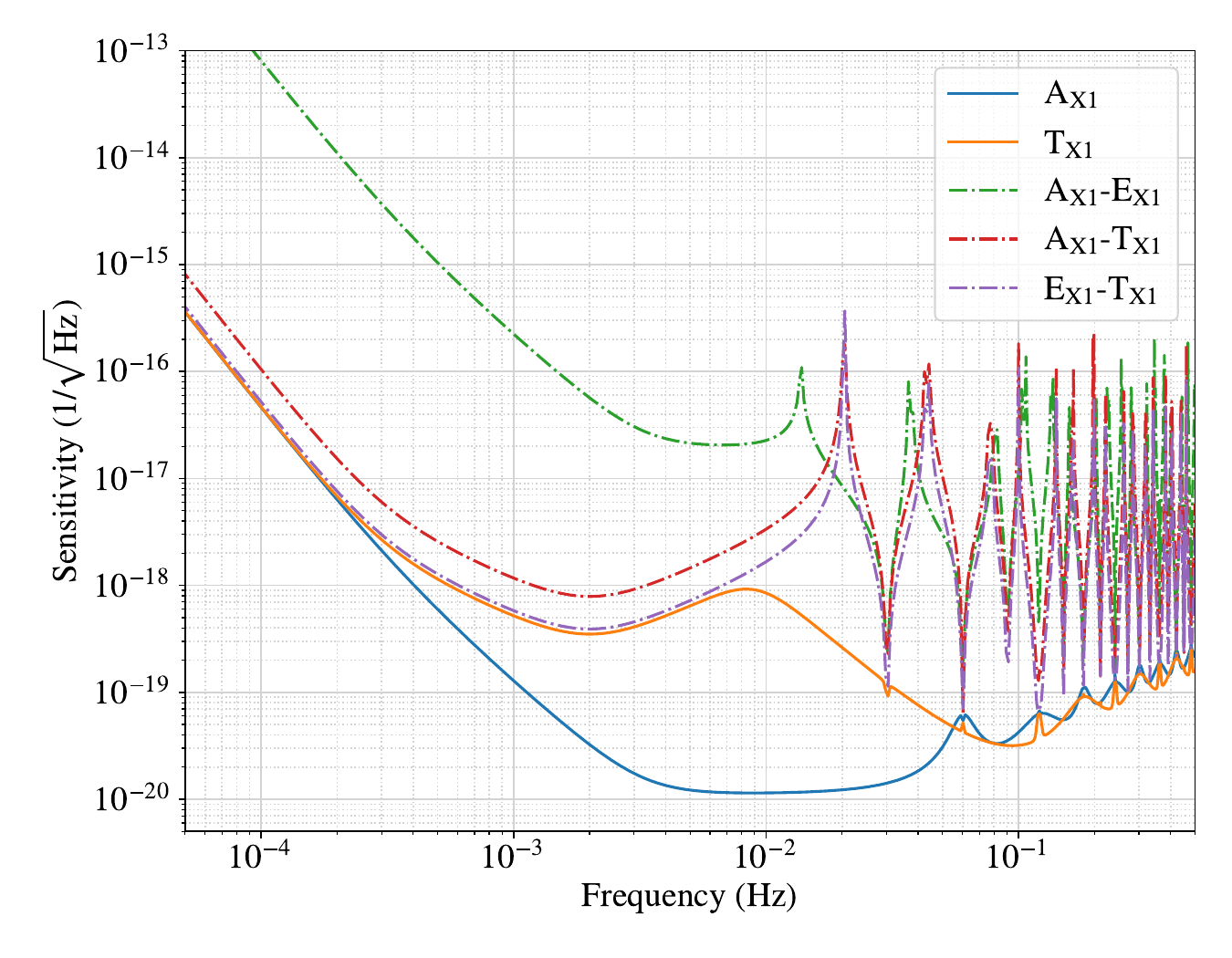}
\includegraphics[width=0.48\textwidth]{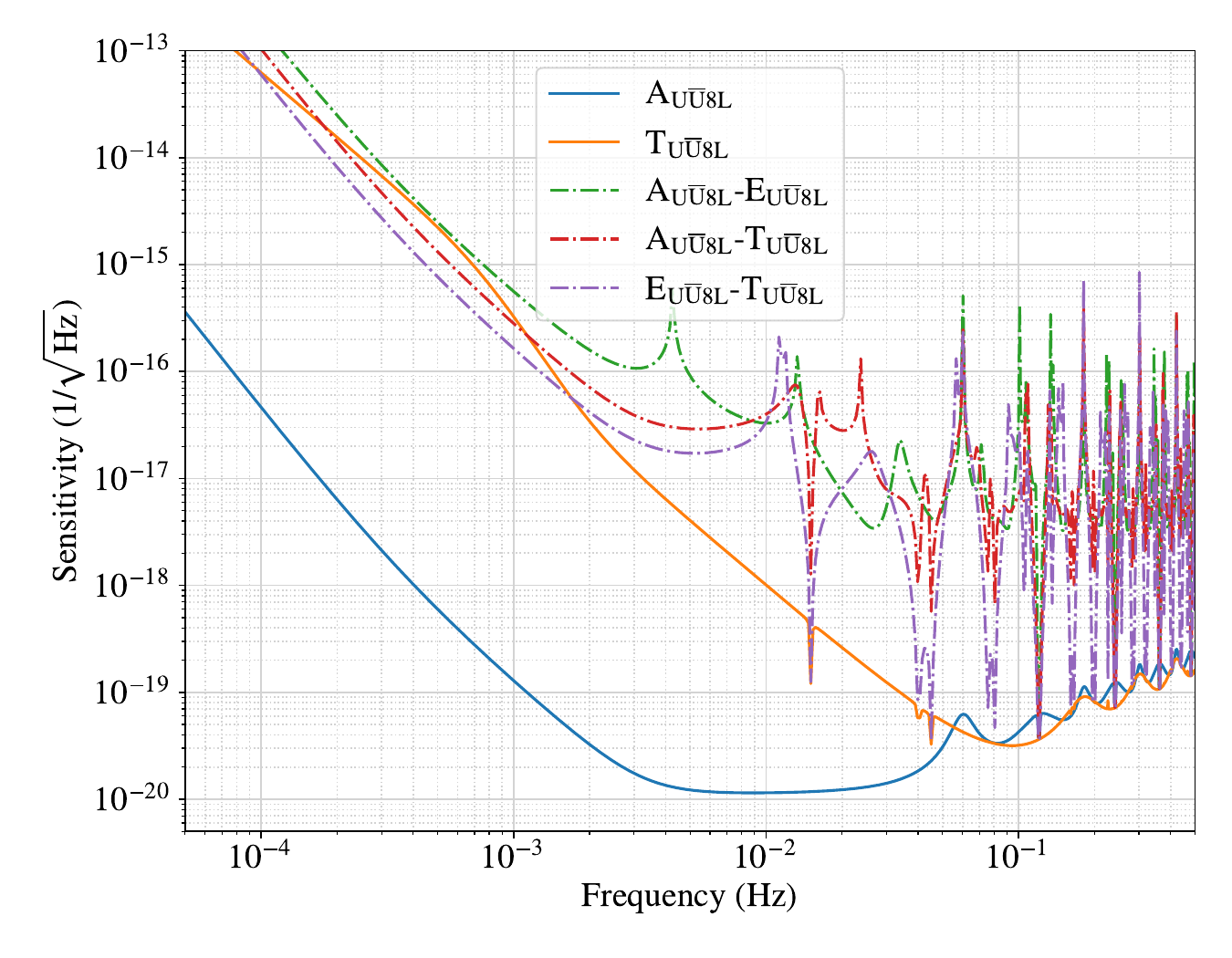}
\includegraphics[width=0.48\textwidth]{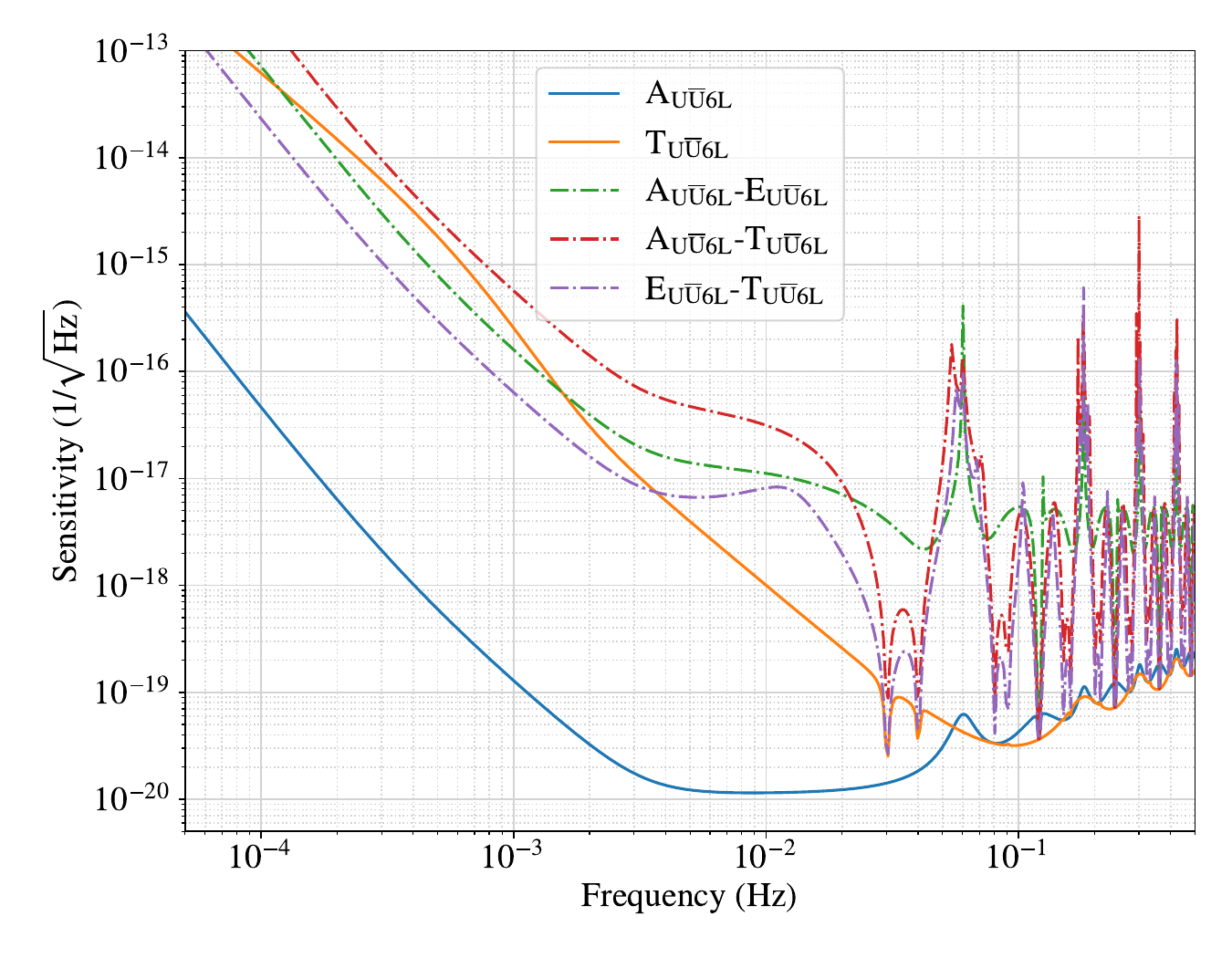}
\includegraphics[width=0.48\textwidth]{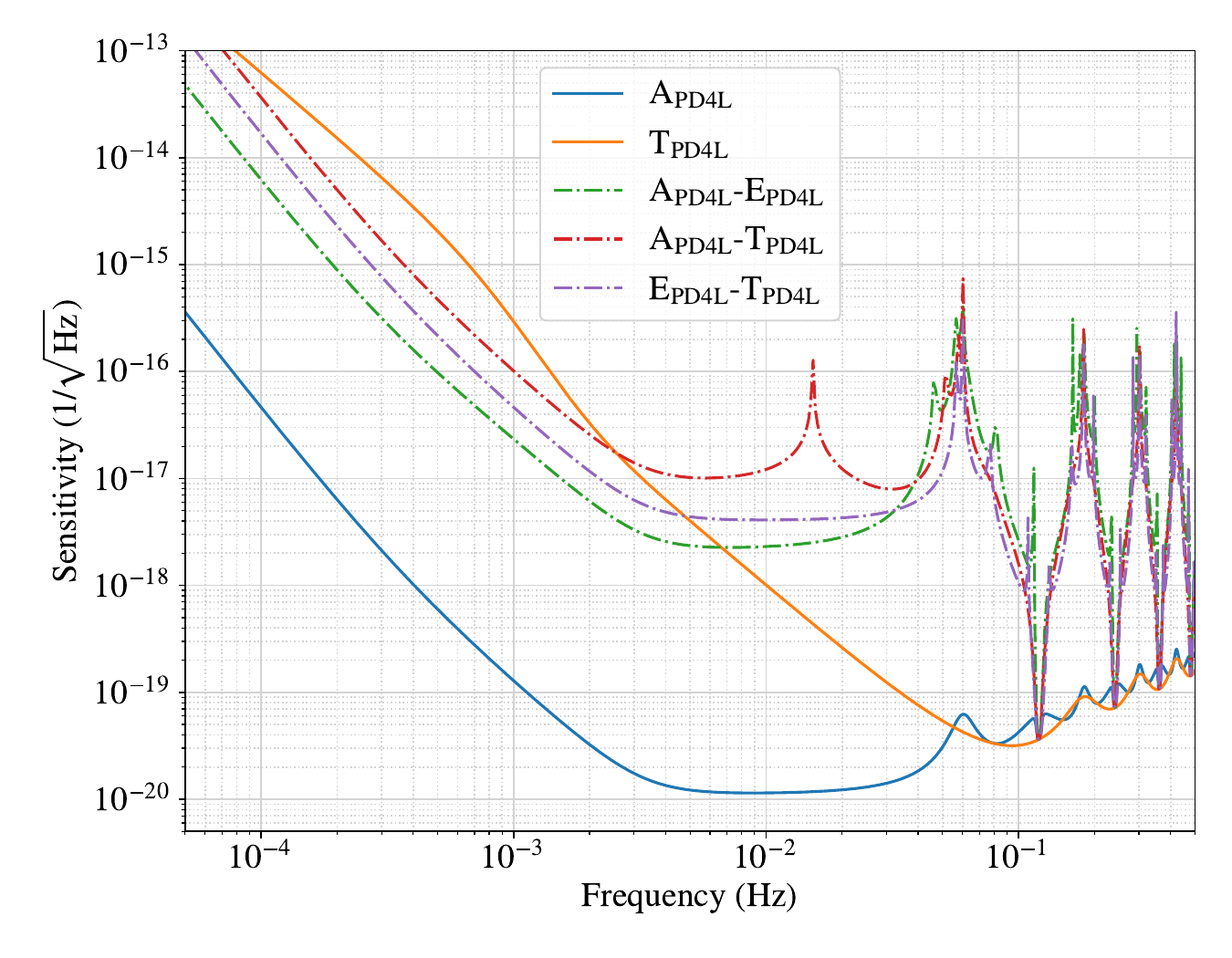}
\caption{\label{fig:sensivity_8L6L4L}
Average sensitivities of optimal channels for TDI configurations: Michelson (upper left), hybrid Relay $8L$ (upper right), hybrid Relay $6L$ (lower left), and PD4L (lower right). Solid curves show sensitivities of single channels; dash-dot curves show those from cross-correlations. Although $R_a$ and $S_a$ differ between configurations, the sensitivities of single science channels are identical (solid blue curves). The cross-correlation sensitivities depend on configuration and frequency. In the low-frequency regime, the cross-channel sensitivities are orders of magnitude worse than diagonal ones, justifying the diagonal approximation. At high frequencies, the off-diagonal terms become more relevant. Cross-correlation sensitivities are also affected by nulls; the PD4L configuration shows the smoothest behavior across the full band.
}
\end{figure*}

The likelihood function for parameter estimation is defined as \cite{Romano:2016dpx}:
\begin{equation} \label{eq:likelihood}
\begin{aligned}
    \ln \mathcal L(d|\vec{\theta}) = \sum_{f_i} \bigg[ & -\frac{1}{2} (\mathbf{\tilde{d}} - \mathbf{\tilde{h}} )^T \mathbf{C}^{-1} (\mathbf{\tilde{d}} - \mathbf{\tilde{h}} )^\ast
      - \frac{1}{2} {\rm ln} (\det 2\pi \mathbf{C} ) \bigg]
\end{aligned}
\end{equation}
where $\mathbf{\tilde{d}}$ is the frequency-domain data, $\mathbf{\tilde{h}}$ represents the signal model vectors derived from frequency-domain waveform (or Fourier-transformed time-domain waveform), and $\mathbf{C}$ is the noise covariance matrix of three optimal TDI channels:
\begin{equation} \label{eq:cov_mat}
\begin{aligned}
 \mathbf{C} = & \frac{T_\mathrm{obs}}{4}
 \begin{bmatrix}
S_\mathrm{A} & S_\mathrm{AE} & S_\mathrm{AT} \\
S_\mathrm{EA} & S_\mathrm{E} & S_\mathrm{ET} \\
S_\mathrm{TA} & S_\mathrm{TE} & S_\mathrm{T}
\end{bmatrix},
\end{aligned}
\end{equation}
with $T_\mathrm{obs}$ the observation time. Since the A, E, and T channels are quasi-orthogonal, the off-diagonal terms are typically small compared to the diagonal ones: $|S_{ab}| \ll S_a, S_b$. Therefore, the inverse matrix can be approximated as \cite{Creighton:2011zz}:
\begin{equation} \label{eq:inv_cov_mat}
\begin{aligned}
 \mathbf{C^{-1}} \simeq & \frac{4}{T_\mathrm{obs}}
 \begin{bmatrix}
\frac{1}{S_\mathrm{A}} & \frac{S_\mathrm{AE}}{S_\mathrm{A} S_\mathrm{E} } & \frac{ S_\mathrm{AT} }{S_\mathrm{A} S_\mathrm{T} } \\
\frac{ S_\mathrm{EA} }{S_\mathrm{E} S_\mathrm{A} } & \frac{1}{S_\mathrm{E} } & \frac{ S_\mathrm{ET} }{S_\mathrm{E} S_\mathrm{T} } \\
\frac{ S_\mathrm{TA} }{S_\mathrm{T} S_\mathrm{A} } & \frac{ S_\mathrm{TE} }{S_\mathrm{T} S_\mathrm{E} } & \frac{1}{ S_\mathrm{T} }
\end{bmatrix}.
\end{aligned}
\end{equation}
The average sensitivity of a single TDI channel is defined as:
\begin{equation} \label{eq:sensitivity_a_avg}
S_{a, \mathrm{avg} } = \left(  \frac{ R_{a}  }{ S_a } \right)^{-1/2},
\end{equation}
and the cross-correlation sensitivity between two channels as:
\begin{equation}
S_{ab, \mathrm{avg} } = \left(  \frac{  \Re[  R_{ab} S^\ast_{ab} ] }{ S_a S_b } \right)^{-1/2}.
\end{equation}

Fig.~\ref{fig:sensivity_8L6L4L} shows the instantaneous sky-average sensitivities of four TDI configurations: Michelson (upper left), hybrid Relay $8L$ (upper right), hybrid Relay $6L$ (lower left), and PD4L (lower right). The sensitivities of individual optimal channels are shown as solid curves, while the sensitivities derived from cross-correlations between two channels are shown as dash-dot curves. Since the A and E channels are expected to be equivalent, only the A-channel sensitivities are plotted.
Despite the differences in GW response functions $R_a$ and noise PSDs $S_a$ among the various TDI configurations, the sensitivities of all A channels are nearly identical. Expect for T$_\mathrm{X1}$, other three T channels also have same sensitivity. The T$_\mathrm{X1}$ may appear more sensitive in the low-frequency band due to arm-length asymmetry \cite{Adams:2010vc}, it is significantly correlated with E$_\mathrm{X1}$ in the same band \cite{Hartwig:2023pft}, and thus may not offer extra signal-to-noise ratio.
In contrast, the cross-correlation sensitivities vary with both the TDI configuration and frequency, reflecting the influence of off-diagonal terms in the covariance matrix. At low frequencies, the sensitivity from cross-correlations can be orders of magnitude worse than that from orthogonal channels, indicating that diagonalizing Eq.~\eqref{eq:cov_mat} is only an approximation. At higher frequencies, the contribution of off-diagonal elements becomes relatively significant. Additionally, the performance of cross-correlation channels is impacted by the presence of null frequencies. Among the configurations, the PD4L observables maintain the smoothest cross-correlations sensitivity curves.

\section{Performance in inferring MBBH signal} \label{sec:inference}

\subsection{Noise-whitened chirp signal}

\begin{figure*}[htbp]
\includegraphics[width=0.95\textwidth]{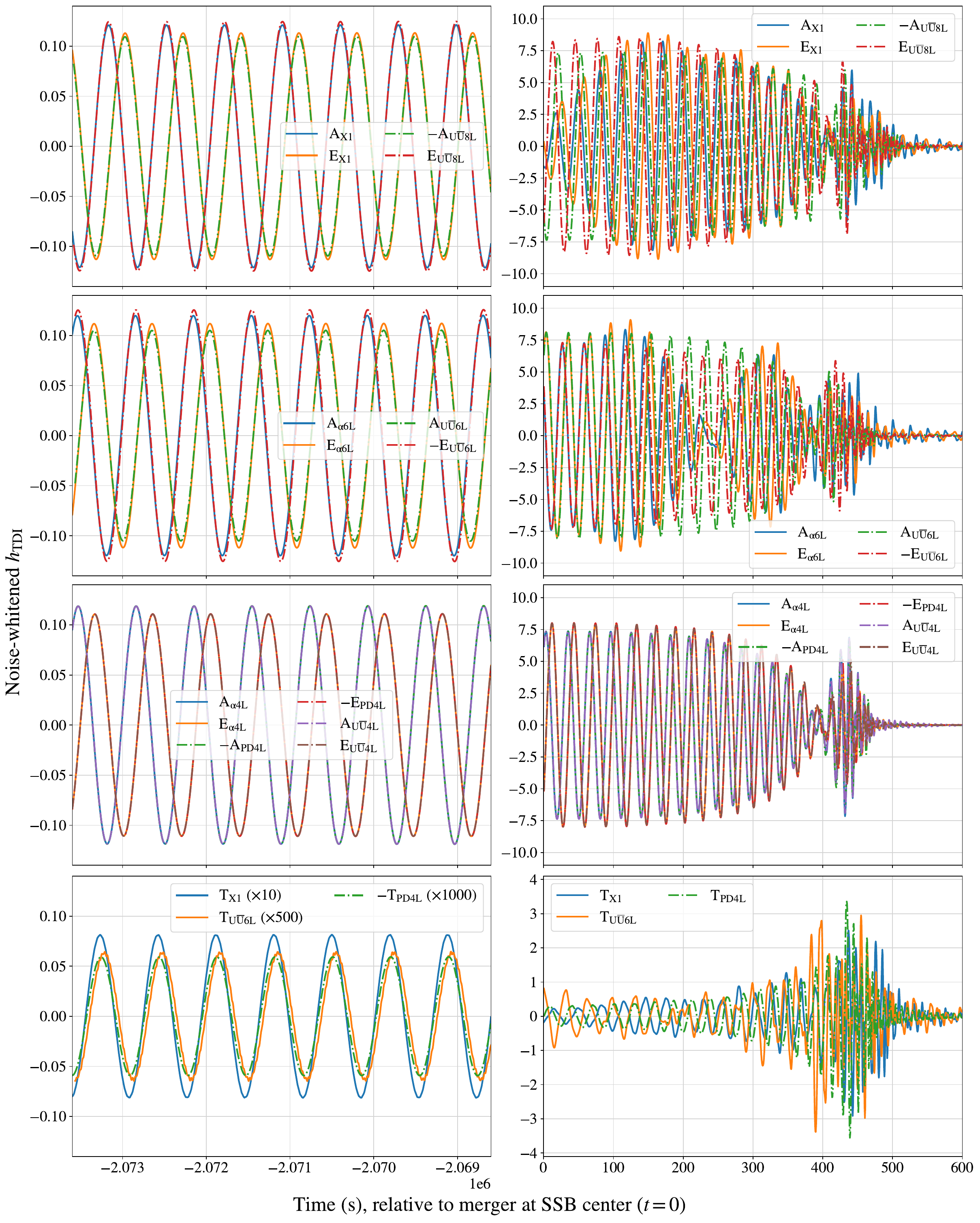}
\caption{\label{fig:TDI_whitened_waveform_td_light} 
Noise-whitened GW signals from a massive binary black hole system ($m_1 = 3 \times 10^4\,M_\odot$, $m_2 = 1 \times 10^4\,M_\odot$, $z = 0.2$), shown in the optimal channels of various TDI configurations. The first to third rows correspond to TDI channels with $8L$ (X1 and U$\overline{\mathrm{U}}$8L), $6L$ ($\alpha$6L and U$\overline{\mathrm{U}}$6L), and $4L$ ($\alpha$4L, PD4L, U$\overline{\mathrm{U}}$4L) time spans, respectively. The bottom row displays the signals in the null channels. The left column highlights the inspiral phase, while the right column shows the merger and ringdown phase. Due to the strong correlation among science channels of different TDI configurations, their waveforms overlap significantly during the inspiral phase. In the high-frequency regime shown on the right, waveform differences become apparent due to frequency aliasing, amplitude modulations from null frequencies, and signal tail effects.
}
\end{figure*}

Although the average sensitivities of the optimal science channels are identical, the response to a specific GW signal can vary across TDI configurations due to differences in transfer functions and noise characteristics. To eliminate such differences and facilitate direct comparison, we define the noise-whitened waveform as
\begin{equation}
 h_\mathrm{TDI, whiten} = \mathcal{F}^{-1} \left[ \frac{ \mathcal{F} [ h_\mathrm{TDI} (t) ] }{ \sqrt{ S_{n,\mathrm{TDI}} (f) / ( 2 \Delta t ) }  }  \right],
\end{equation}
where $\Delta t$ is the sampling interval, and $\mathcal{F}$ and $\mathcal{F}^{-1}$ denote the Fourier and inverse Fourier transforms. The noise PSD $S_{n,\mathrm{TDI}} (f)$ is computed using an analytical model in which the arm lengths are fixed at their instantaneous values at the midpoint of the observation. This approximation does not exactly match the time-dependent delays involved in constructing the GW signal $h_\mathrm{TDI}(t)$, particularly near response nulls, and therefore may lead to incomplete cancellation of the null-induced features after whitening. In practice, the PSD is typically estimated directly from the data, often by segmenting the time series and averaging over multiple segments. Several effects—such as stochastic fluctuations in the spectrum, finite frequency resolution, time-varying arm lengths, averaging procedures, and spectral leakage—can further limit the accuracy of the whitening process. As a result, the residual modulations observed around the null frequencies can also reflect limitations in estimating the PSD of the TDI channels.
We set $\Delta t = 0.5$ s, with interpolation applied during the TDI process \cite{Shaddock:2004ua,Bayle:2019dfu}.
Figure~\ref{fig:TDI_whitened_waveform_td_light} presents the whitened waveforms obtained from various TDI optimal channels. The GW signal is generated using the SEOBNRv5HM waveform model \cite{Pompili:2023tna}, and TDI operations are carried out using the \textsf{SATDI} \cite{Wang:2024ssp}. The source is a binary with component masses $m_1 = 3 \times 10^4\,M_\odot$ and $m_2 = 10^4\,M_\odot$ (in the source frame). It is placed at redshift $z = 0.2$, corresponding to a luminosity distance of $d_L = 1012.3$ Mpc \cite{Planck:2018vyg,Astropy:2013muo}. This relatively nearby distance—though lower than typical astrophysical expectations—is chosen to boost the signal-to-noise ratio and highlight differences across TDI configurations. GW emitted in its merger and ringdown phases could reach frequencies above 0.1 Hz, thereby falling into high-frequency regime of LISA sensitive band.

The first to third rows of Fig.~\ref{fig:TDI_whitened_waveform_td_light} display the whitened signals from TDI science channels with time delays of $8L$ (A$_\mathrm{X1}$, E$_\mathrm{X1}$, A$_\mathrm{U\overline{U}8L}$, E$_\mathrm{U\overline{U}8L}$), $6L$ (A$_\mathrm{\alpha6L}$, E$_\mathrm{\alpha6L}$, A$_\mathrm{U\overline{U}6L}$, E$_\mathrm{U\overline{U}6L}$), and $4L$ (A$_\mathrm{\alpha4L}$, E$_\mathrm{\alpha4L}$, A$_\mathrm{U\overline{U}4L}$, E$_\mathrm{U\overline{U}4L}$, A$_\mathrm{PD4L}$, E$_\mathrm{PD4L}$), respectively. The final row shows the whitened signals in the corresponding null channels: T$_\mathrm{X1}$, T$_\mathrm{U\overline{U}6L}$, and T$_\mathrm{PD4L}$. For visual clarity, the inspiral amplitude of T$_\mathrm{X1}$, T$_\mathrm{U\overline{U}6L}$ and T$_\mathrm{PD4L}$ are scaled by factors of 10, 500 and 1000, respectively, to bring them within the same range as the other science channels.
The left column focuses on a segment of the inspiral waveform near 1.4 mHz ($u \simeq 0.01$), while the right column shows the merger-ringdown phase above $\sim$30 mHz ($u \simeq 0.25$). The x-axis is expressed in Barycentric Dynamical Time (TDB), with $t = 0$ aligned to the merger time at the solar system barycenter (SSB). During the inspiral phase, waveforms from different TDI configurations align closely, indicating strong mutual correlation. Minor phase offsets, caused by different arm lengths, have been corrected via time shifts. In the merger-ringdown phase, differences emerge when the GW wavelength becomes comparable to or shorter than the TDI time span. These effects include \cite{Wang:2025mee}:
\begin{itemize}
\item TDI configurations with longer time delays (e.g. Michelson and hybrid Relay with $8L$) are more susceptible to high-frequency aliasing;
\item Longer delays operation also lead to prolonged tails at the end of signal;
\item Configurations with more null frequencies (e.g., Michelson and Sagnac with $6L$) exhibit stronger amplitude modulations.
\end{itemize}
A common frequency-domain treatment models the TDI responded waveform as \cite[and references therein]{Vallisneri:2012np,Katz:2020hku}
\begin{equation}
h_\mathrm{TDI} (f, t) = R_\mathrm{TDI}(f, t) h(f).
\end{equation}
However, this factorized form can fail to capture aliasing and the rapid frequency evolution that occur over the finite delay span; longer spans exacerbate these effects and increase modeling error. Near merger, the GW frequency can change appreciably across the TDI delay span $\tau$, so the effective response is time-dependent and mixes neighboring frequencies, leading to frequency-domain leakage/aliasing when $\Delta f \sim (\mathrm{d}f / \mathrm{d} t) \tau \gtrsim 1/\tau$. Using shorter-span TDI (e.g., PD4L with $\tau = 4L$) reduces $\Delta f$ and mitigates the modeling error.
Unequal link delays also produce tails: at a given sample some links carry GW signal while others do not, injecting edge-like components whose Fourier transforms introduce sidelobes and ripples. Shorter temporal footprints suppress these artifacts and thereby improve spectral modeling.
Overall, TDI configurations with shorter time delays and fewer null frequencies are preferred. The observables of PD4L, (A$_\mathrm{PD4L}$, E$_\mathrm{PD4L}$, T$_\mathrm{PD4L}$), exhibit $4L$ time span and the fewest nulls, making it particularly suitable for data analysis in high frequency band.

\subsection{Performance comparison in parameter inference}

To assess the performance of different TDI configurations in analyzing GW signals from MBBHs, we conduct parameter inference using the simulated waveform shown in Fig.~\ref{fig:TDI_whitened_waveform_td_light}. The noise-whitened GW waveforms exhibit strong overlap in the low-frequency band, suggesting that all TDI configurations should yield comparable parameter inference performance for such signals. This expectation has been verified in \cite{Wang:2025mee} using both the hybrid Relay $8L$ and PD4L schemes for a MBBH with masses $(m_1=6 \times 10^5, m_2=2 \times 10^5) \mathrm{M}_\odot$ at redshift $z=0.2$.
However, at higher frequencies, TDI configurations with shorter effective delays facilitate signal modeling in the frequency domain. In this analysis, the frequency waveform is generated using the SEOBNRv5HM reduced-order model \cite{Pompili:2023tna}, and the response function of TDI is computed using the SATDI \cite{Wang:2024ssp}. Parameter inference is performed using the nested sampler \textsf{MultiNest} \cite{Feroz:2008xx,Buchner:2014nha}.

\begin{figure*}[htb]
\includegraphics[width=0.94\textwidth]{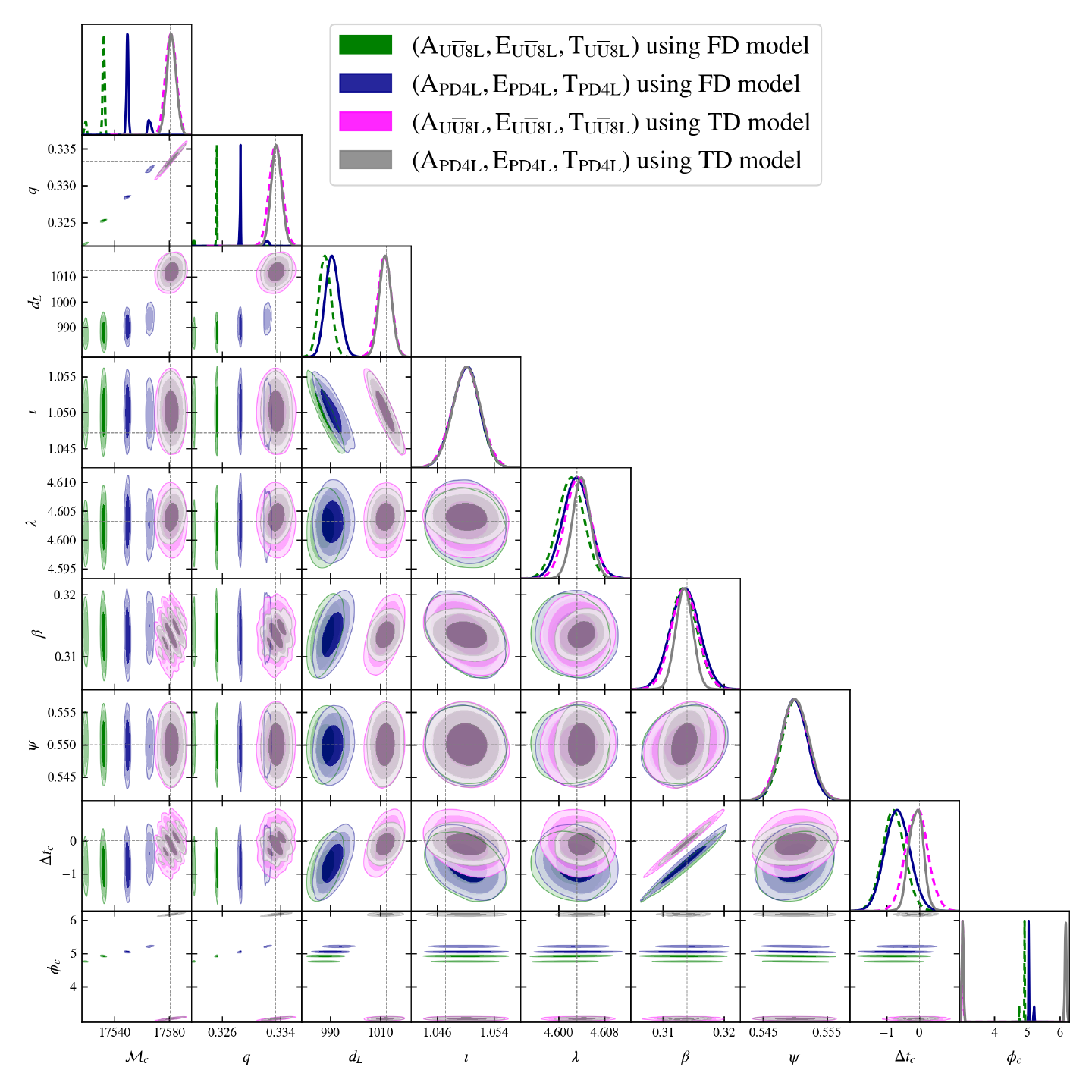}
\caption{\label{fig:corner_mbbh}
Posterior parameter distributions inferred using the hybrid Relay 8L (A$_\mathrm{U\overline{U}8L}$, E$_\mathrm{U\overline{U}8L}$, T$_\mathrm{U\overline{U}8L}$) and PD4L (A$_\mathrm{PD4L}$, E$_\mathrm{PD4L}$, T$_\mathrm{PD4L}$) TDI schemes for a GW signal from a coalescing MBBH with $(m_1=3 \times 10^4, m_2 = 10^4)\ M_\odot$ at redshift $z=0.2$. The low-frequency cutoff is $f_\mathrm{min} = 22$~mHz, corresponding to the last $\sim$1700 seconds before merger. Green and blue contours show results obtained using a frequency-domain waveform model; magenta and grey contours are based on time-domain waveforms. Shaded regions correspond to 1$\sigma$, 2$\sigma$, and 3$\sigma$ confidence intervals, with darker shades indicating higher confidence. Grey dashed lines mark the injected parameter values. To highlight TDI limitations at high frequencies, the noise PSD is artificially reduced by a factor of 10.
}
\end{figure*}

We compare the performance of the hybrid Relay $8L$ (A$_\mathrm{U\overline{U}8L}$, E$_\mathrm{U\overline{U}8L}$, T$_\mathrm{U\overline{U}8L}$) and PD4L (A$_\mathrm{PD4L}$, E$_\mathrm{PD4L}$, T$_\mathrm{PD4L}$) schemes, with posteriors shown in Fig.~\ref{fig:corner_mbbh}. Using the frequency-domain signal model, the green and blue contours represent results from the hybrid Relay and PD4L configurations, respectively. Marginal distributions from the hybrid Relay channels are shown as dashed curves, and those from PD4L as solid curves.
For mass parameters, PD4L yields posterior distributions closer to the injected values compared to the hybrid Relay configuration, although neither exactly recovers the true values, highlighting the benefits of reduced aliasing due to its shorter delay length. In addition, the inferred luminosity distance is biased relative to the injected value, primarily due to inaccuracies in recovered mass parameters.
This indicates that the underlying modeling limitations, particularly the breakdown of the factorized frequency-domain approximation for rapidly evolving signals, can be mitigated but not completely removed by the choice of TDI configuration.
It is important to emphasize that the advantage of shortened time-span TDI configurations identified here is not fundamental. In principle, more complete frequency-domain models that properly account for time-dependent delays, such as those developed in \cite{Marsat:2020rtl}, can mitigate aliasing and leakage effects for any TDI scheme. However, within the commonly adopted factorized frequency-domain framework, shorter-span configurations exhibit improved robustness, as they reduce the breakdown of the approximation when the GW frequency evolves significantly over the delay time.

In addition to the frequency-domain analysis, we also conduct inference using time-domain signal modeling, with results shown in magenta (hybrid Relay $8L$) and grey (PD4L) in Fig.~\ref{fig:corner_mbbh}. The significant overlap between the two sets of contours confirms that both TDI configurations extract consistent information from the GW signal. In this case, the injected values are well recovered across all parameters. Time-domain modeling is achieved by directly implementing TDI operations in the time series, which avoids approximations inherent in frequency-domain modeling. This approach is more accurate for short-duration signals and remains computationally feasible for such cases. However, for long-duration signals, the computational cost of time-domain inference becomes substantial.

In this inference, we compare TDI configurations with effective delays of $8L$ and $4L$, but similar conclusions are expected to apply to other schemes. While shorter-delay TDI configurations help suppress aliasing at high frequencies, they cannot eliminate it entirely. Frequency aliasing becomes significant when the dimensionless frequency parameter $u > 0.1$, due to fundamental limitations in the TDI formalism. This degradation is especially relevant for fast-evolving signals and can adversely affect frequency-domain inference accuracy. In such regimes, GPU-accelerated time-domain inference \cite{Garcia-Quiros:2025usi} may provide more efficient and robust solutions.

\section{Conclusion and Discussion} \label{sec:conclusions}

In this study, we revisit the correlations among orthogonal channels in representative second-generation TDI configurations. While generator-based decompositions suggest that different TDI observables are interdependent and derivable from a common set of generators in the idealized limit, we extend this framework by providing a quantitative assessment that incorporates realistic unequal arm lengths. By evaluating sky-averaged GW response functions, instrumental noise spectra, and responses to a representative MBBH signal, we demonstrate that the near-complete correlations expected in the ideal case are partially broken under realistic conditions. Our results not only clarify the extent of generator-level redundancy in practice but also highlight that highly correlated TDI configurations can exhibit distinct behaviors in practical data analysis.

Due to variations in TDI path geometries, different configurations exhibit distinct null frequencies and effective time spans. In the high-frequency regime, these differences lead to variations in modulation patterns and aliasing effects in the GW waveforms. Among the configurations considered, PD4L stands out as particularly advantageous for frequency-domain signal modeling due to its compact $4L$ time span and minimal null frequencies across all three optimal channels. Our parameter inference study confirms that PD4L yields relatively more accurate parameters compared to the hybrid Relay $8L$ configuration. This superior performance arises from PD4L’s shorter span, which reduces high-frequency artifacts and improves waveform fidelity, although it cannot completely eliminate the frequency aliasing inherent in the TDI formalism. Nevertheless, our time-domain analysis indicates that both configurations preserve the essential physical information, underscoring the potential of time-domain methods for analyzing rapidly evolving GW signals at high frequencies and revealing the redundancy among TDI configurations.

It is important to note that our study assumes all six inter-S/C links are fully functional, allowing construction of three orthogonal channels that capture the complete GW signal. However, in realistic scenarios where one or more links may fail, orthogonal channels can no longer be constructed. In such cases, alternative TDI combinations using the remaining functional links must be developed to suppress laser noise and recover the signal. This highlights the need for resilient and flexible TDI designs.

In summary, our results underscore the critical role that TDI configuration plays in both signal modeling and parameter estimation. Configurations with shorter delay structures are particularly effective in the high-frequency regime, while time-domain methods offer complementary advantages in preserving signal integrity. Future work should explore the impact of longer-duration signals, hybrid TDI schemes tailored for specific frequency bands, and the use of Bayesian model selection to optimize data analysis strategies in space-based GW detection.

\begin{acknowledgments}

G.W. acknowledges Alex Nitz and Shichao Wu for helpful discussions.
G.W. was supported by the National Key R\&D Program of China under Grant No. 2021YFC2201903 and NSFC Grant No. 12575058.
This work are performed by using the python packages \textsf{numpy} \cite{harris2020array}, \textsf{scipy} \cite{2020SciPy-NMeth}, \textsf{pandas} \cite{pandas}, \textsf{MultiNest} \cite{Feroz:2008xx} and \textsf{PyMultiNest} \cite{Buchner:2014nha}, and the plots are make by utilizing \textsf{matplotlib} \cite{Hunter:2007ouj}, \textsf{GetDist} \cite{Lewis:2019xzd}, and \textsf{HEALPix} \cite{2005ApJ...622..759G,Zonca2019}. The GW signals are generated using \textsf{LALSuite} \cite{lalsuite,swiglal} and \textsf{PyCBC} \cite{alex_nitz_2024_10473621}.

\end{acknowledgments}

\appendix

\section{TDI observables from self-splicing} \label{sec:tdi_in_Vallisneri_2005}

\begin{figure*}[htbp]
\includegraphics[width=0.48\textwidth]{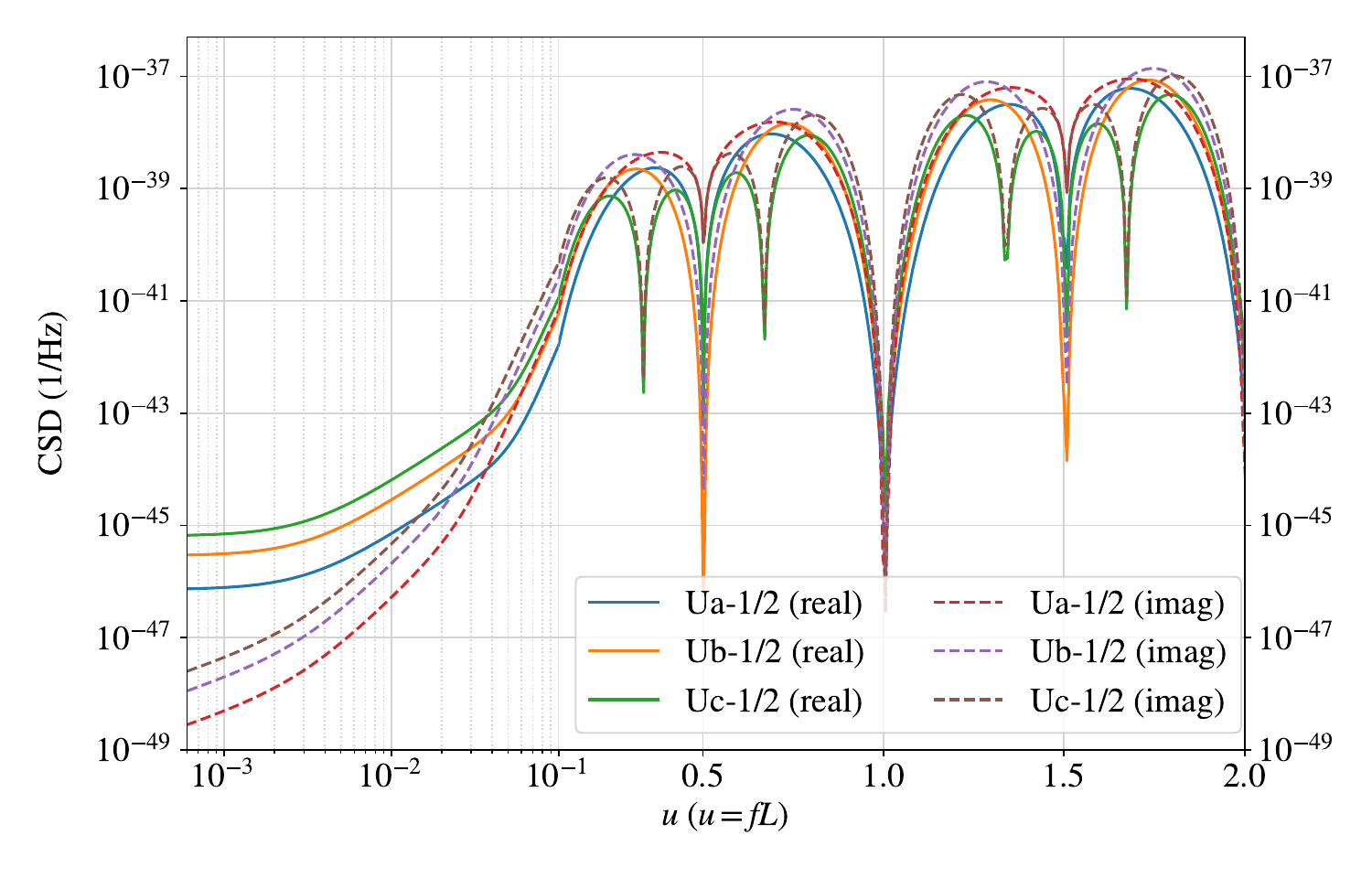}
\includegraphics[width=0.48\textwidth]{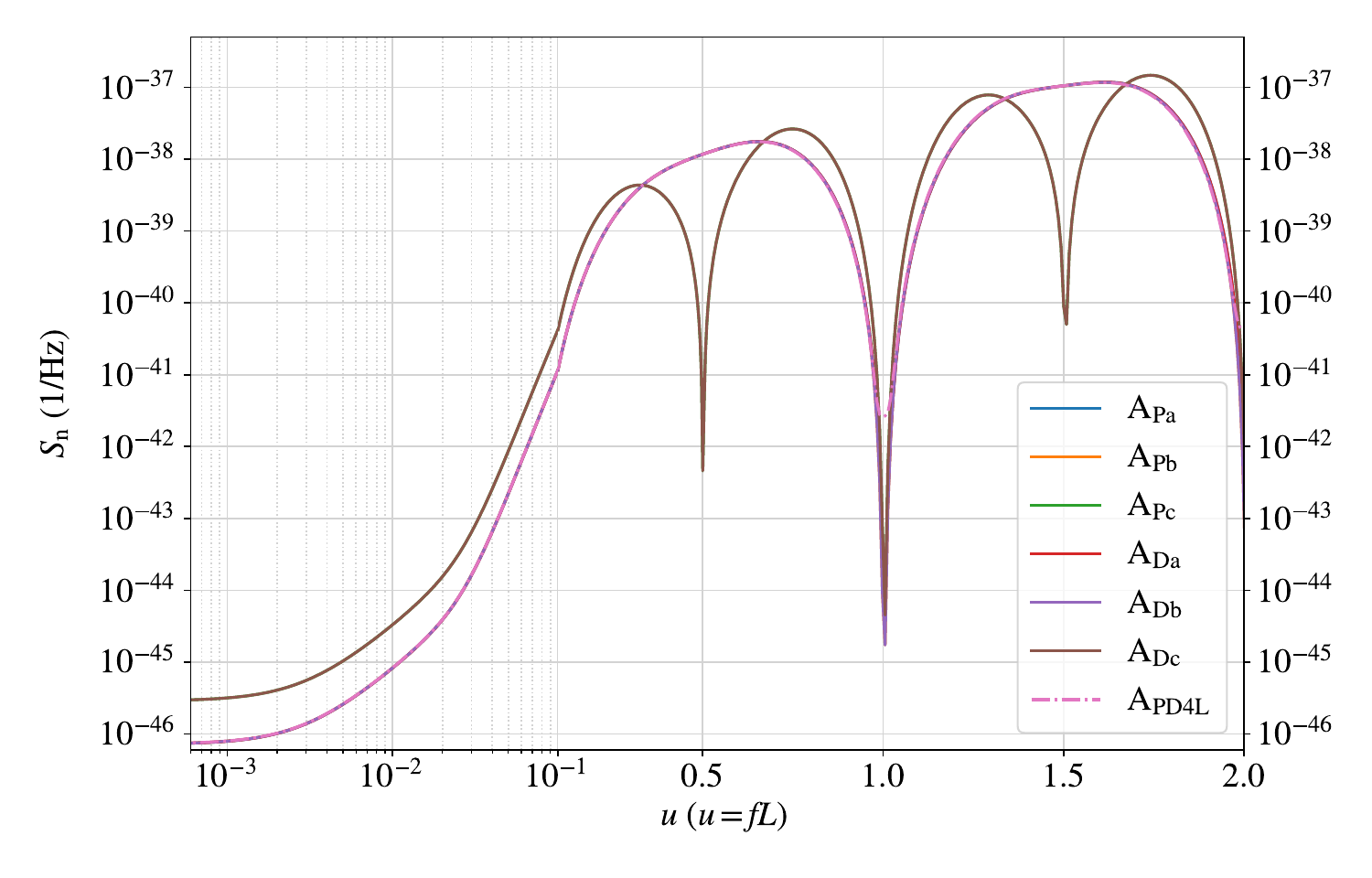}
\includegraphics[width=0.48\textwidth]{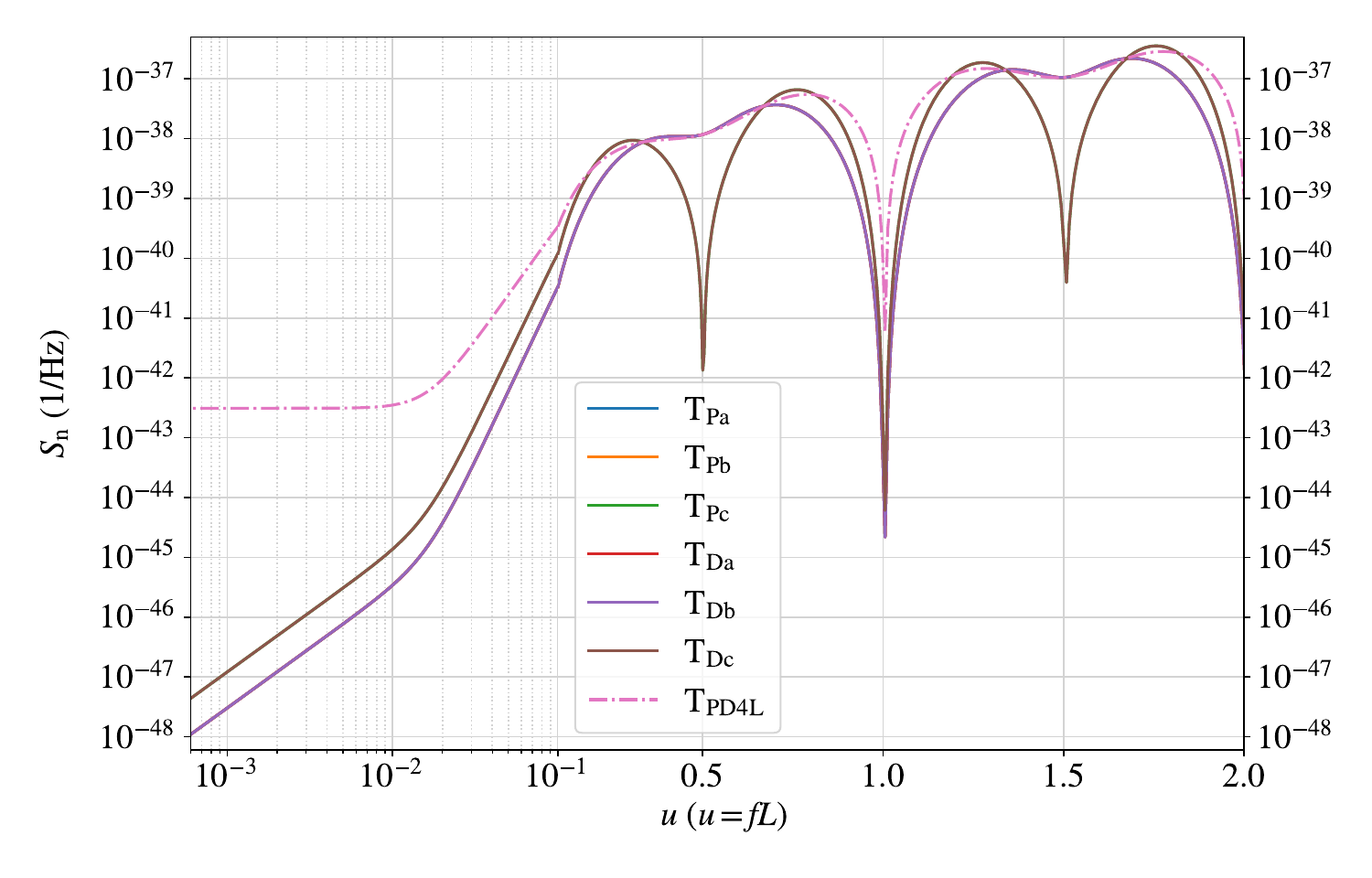}
\includegraphics[width=0.48\textwidth]{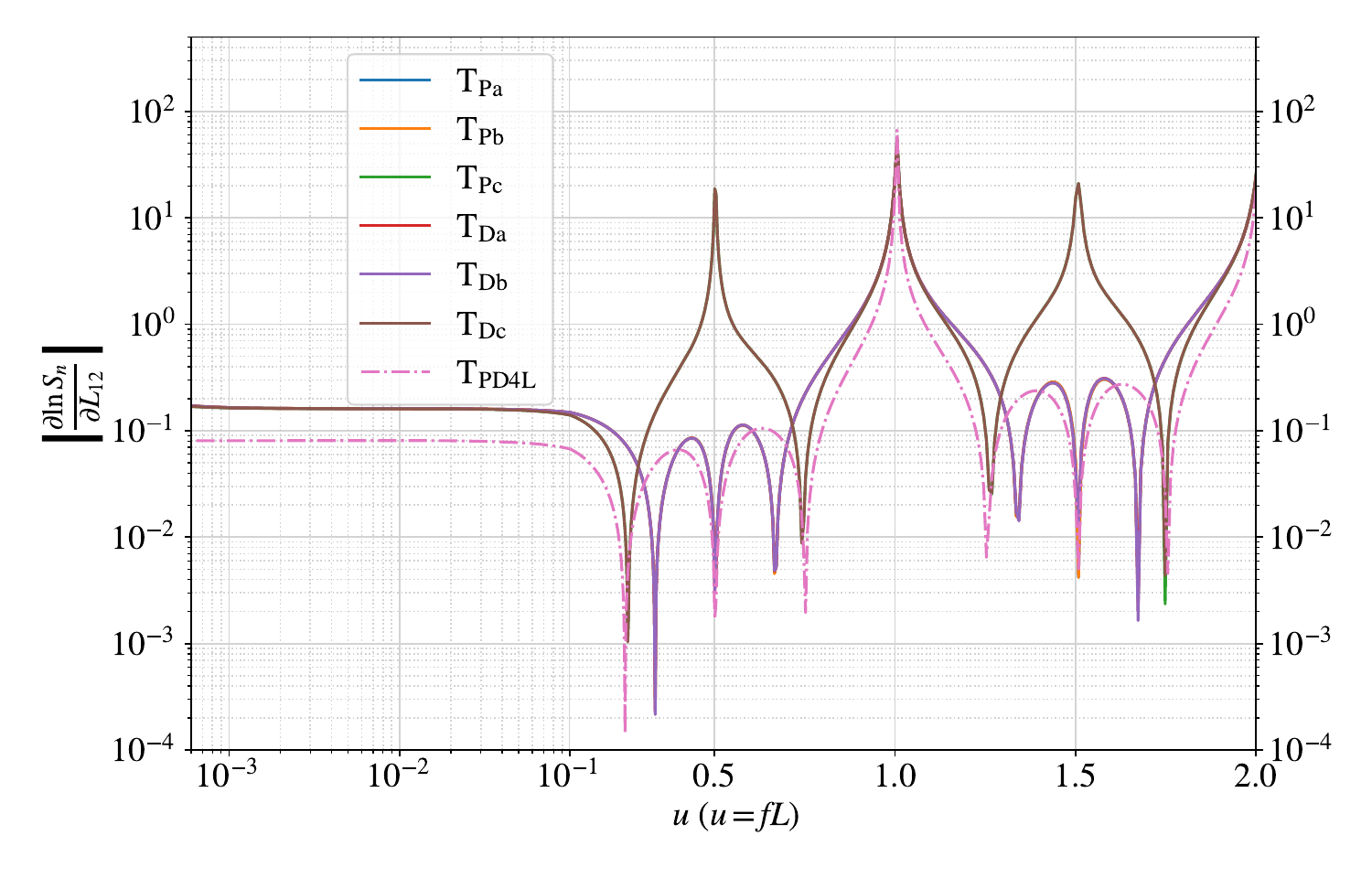}
\caption{\label{fig:dSn_dLij_U_P_D_abc}
Noise properties of selected second-generation TDI configurations from \citet{Vallisneri:2005ji}. Upper left: real and imaginary components of the CSDs for U-type observables. Upper right: PSDs of the science (A) channels for E-type and P-type configurations, compared with PD4L. Lower left: PSDs of the corresponding T channels. Lower right: the derivatives of the T-channel PSDs with respect to arm length $L_{12}$.
}
\end{figure*}

\citet{Vallisneri:2005ji} developed a suite of second-generation TDI configurations by self-splicing first-generation TDI observables. Among them, alternative Michelson-type observables can reduce their minimal null frequencies to $u = m/(2L)$, although they do not achieve the optimal suppression of nulls.
Three Relay-type (U-type) second-generation TDI configurations are provided in Eq. (A6) of \cite{Vallisneri:2005ji}:
\begin{eqnarray}
\text{Ua-1:} (5L) \ & \overrightarrow{21323} \ \overleftarrow{3 1 2 3 1 2} \ \overrightarrow{2 3 2 1 3 } \ \overleftarrow{3 2 3 2}, \label{eq:Ua-1_path} \\
\text{Ub-1:} (6L) \ & \overrightarrow{2 1 3 2 3} \ \overleftarrow{3 1 2} \ \overrightarrow{2 3 2 1 3} \ \overleftarrow{3 2 3 1 2 3 2}, \\
\text{Uc-1:} (7L) \ &  \overrightarrow{2 1 3 2 3 2 1 3} \ \overleftarrow{3 2 3 1 2} \ \overrightarrow{2 3} \ \overleftarrow{3 1 2 3 2}
\end{eqnarray}
Note that we use S/C indices here, rather than arm indices as in the original reference. The three ordinary observables are obtained by cyclic permutation of the S/C indices. For example, Ua-2 and Ua-3 can be generated from Ua-1 by applying the substitutions $(1 \rightarrow 2, 2 \rightarrow 3, 3 \rightarrow 1)$ and $(1 \rightarrow 3, 2 \rightarrow 1, 3 \rightarrow 2)$, respectively. The Ua group features minimal nulls at $u = m/L$. However, similar to their first-generation Relay scheme, the CSD between the ordinary Ua channels shows significant imaginary components. These components are nonzero even in the equal-arm limit and are comparable to the real parts at high frequencies, as shown in the upper-left panel of Fig. \ref{fig:dSn_dLij_U_P_D_abc}. As a result, orthogonalization of the Ua observables is not feasible, making these configurations less favorable.

Examples of E-type configurations are provided in Eq. (A10) of \cite{Vallisneri:2005ji}, including:
\begin{eqnarray}
\text{Da-1:} (4L) \ &  \overrightarrow{3 2 3 2 1} \ \overleftarrow{1 3 2 3} \ \overrightarrow{3 1} \ \overleftarrow{1 2 } \  \overrightarrow{2 3 1} \ \overleftarrow{1 2 3 2}  \ \overrightarrow{ 2 1 } \ \overleftarrow{ 1 3 } \\
\text{Db-1:} (4L) \ & \overrightarrow{ 2 3 2 3 1} \ \overleftarrow{ 1 2 3 2 } \ \overrightarrow{ 2 1 } \ \overleftarrow{ 1 3 } \ \overrightarrow{ 3 2 1} \overleftarrow{ 1 3 2 3 } \ \overrightarrow{ 3 1 } \ \overleftarrow{ 1 2 }, \\
\text{Dc-1:} (5L) \ & \overrightarrow{3 2 3 1 } \ \overleftarrow{ 1 2 } \ \overrightarrow{2 3 2 1 } \ \overleftarrow{1 3 2 3 } \ \overrightarrow{3 1 } \ \overleftarrow{1 2 3 2 } \ \overrightarrow{ 2 1 } \ \overleftarrow{ 1 3 }.
\end{eqnarray}
Similarly, P-type configurations are described in Eq. (A14):
\begin{eqnarray}
\text{Pa-1:} (4L) \ &  \overrightarrow{1 2 3 2 3} \ \overleftarrow{3 1} \ \overrightarrow{1 2} \ \overleftarrow{2 3 2 1} \  \overrightarrow{1 3 2} \ \overleftarrow{2 1}  \ \overrightarrow{1 3} \ \overleftarrow{3 2 3 1}, \\
\text{Pb-1:} (4L) \ & \overrightarrow{1 3 2 3 2} \ \overleftarrow{2 1} \ \overrightarrow{1 3} \ \overleftarrow{3 2 3 1} \ \overrightarrow{1 2 3} \ \overleftarrow{3 1} \ \overrightarrow{1 2} \ \overleftarrow{2 3 2 1}, \\
\text{Pc-1:} (5L) \ & \overrightarrow{1 3 2 3} \ \overleftarrow{3 1} \ \overrightarrow{1 2 3 2} \ \overleftarrow{2 1} \ \overrightarrow{1 3} \ \overleftarrow{3 2 3 1} \ \overrightarrow{1 2} \ \overleftarrow{2 3 2 1} .
\end{eqnarray}
The noise PSDs of the science (A) channels for both E-type and P-type observables are either identical to or worse than those of the PD4L configuration (exhibiting more nulls at higher frequencies), as shown in the upper-right panel of Fig. \ref{fig:dSn_dLij_U_P_D_abc}. However, their T channels show significantly lower noise than the T$\mathrm{PD4L}$ channel in lower frequency band (lower-left panel). The derivatives of these T-channel PSDs with respect to the arm length $L{12}$, shown in the lower-right panel, indicate that PD4L exhibits more stable spectral behavior across the frequency band.
In summary, the PD4L configuration outperforms the U-type, E-type, and P-type groups in both noise robustness and spectral stability.

\section{Null frequencies in the first-generation TDI observables} \label{sec:null_in_1st_TDI}

\begin{figure}[htbp]
\includegraphics[width=0.48\textwidth]{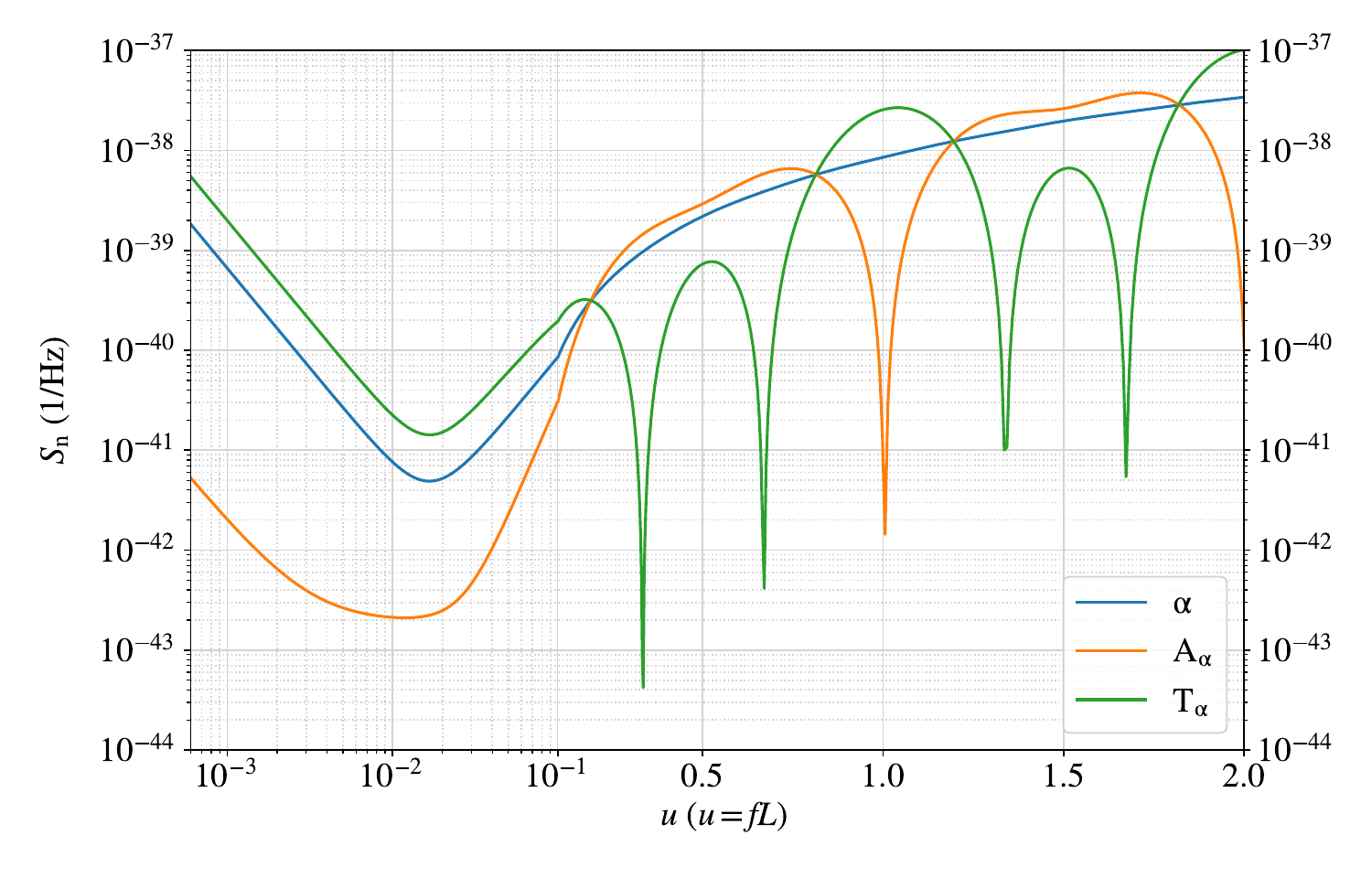}
\includegraphics[width=0.48\textwidth]{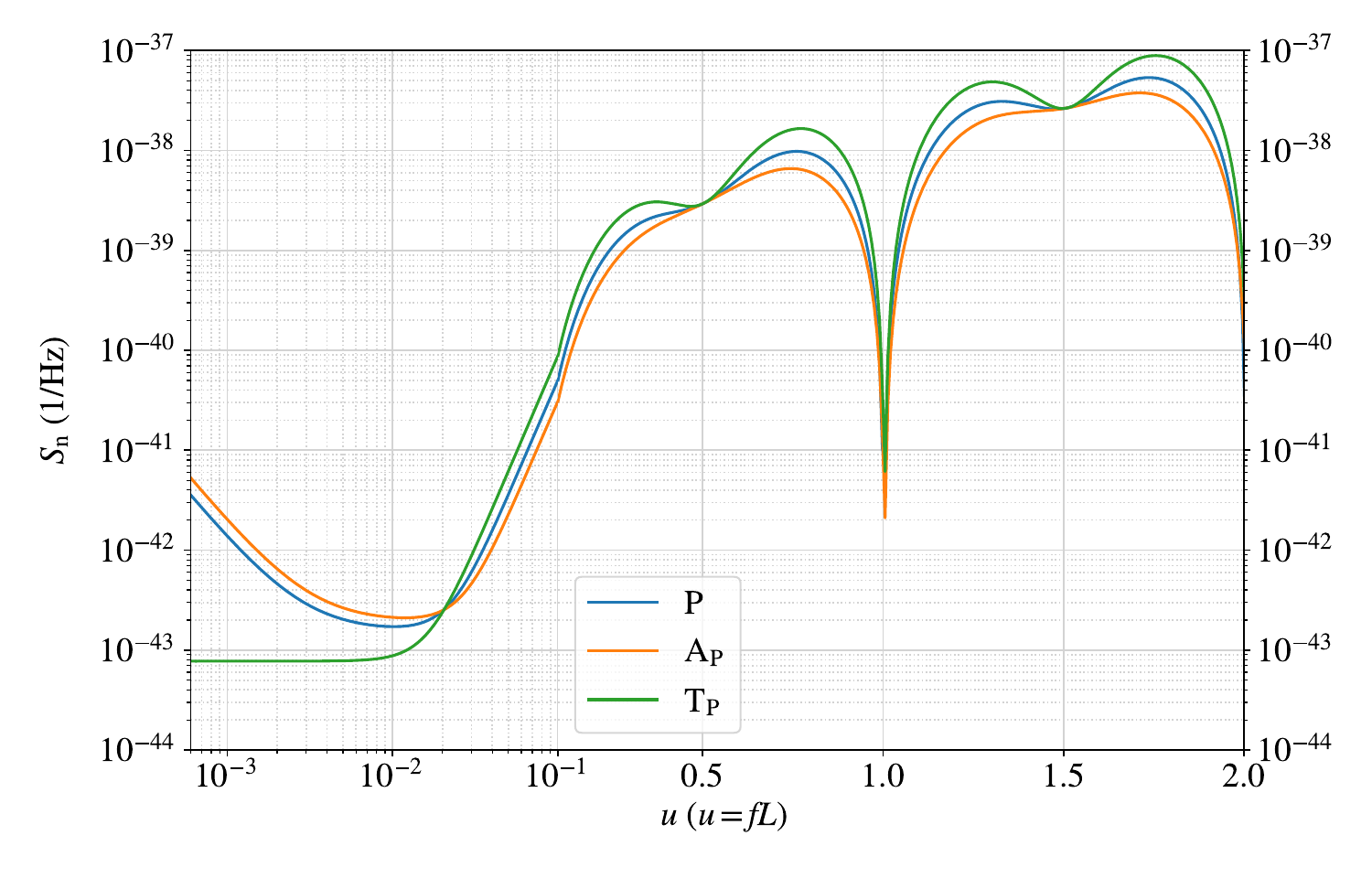}
\caption{\label{fig:Sn_alpha_Monitor_AT} Noise PSDs of the first ordinary channel and its orthogonal combinations for the Sagnac configuration (upper panel) and Beacon (lower panel). }
\end{figure}

The first-generation TDI can be categorized into five configurations comprising 18 channels: Michelson (X, Y, Z), Sagnac ($\alpha$, $\beta$, $\gamma$), Relay (U, V, W, $\mathrm{\overline{U}}$, $\mathrm{\overline{V}}$, $\mathrm{\overline{W}}$), Beacon (P, Q, R), and Monitor (D, F, G) \cite{1999ApJ...527..814A,2000PhRvD..62d2002E,Tinto:2020fcc}. Except for the Michelson configuration, the first channel of each of the remaining four can be expressed as \cite{Vallisneri:2005ji}:
\begin{align} 
  \text{Sagnac-}\mathrm{\alpha:} & \ \overrightarrow{1 2 3 1} \ \overleftarrow{1 2 3 1}, \label{eq:alpha_path}  \\
  \text{Relay-U:} & \ \overrightarrow{3 1 2 3 2} \ \overleftarrow{2 1 3 2 3}, \label{eq:U_path} \\
  \text{Relay-} \overline{\rm U} \mathrm{:} & \ \overrightarrow{ 2 3 2 1 3 } \ \overleftarrow{ 3 2 3 1 2 }. \label{eq:UUbar_path} \\
  \text{Beacon-P:} & \ \overrightarrow{1 2 3 2} \ \overleftarrow{2 1} \ \overrightarrow{1 3} \ \overleftarrow{3 2 3 1}, \label{eq:P_path} \\
 \text{Monitor-D:} & \ \overrightarrow{2 3 2 1} \ \overleftarrow{1 3 2 3} \ \overrightarrow{3 1} \ \overleftarrow{1 2}. \label{eq:D_path}
\end{align}
The orthogonal transformation is not applicable to the Relay configuration because the imaginary parts of the cross-correlations between its ordinary channels are nonzero in the equal-arm case and not significantly smaller than the real parts in the unequal-arm case \cite{Vallisneri:2007xa}. The Beacon and Monitor configurations are essentially equivalent, as discussed in \cite{Wang:2020fwa}.
The PSDs of the Sagnac-$\alpha$ channel and its associated orthogonal channels A$_\alpha$ and T$_\alpha$ are shown in the upper panel of Fig.~\ref{fig:Sn_alpha_Monitor_AT}, while the PSDs for the Beacon channel P and its orthogonal counterparts A$\mathrm{P}$ and T$\mathrm{P}$ are displayed in the lower panel. The Sagnac-$\alpha$ channel, shown as the blue curve, exhibits no null frequencies and is dominated by OMS noise across the entire frequency band. This behavior arises from the asymmetry in OMS noise between links from S/C$i$ to S/C$j$ and their reverse directions, as well as from the non-repeating paths of the clockwise and counterclockwise loops. However, null frequencies emerge in its orthogonal channels: the science channels (A$_\alpha$) exhibit nulls at normalized frequencies $u = m$ ($m = 1, 2, 3, \ldots$), while the null channel (T$_\alpha$) has nulls at $u = m/3$ for non-integer $u$. Similarly, the Beacon (P, Q, R) and Monitor (D, F, G) configurations also show null frequencies at $u = m$ in both their ordinary and optimal channels, as illustrated in the lower panel of Fig.~\ref{fig:Sn_alpha_Monitor_AT}.

\section{Angular GW Response Patterns Across Frequencies} \label{sec:Mollweide_maps}

\begin{figure*}[htb]
\includegraphics[width=0.48\textwidth]{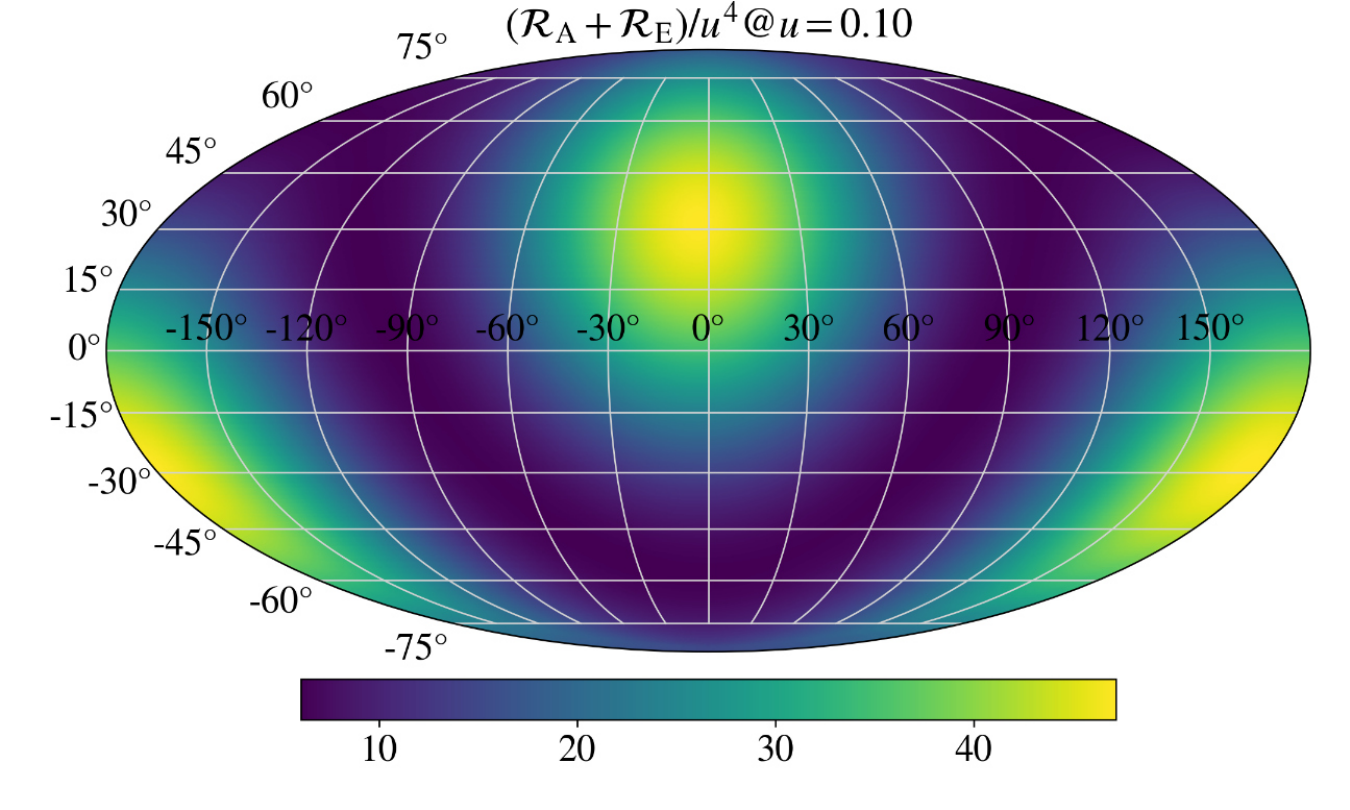}
\includegraphics[width=0.48\textwidth]{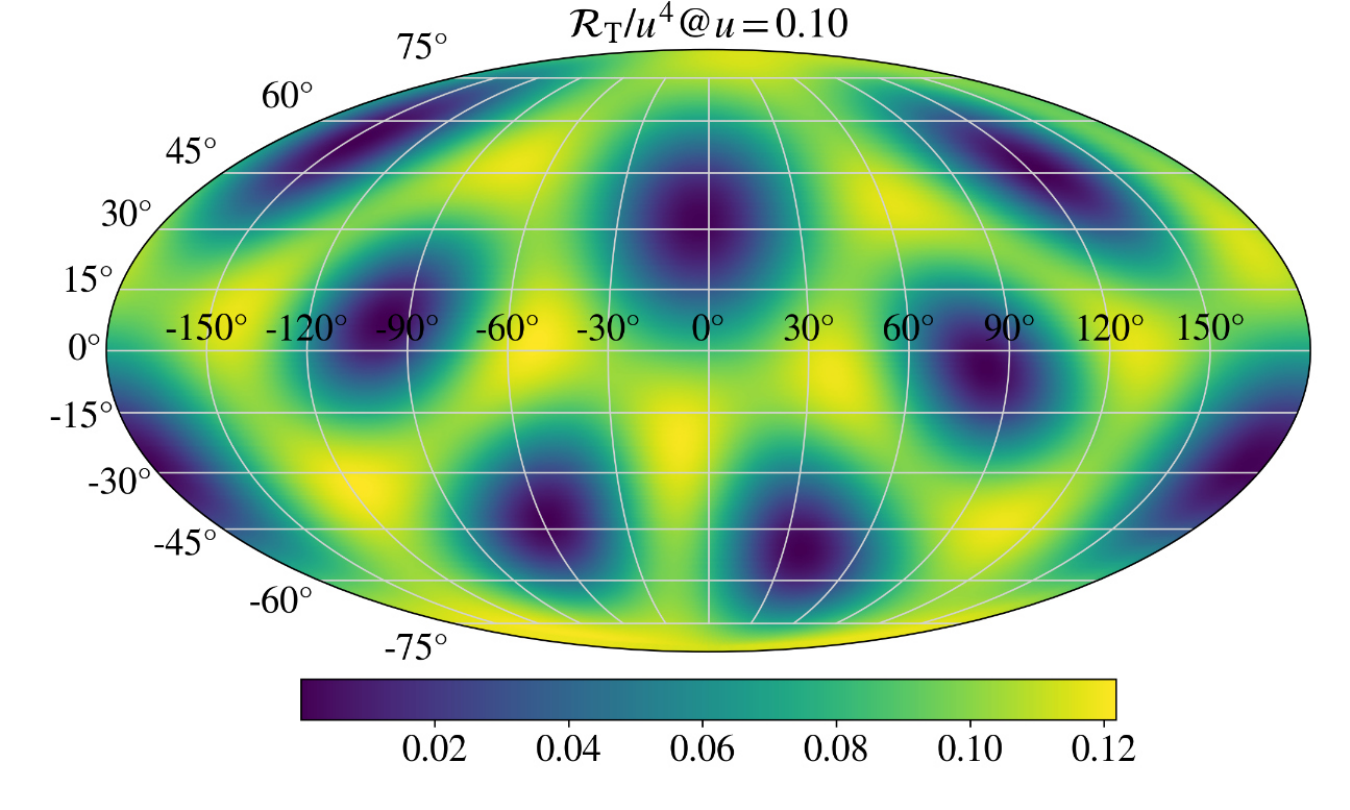}
\includegraphics[width=0.48\textwidth]{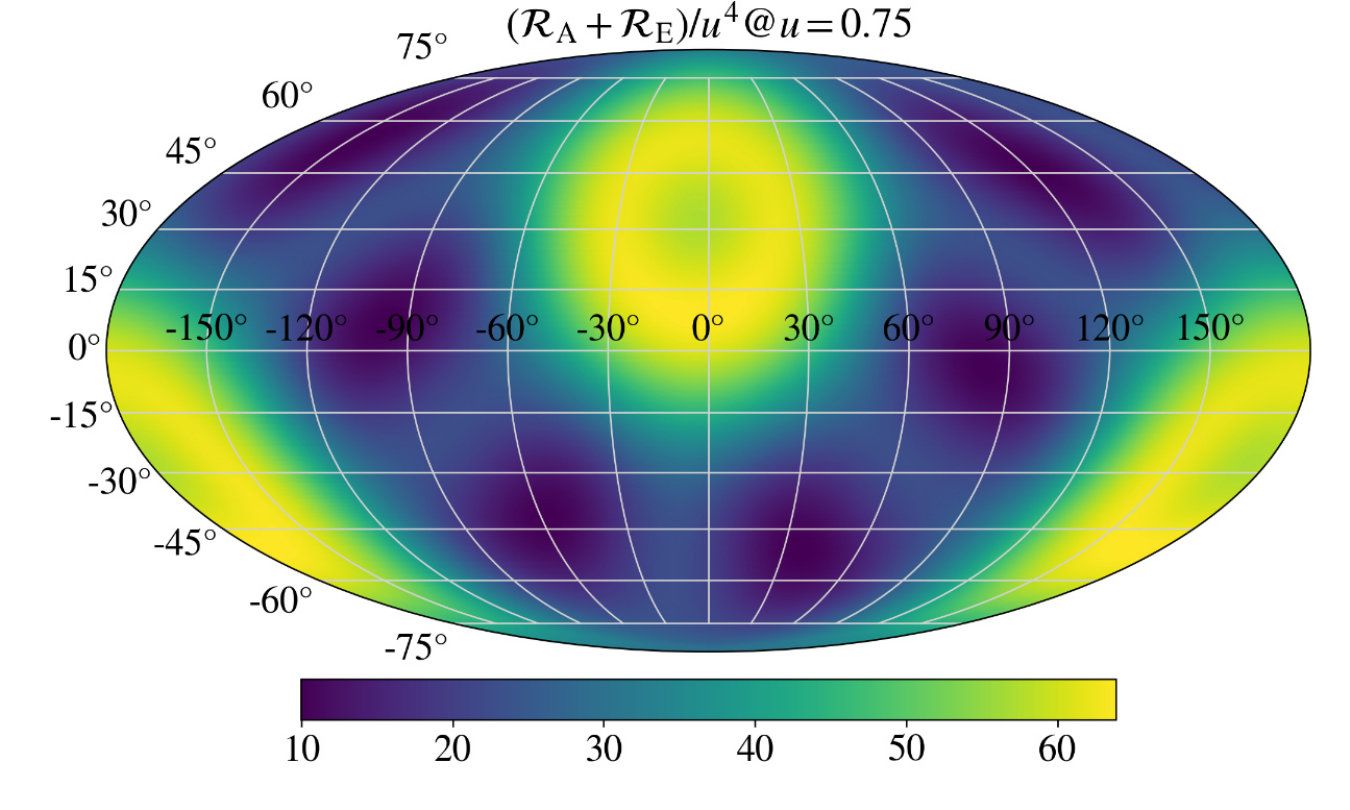}
\includegraphics[width=0.48\textwidth]{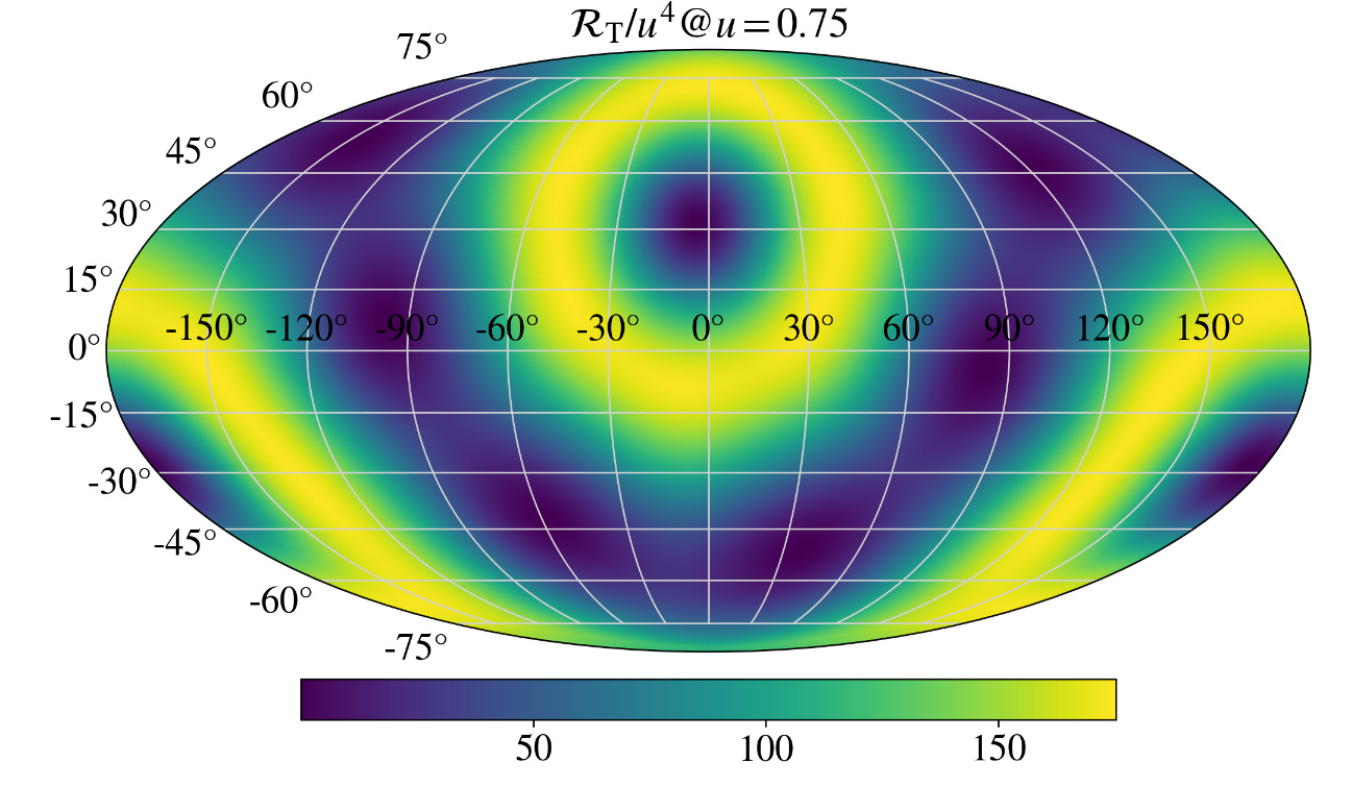}
\includegraphics[width=0.48\textwidth]{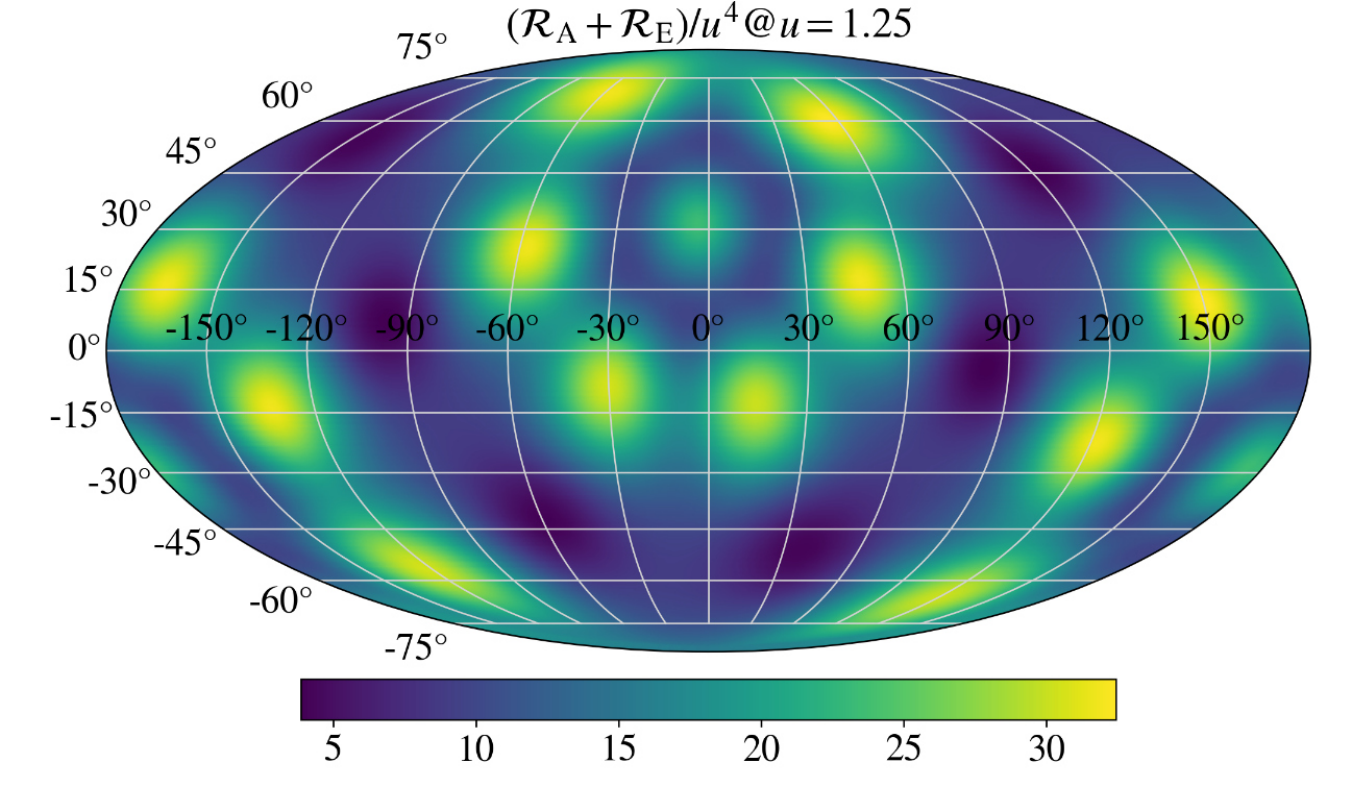}
\includegraphics[width=0.48\textwidth]{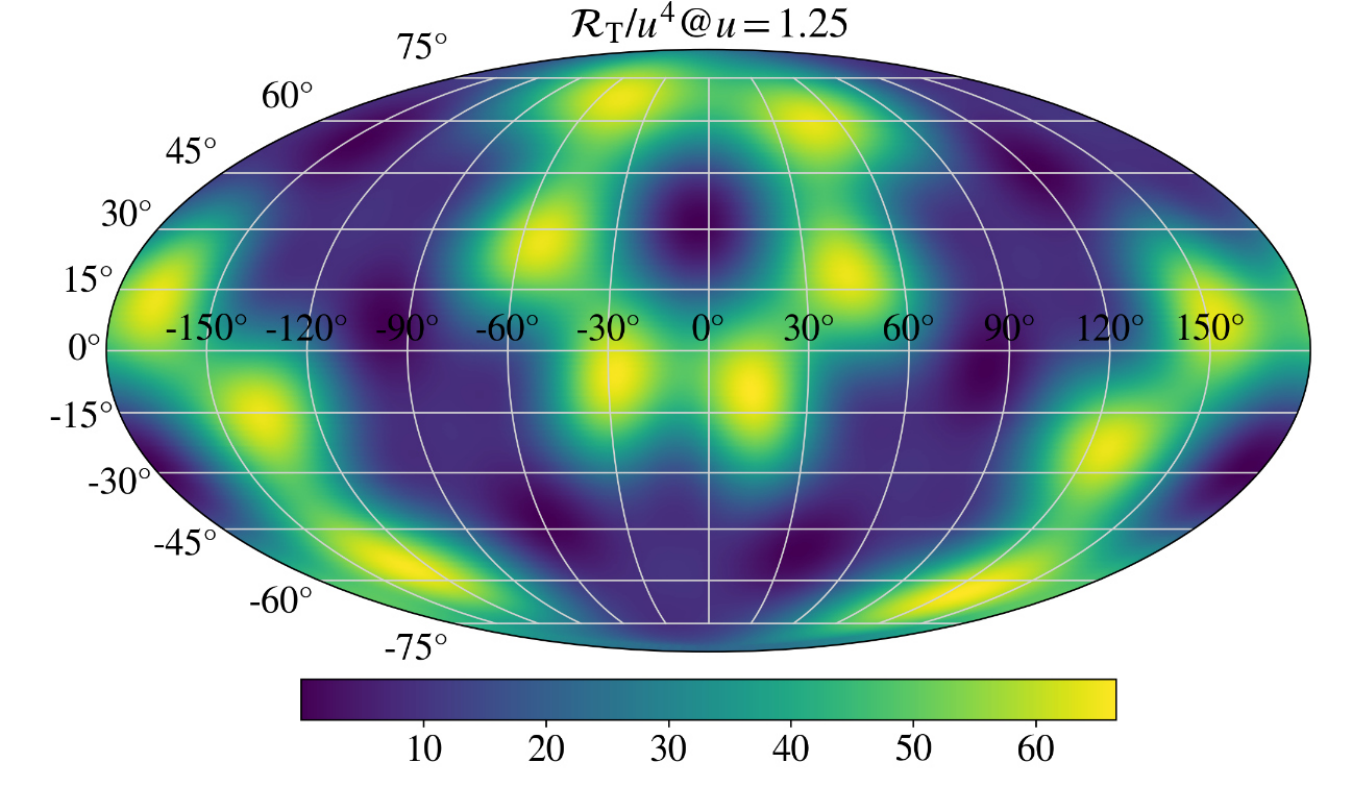}
\caption{\label{fig:resp_Mollweide} Mollweide sky maps (ecliptic coordinates) of the instantaneous GW response of PD4L configuration at three dimensionless frequencies, $u = fL \in \{0.10,\,0.75,\,1.25\}$. The left column shows the combined science channel response $\left(R_\mathrm{A}+R_\mathrm{E} \right)/u^4$; the right column shows the null channel response $R_\mathrm{T}/u^4$.
}
\end{figure*}

\begin{figure*}[htbp]
\includegraphics[width=0.45\textwidth]{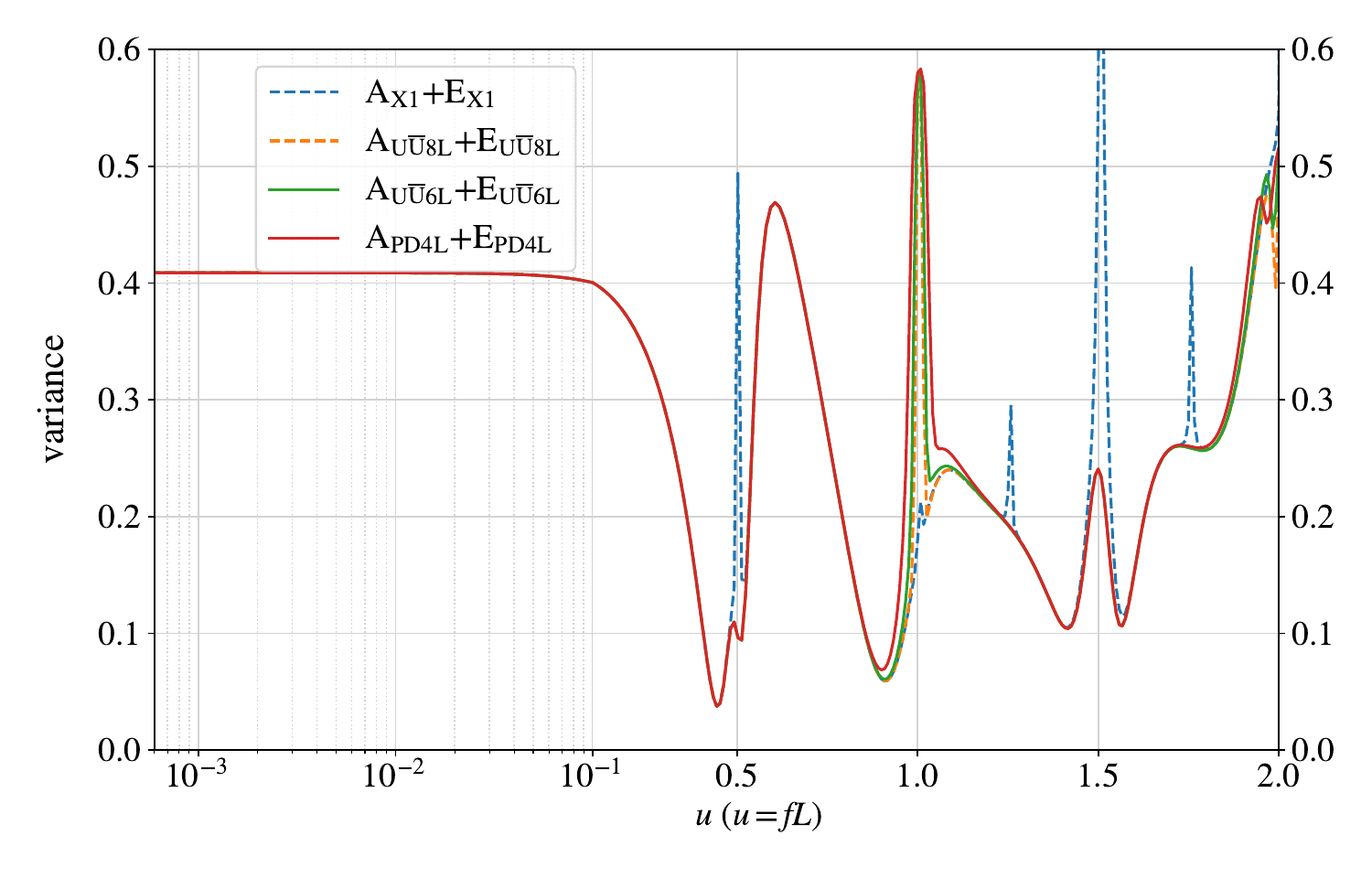}
\includegraphics[width=0.45\textwidth]{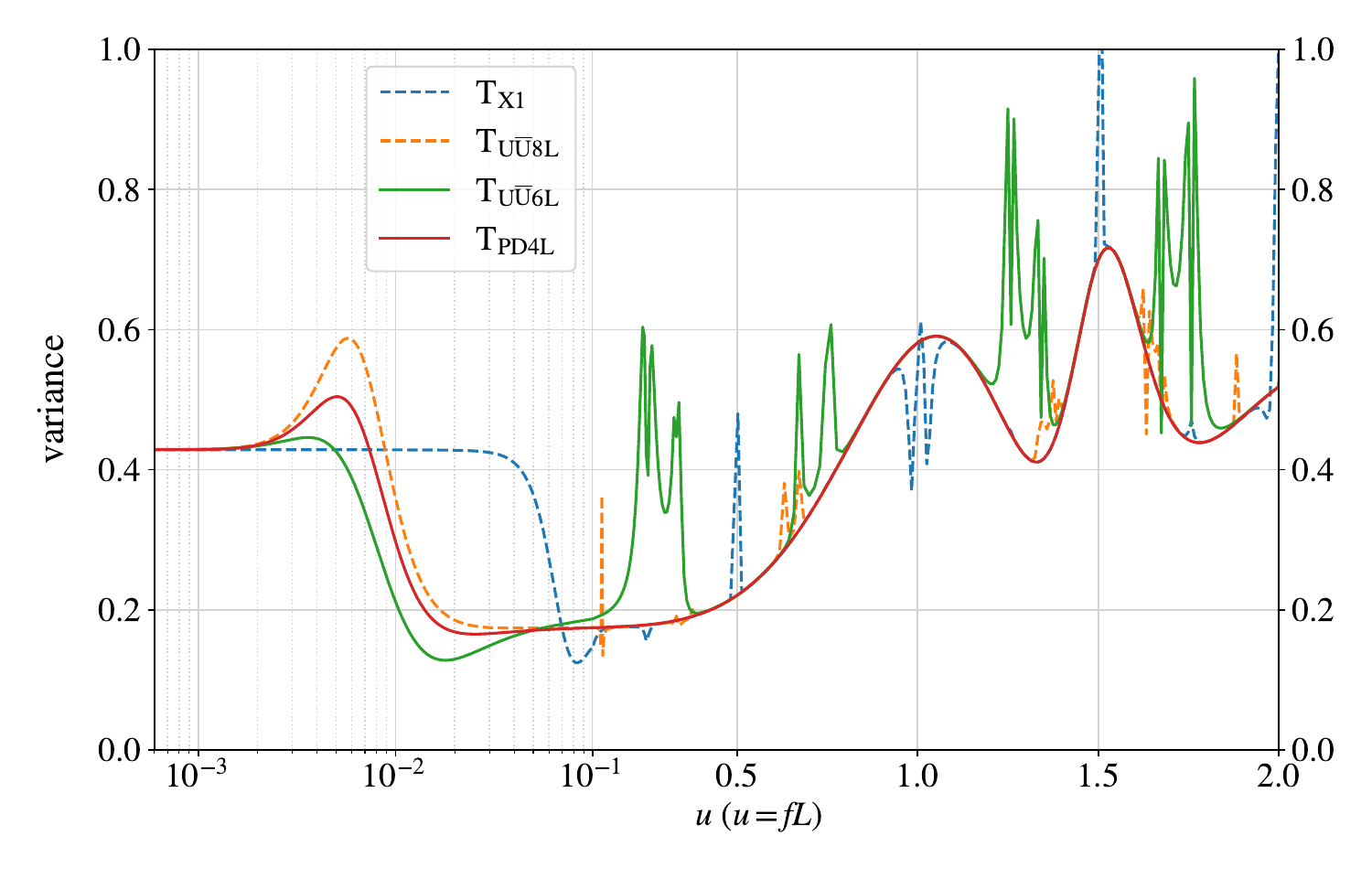}
\includegraphics[width=0.45\textwidth]{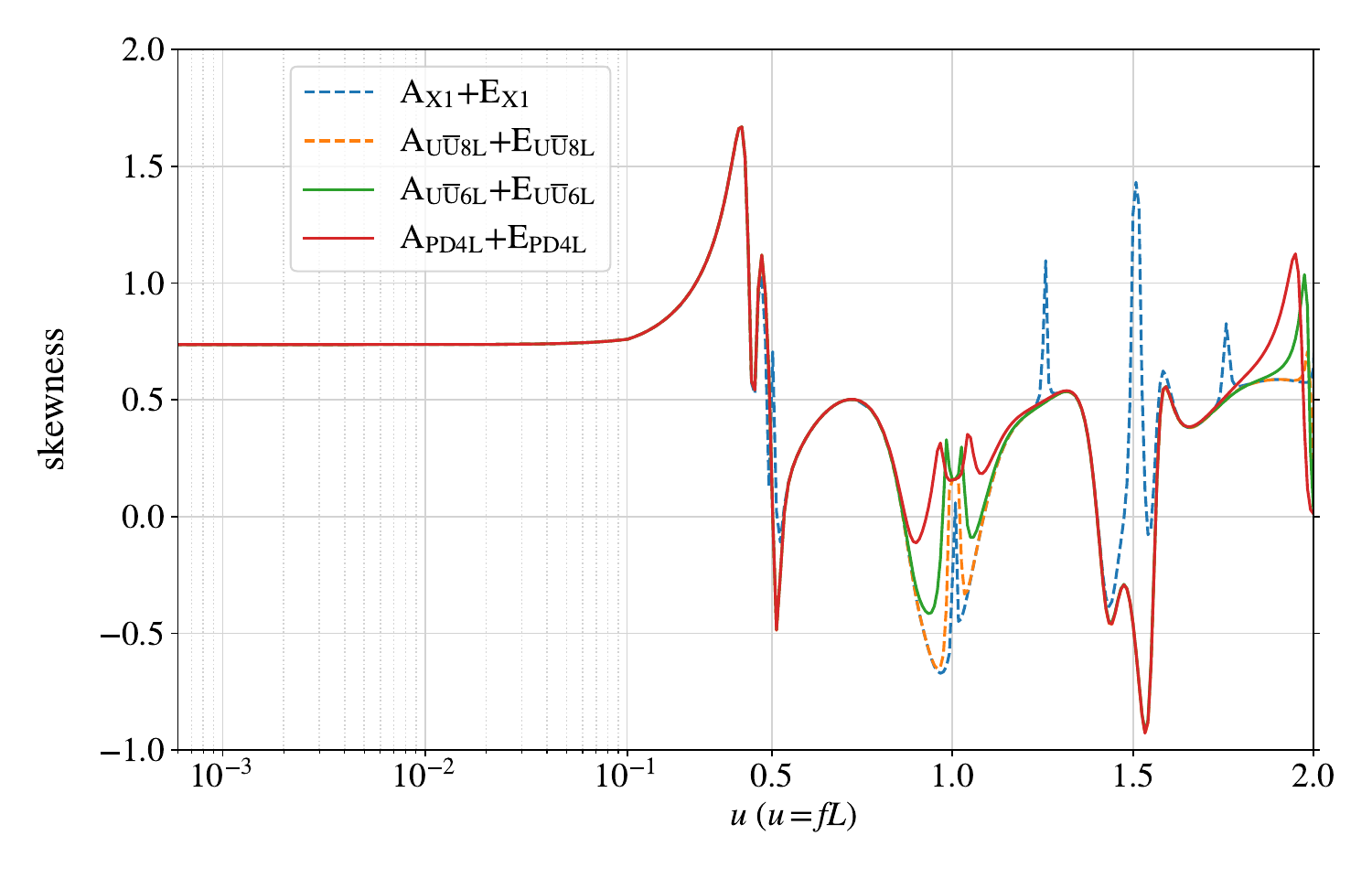}
\includegraphics[width=0.45\textwidth]{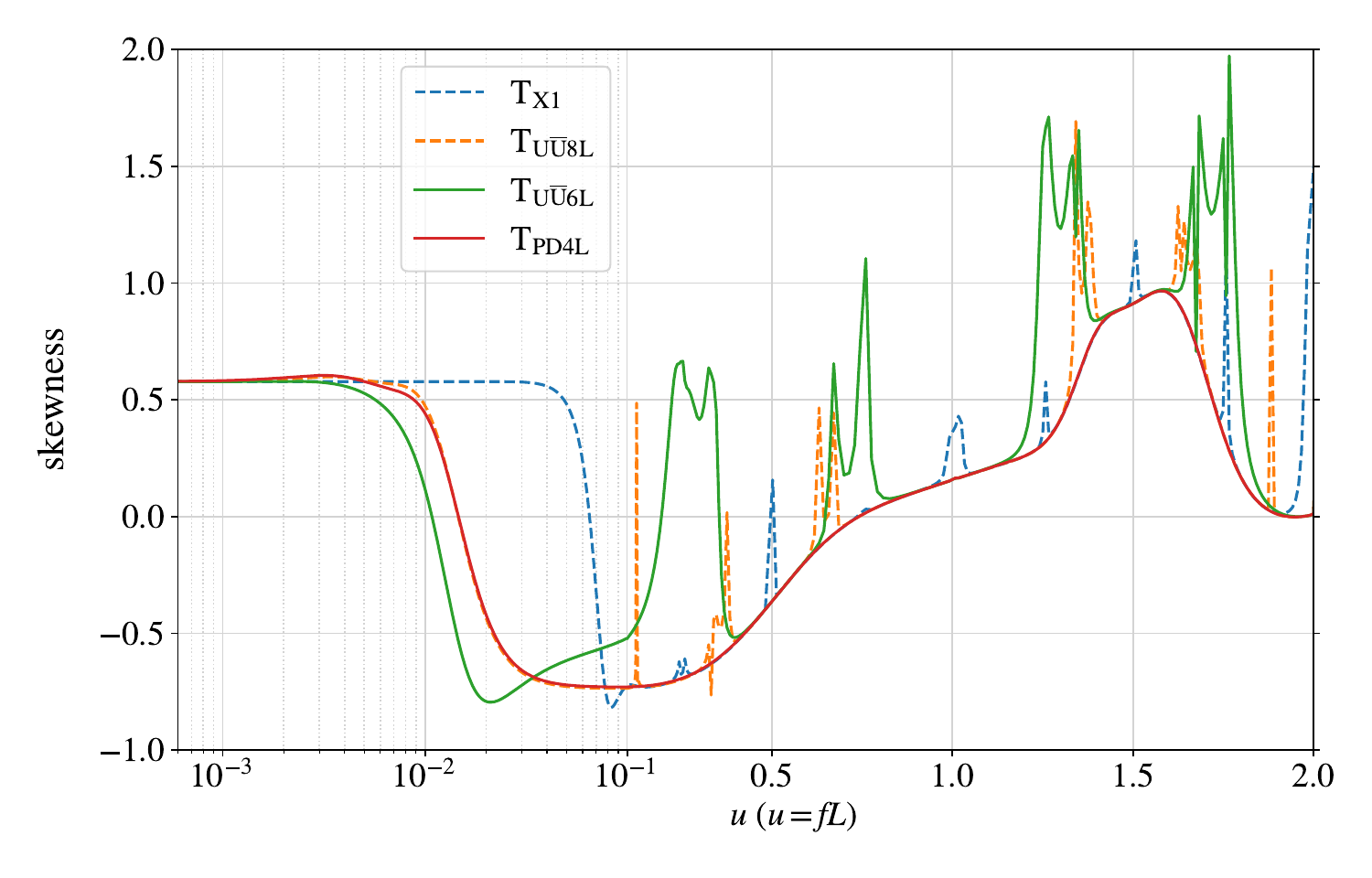}
\includegraphics[width=0.45\textwidth]{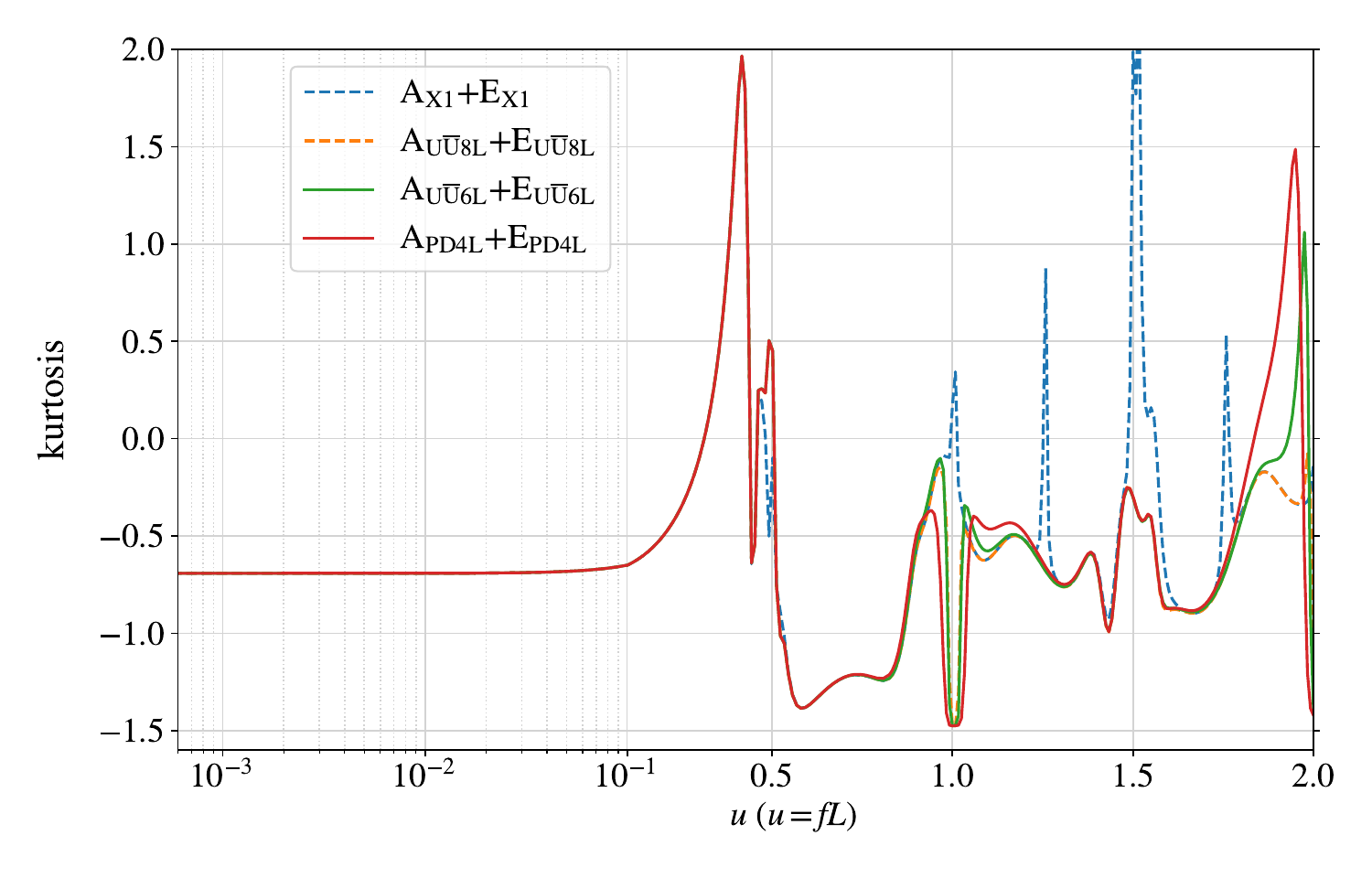}
\includegraphics[width=0.45\textwidth]{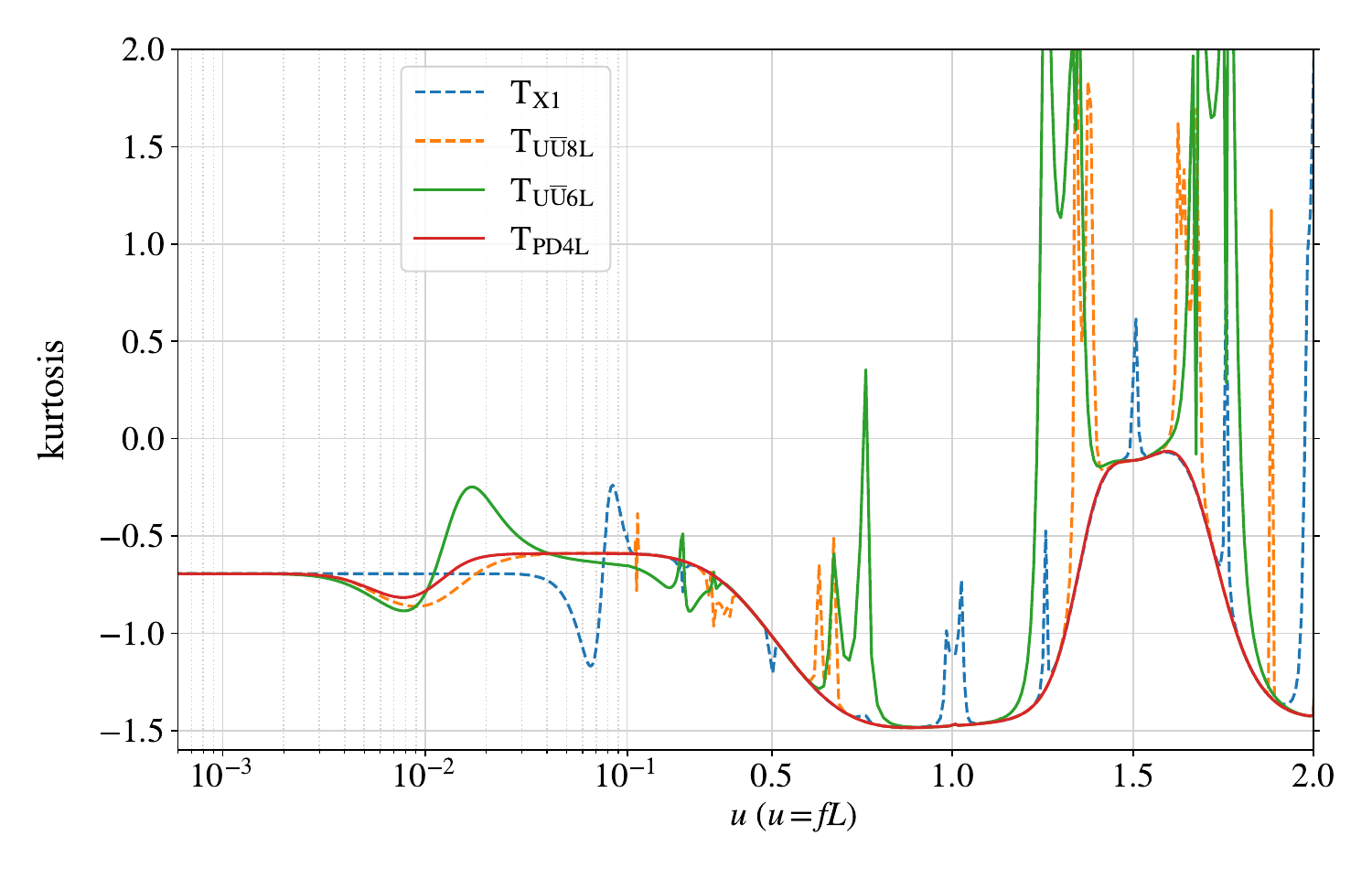}
\caption{\label{fig:resp_variance_skewness_kurtosis}
Normalized variance (top), skewness (middle), and kurtosis (bottom) of selected TDI channels. Left panels show the combined science channels (A+E), while right panels show the null (T) channels. These results complement Fig.~\ref{fig:resp_avg} and exhibit consistent trends. At low frequencies ($u < 0.1$), all science channels display nearly constant variance, as expected from the long-wavelength approximation. At higher frequencies, strong frequency dependence emerges, particularly near response nulls. The spikes observed around nulls are primarily due to numerical artifacts.
}
\end{figure*}

Fig. \ref{fig:resp_Mollweide} shows the sky maps of the instantaneous GW response at three dimensionless frequencies, $u = fL \in \{0.10,\,0.75,\,1.25\}$. The left column shows the combined science channel response $\left(R_\mathrm{A}+R_\mathrm{E} \right)/u^4$; the right column shows the null-channel $R_\mathrm{T}/u^4$.
At low frequency ($u=0.10$, long-wavelength regime), the A+E response is close to a quadrupolar pattern and sensitive to normal direction of constellation, while the $T$ shows more irregular pattern with response is significantly suppressed, illustrating T is noise-dominated at small $u$. At intermediate frequency ($u=0.75$), finite-arm-length effects broaden sensitive areas and slightly decrease of the sky-response variance as shown in Fig. \ref{fig:resp_variance_skewness_kurtosis}. And T channel become sensitive to the GW, especially for the direction around the normal direction of constellation, At higher frequency ($u=1.25$), both patterns of A+E and T become more intricate, with additional nodes and alternating bright/dim regions. 

The normalized variance of the sky-averaged GW response is defined to assess directional sensitivity variation as:
\begin{align}
\mathrm{Var}_\mathrm{A+E} = & \frac{ \mathrm{var} \left[  F^2_\mathrm{A} (f, \beta, \lambda) + F^2_\mathrm{E} (f, \beta, \lambda) \right] }{ \mathrm{mean} \left[ F^2_\mathrm{A} (f, \beta, \lambda) + F^2_\mathrm{E} (f, \beta, \lambda) \right] }, \\
\mathrm{Var}_\mathrm{T} = & \frac{ \mathrm{var} \left[  F^2_\mathrm{T} (f, \beta, \lambda) \right] }{R_\mathrm{T}}
\end{align}
The results are shown in the upper row of Fig.~\ref{fig:resp_variance_skewness_kurtosis}. For the science channels (left panels), the normalized variance of A+E is nearly identical across all configurations over most of the frequency range. Noticeable differences only appear near null frequencies, where numerical artifacts introduce sharp fluctuations. In the low-frequency regime ($u < 0.1$), the variance remains nearly constant, consistent with the long-wavelength approximation. At higher frequencies, however, the variance develops a strong frequency dependence.
For the null channels (right panels), the behavior is more distinct among different configurations. Larger variations and more frequent spikes are observed near nulls, again mainly caused by numerical artifacts.
The skewness and kurtosis of the GW response are shown in the middle and bottom rows, respectively. These higher-order statistics provide complementary diagnostics of the distribution shape of the directional response. In general, TDI channels with fewer null frequencies exhibit smoother behavior in all three statistics (variance, skewness, and kurtosis).

\nocite{*}
\bibliography{apsref}

\end{document}